\documentclass[reprint,superscriptaddress,
longbibliography,aps,prx,twocolumn]{revtex4-2}
\usepackage[T1]{fontenc}
\usepackage{graphicx}
\usepackage{dcolumn}
\usepackage{amsmath}
\usepackage{amssymb}
\usepackage{bm}
\usepackage{mathtools} 
\usepackage{mathrsfs}
\usepackage{bbm}
\usepackage[normalem]{ulem}
\usepackage{physics}
\usepackage[dvipsnames,x11names]{xcolor}
\usepackage{xr-hyper}
\usepackage{footnote}
\usepackage{csquotes}
\usepackage[colorlinks=true,allcolors=blue]{hyperref}
\usepackage{comment}
\excludecomment{temphide}

\usepackage{orcidlink}

\newcommand{\fref}[2][]{\hyperref[#2]{\ref*{#2}#1}}

\newcommand{\externalPotential}{\Phi}
\newcommand{\jointpdf}{{P}}
\newcommand{\marginaldistro}{\rho} 
\newcommand{\allxs}{\mathbf{X}} 
\newcommand{\allus}{\mathbf{U}} 
\newcommand{\allthetas}{\boldsymbol{\Theta}}

\newcommand{\qsspeed}{v_{\mathrm{QS}}}
\newcommand{\tir}{t_\rho}
\newcommand{\ellr}{\ell_\rho}
\newcommand{\vloc}{\bar{v}}
\newcommand{\rcut}{r_\mathrm{cut}}
\newcommand{\cms}{c_\mathrm{MS}}
\newcommand{\xpe}{{x_\perp}}
\newcommand{\xvpa}{{\xv_\parallel}}
\newcommand{\gradpa}{{\grad_\parallel}}
\newcommand{\gradpe}{\partial_\perp}
\newcommand{\lappa}{{\laplacian_\parallel}}
\newcommand{\kernel}{\mathcal{K}}
\newcommand{\nuF}{\nu_\mathrm{F}}

\newcommand{\id}{\vb{1}}
\newcommand{\Jv}{\vb{J}}
\newcommand{\pol}{\vb{p}}
\newcommand{\Qv}{\vb{Q}}
\newcommand{\qv}{{\vb{q}}}
\newcommand{\Tv}{\vb{T}}

\newcommand{\Sv}{\vb{S}}
\newcommand{\xv}{{\vb{x}}}
\newcommand{\uv}{{\vb{u}}}

\newcommand{\ofour}{\mathcal{O}(\nabla^4)}
\newcommand{\osix}{\mathcal{O}(\nabla^6)}
\newcommand{\Lv}{\boldsymbol{\Lambda}}
\newcommand{\Lvt}{\boldsymbol{\tilde{\Lambda}}}
\newcommand{\mob}{\mathcal{M}}
\newcommand{\aoupmob}{m}
\newcommand{\epslliq}{\varepsilon_1}
\newcommand{\epslvap}{\varepsilon_0}
\newcommand{\alphag}{\alpha_\gamma}
\newcommand{\alphagt}{\tilde{\alpha}_\gamma}
\newcommand{\epst}{\varepsilon_t}
\newcommand{\etaTsq}{\eta_T^2}
\newcommand{\etaT}{\eta_T}

\newcommand{\epsfac}{s}
\newcommand{\mub}{\mu_0}
\newcommand{\rhoV}{\rho_\mathrm{V}}
\newcommand{\rhoL}{\rho_\mathrm{L}}
\newcommand{\rhoVL}{\rho_\mathrm{V,L}}
\newcommand{\rhoLV}{\rho_\mathrm{L,V}}
\newcommand{\sigmaV}{\sigma_\mathrm{V}}
\newcommand{\sigmaL}{\sigma_\mathrm{L}}
\newcommand{\sigmaVL}{\sigma_\mathrm{V,L}}
\newcommand{\sigmacw}{\sigma_\mathrm{cw}}
\newcommand{\psiV}{\psi_\mathrm{V}}
\newcommand{\psiL}{\psi_\mathrm{L}}
\newcommand{\psiVL}{\psi_\mathrm{V,L}}
\newcommand{\psicw}{\psi_\mathrm{cw}}
\newcommand{\rhom}{\rho_\mathrm{m}}
\newcommand{\rhow}{\rho_\mathrm{w}}
\newcommand{\rhoc}{\rho_\mathrm{c}}
\newcommand{\symm}{\mathsf{S}}
\newcommand{\TS}{\mathsf{TS}}

\begin{document}

\clearpage
\title{Active Cahn--Hilliard theory for nonequilibrium phase separation:\texorpdfstring{\\}{}
quantitative macroscopic predictions and a microscopic derivation}

\author{Sumeja Burekovi\'c\,\orcidlink{0009-0004-1600-2112}
}
\thanks{These authors contributed equally to this work.}
\affiliation{Service de Physique de l'\'Etat Condens\'e, CEA, CNRS Universit\'e Paris-Saclay, CEA-Saclay, 91191 Gif-sur-Yvette, France}

\author{Filippo De Luca\,\orcidlink{0000-0003-3849-343X}
}
\thanks{These authors contributed equally to this work.}
\affiliation{DAMTP, Centre for Mathematical Sciences, University of Cambridge, Wilberforce Road, Cambridge CB3 0WA, United Kingdom}

\author{Michael E. Cates\,\orcidlink{0000-0002-5922-7731}
}
\affiliation{DAMTP, Centre for Mathematical Sciences, University of Cambridge, Wilberforce Road, Cambridge CB3 0WA, United Kingdom}

\author{Cesare Nardini\,\orcidlink{0000-0002-0466-1418}
}
\affiliation{Service de Physique de l'\'Etat Condens\'e, CEA, CNRS Universit\'e Paris-Saclay, CEA-Saclay, 91191 Gif-sur-Yvette, France}
\affiliation{Sorbonne Universit\'e, CNRS, Laboratoire de Physique Th\'eorique de la Mati\`ere Condens\'ee, 75005 Paris, France}

\begin{abstract}
Phase-separating active systems can display phenomenology that is impossible in equilibrium. The binodal densities are not solely determined by a bulk (effective) free energy, but also affected by gradient terms, while capillary waves and Ostwald processes are determined by three distinct interfacial tensions. These and related phenomena were so far explained at continuum level using a top-down minimal theory (Active Model B+). This theory, by Taylor-expanding in the scalar order parameter (or density), effectively assumes that phase separation is weak, which is not true across most of the phase diagram. Here, we develop a quantitative account of active phase separation, by introducing the active counterpart of Cahn--Hilliard theory, constructing the density current from all possible terms with up to four spatial derivatives {\em without} Taylor-expanding in the density.
From this $\ofour$ theory, we show how to compute binodals and interfacial tensions for arbitrary choices of the five density-dependent `coefficient functions' that specify the theory (replacing the four constant coefficients of Active Model B+).  
We further show how a well-studied particle model for active phase separation, involving thermal quorum-sensing active particles (tQSAPs), yields a fully specified example of the $\ofour$ theory upon coarse-graining. We find that to coarse-grain consistently at $\ofour$ requires a systematic procedure, based on multiple-scale analysis, to eliminate fast-evolving orientational moments. Using this, we calculate from microscopic physics all five coefficient functions of the active Cahn--Hilliard theory for tQSAPs. We identify contributions that were missed in previous continuum theories, and show how neglecting them becomes justified only in the limit of large quorum-sensing range parameter $\gamma$. Comparison with particle-based simulations of tQSAPs shows that, both qualitatively and quantitatively, our $\ofour$ theory improves on previous continuum models, 
predicting not only that binodal densities depend on $\gamma$, but also---unexpectedly---that phase boundaries become reentrant at small thermal diffusivity. A further surprising prediction of the wider theory is that in certain regimes of parameter space, bulk phase separation exhibits an instability in which, at $\ofour$ level, the interfacial width shrinks to zero. We link this to the runaway effects of active pumping terms that cause positive feedback of a sharpening interface.
\end{abstract}

\maketitle
\section{Introduction}
\label{sec:intro}
Phase separation in systems far from thermal equilibrium is commonly observed across a wide range of spatial and temporal scales~\cite{cates-2025}. Examples arise in suspensions of self-propelled particles~\cite{cates-2015}, biological tissues~\cite{steinberg1963reconstruction,foty2005differential}, biomolecular condensates in cells~\cite{hyman2014liquid,banani2017,Weber2019review,berry2018physical}, chromatin organization in the nucleus~\cite{erdel2018formation,cook2018transcription,mirny2019two}, vibrated granular materials~\cite{oyarte2013phase,clerc2008liquid,prevost2004nonequilibrium}, social dynamics~\cite{schelling1971dynamic,sen2014sociophysics}, and ecology~\cite{rietkerk2008regular}. All these systems break detailed balance (time-reversal symmetry) at the scale of their individual constituents and can thus be classed as active systems~\cite{ramaswamy-2010,marchetti-2013,bechinger-2016,vrugt-2025,gompper-2020}. Besides their relevance in understanding biological systems, the study and control of active materials opens further avenues for engineering rheological and mechanical properties~\cite{gompper-2020}. 
\par 
Because active systems are not constrained by thermodynamics, active phase separation exhibits unique phenomenologies that are not allowed in equilibrium systems. Examples include: motility-induced phase separation (MIPS), where phase separation occurs in the absence of attractive forces \cite{tailleur-2008, fily-2012, redner-2013a, stenhammar-2014, cates-2015}; demixing of two species of particles induced by their differential growth and death~\cite{hupe-2026}; demixing by coupling different particle species to unequal thermal fluctuations~\cite{weber2016binary,grosberg2015nonequilibrium,ilker2020phase,smrek2017small,mccarthy2023demixing} (a mechanism that was proposed to have a role in the spatial organization of chromatin in the cell nucleus~\cite{ganai2014chromosome,agrawal2017chromatin}); microphase separation driven by  nonequilibrium chemical reactions, which is a candidate mechanism for the formation of finite-size biomolecular condensates in cells~\cite{julicher2024droplet,zwicker2017growth}; bubbly phase separation, where a flux of material persists in the statistical steady state~\cite{fausti-2024,tjhung-2018}; and traveling bands of unequal density arising in the presence of nonreciprocal interactions~\cite{saha-2020,you2020nonreciprocity,frohoff2023non,frohoff2021suppression,brauns-2024}. Several local active mechanisms were also discovered that lead to microphase separation~\cite{tjhung-2018,fausti-2024,li2021hierarchical,singh2019hydrodynamically,zwicker2017growth}, which instead requires long-range interactions in equilibrium systems. In this work, we restrict our focus to bulk phase separation in systems with a single, conserved density or scalar order parameter (see Ref.~\cite{cates-2025} for a review of this and other cases).
\par 
Continuum descriptions of such scalar active systems have provided significant understanding of the generic properties of active phase separation. Such understanding goes beyond describing the observed phenomenology, highlighting the physical mechanisms at work; continuum descriptions can also point to as-yet unseen phenomena by charting out phase diagrams for experimental investigation.
Their construction, via conservation laws and symmetry arguments, follows a `top-down' path first introduced with Models A--J for passive phase ordering~\cite{hohenberg-1977,chaikin2000principles,bray-1994}. In these Landau--Ginzburg type theories, a small miscibility gap is (in effect) assumed, so that phase separation is weak, and only low-order terms in powers of the order parameter are retained. The order parameter is the deviation in density away from its value at the mean-field critical point where phase separation first appears on changing a temperature-like variable. An expansion in spatial derivatives is also intrinsic to this approach.

So-called `active field theories' differ crucially from Models A--J because locally broken time-reversal symmetry implies that new nonlinearities are allowed~\cite{cates-2025}. Consistent with the Landau--Ginzburg approach, these are added to lowest nontrivial order in an expansion in both density and spatial gradients.

The case relevant to us is Active Model B+ (AMB+). This describes active phase separation of a single, scalar conserved order parameter~\cite{tjhung-2018,nardini-2017}, without coupling to vector or tensor fields such as those for fluid flow~\cite{tiribocchi2015active,singh2019hydrodynamically,caballero2024interface} or liquid crystallinity~\cite{tjhung2012spontaneous}. In this paper, we aim
to advance our understanding and control of phase-separating active systems, by developing continuum theories that do not rely on expanding in powers of the order parameter, so that they can quantitatively apply beyond the formal limit of weak phase separation. The need for such a development is clear from nearly a century of studies on equilibrium phase-separating systems~\cite{onuki-2002}. There, while the Ginzburg-Landau quartic free energy of Model B is sufficient to describe equilibrium phase separation qualitatively, it is quantitative only for weak phase separation. It also predicts some universal quantities, such as the scaling exponents for coarsening dynamics, because these govern both weak and strong cases~\cite{bray-1994}. 

Accurate prediction of phase boundaries and/or interfacial tensions requires more detailed theories. These range in complexity from Flory--Huggins theory~\cite{huggings-1941, flory-1942}, via scaled particle theory \cite{lebowitz-1965}, to modern approaches combining, {\em e.g.} statistical associating fluid theory and density functional theory~\cite{llovell-2010}. All of these provide local equations of state whose equilibrium binodals exhibit saturating densities at low temperature (unlike Model B). To address interfacial tensions in equilibrium, finite-range interactions must be allowed for and, at its simplest, this can be done by adding square gradient contributions to the local free-energy density $f(\rho)$ written as a function of the coarse-grained particle density $\rho$ (see Eq.~\eqref{eq:fenergy_pass} below), resulting in a dynamical description for the density that contains terms up to order $\ofour$ in gradients (this assumes the interactions to be sufficiently short-ranged). This description, allowing for arbitrary $f(\rho)$, is known as Cahn--Hilliard theory~\cite{cahn-1958}. Within this term, we include extensions that allow a density-dependent collective mobility $\mathcal{M}(\rho)$ and square gradient coefficient $K(\rho)$~\cite{bray-1994}. 

In a nonequilibrium context, the breaking of detailed balance allows for a wider freedom in constructing the dynamics of the density field. At the Cahn--Hilliard level, each active gradient term brings, through its prefactor, another function of density into the equations of motion, which we call a `coefficient function'. Including the counterpart of the equilibrium chemical potential $\mub(\rho)$ and also $K(\rho)$, there are five such functions in total, three of which are inherently active (see Eq.~\eqref{eq:o4} below), which define the active counterpart of Cahn--Hilliard theory at $\ofour$ level.

In this active setting, going beyond the weak separation limit creates qualitatively, as well as quantitatively, new behavior. For instance, in the weak separation limit of AMB+, it is known 
from the {\em uncommon tangent construction}~\cite{wittkowski-2014,solon-2018,tjhung-2018} that the binodals do not solely depend on $f(\rho)$ but are affected by the higher gradient terms in the dynamics~\cite{wittkowski-2014,tjhung-2018,cates-2025}, even though in equilibrium these affect {\em only} interfacial tensions and {\em not} binodals. Below, we shall show that even the uncommon tangent construction is replaced at Cahn--Hilliard level by a more general one. This prescribes mismatches not only in intercepts, but also in slopes, of the two tangents to $f(\rho)$ at the binodal densities. (Recall that in all equilibrium models, both mismatches vanish.)

Even in equilibrium, a challenge with models formulated at Cahn--Hilliard level is to systematically link their coefficient functions to either experimental data or microscopic particle models. Such models are advantageous for mechanistic interpretation, encoding the properties of the individual constituents and interactions directly in their parameters. While they often entail significant idealizations when compared to experiments (especially for biological systems), minimal particle models, such as quorum-sensing (QS) active particles~\cite{cates-2013, cates-2015, solon-2015, solon-2018, solon-2018-njp, martin-2021,dinelli-2024} or self-propelled particles with simple two-body interaction forces~\cite{romanczuk-2012,fily-2012,redner-2013a, stenhammar-2014},
allow us to build intuition for which microscopic ingredients lead to which macroscopic outcomes. 
Thus far though, even when the novel phenomenology predicted at field-theoretic level is indeed observed in particle models~\cite{cates-2025}, it can be unclear which microscopic ingredients are causing it. For example, spherical particles with hard-core repulsion are observed to undergo bubbly phase separation in $d=2$~\cite{stenhammar-2014,patch2018curvature,caporusso2020micro} while equilibrium-like phase separation is observed both in $d=3$ and for dumbbells~\cite{caporusso2024phase}. Likewise, it is not yet clear what microscopic factors induce unstable capillary waves and the associated active foam state~\cite{cates-2025}.

To connect particle models undergoing MIPS with field-theory descriptions, several coarse-graining methods have been employed. These include density functional theory and interaction-expansion methods~\cite{wittkowski-2011,vrugt-2023}, the diffusion-drift approximation~\cite{cates-2013, cates-2015, solon-2015, solon-2018, solon-2018-njp,dinelli-2024,dean-1996}, weakly nonlinear analysis~\cite{speck-2014-weaknl-prl, speck-2014-weaknl-long,bergmann-2018,rapp-2019,neville-2025}, geometrical methods~\cite{bruna-2012, kalz-2024}, liquid-state theory~\cite{li-2023-lst}, mechanical arguments involving the stress tensor~\cite{omar-2023,langford-2024-interface,evans-2025},
Doi--Peliti field theories \cite{pruessner-2025}, and kinetic theory \cite{soto-2024, pinto-goldberg-2025}. However, we argue here that in obtaining a quantitative theory (beyond weak separation) via coarse-graining, not only should Taylor-expanding the density be avoided, but for consistency all terms up to a specified order in gradients should be kept. Thus, to offer a reliable improvement on AMB+, one should keep in the equation of motion for $\rho$ all terms that are $\ofour$ in gradients, not just a subset. A systematic coarse-graining procedure that respects this requirement is developed for the first time in this Paper, which is organized as follows. 

We first introduce (in Sec.~\ref{sec:ofour_intro}) the most generic field theory to fourth order in gradients (the `$\ofour$ theory', which is the active counterpart of Cahn--Hilliard theory), for describing active phase separation of a single, conserved scalar order parameter (the density $\rho$), wherein rotational and translational invariance are respected, and chirality is absent.
The $\ofour$ theory is an important step towards a complete description of active phase-separating systems, allowing quantitative predictions once all its coefficient functions are either determined by coarse-graining a microscopic model or inferred from observational data. 
In principle, a fully predictive theory would also require one to avoid expanding in gradients, since interfaces far from the critical point can cease to be diffuse (on the scale set by microscopic interactions), as we discuss quantitatively later on. Whereas in equilibrium models this is perhaps best addressed via (non-square-gradient) density functional theories, such approaches are still in their infancy for active matter~\cite{wittkowski-2011,te2020classical}.
Meanwhile, the wide success of square-gradient, Cahn--Hilliard-type theories in the equilibrium domain~\cite{bray-1994,onuki-2002} suggests that much can be learned by generalizing them to active matter, as done here.

In Sec.~\ref{sec:ofour_framework}, we show how to obtain semi-analytically from the $\ofour$ theory the crucial observables determining the properties of phase-separating systems. These are the binodal densities, two nonequilibrium interfacial tensions governing the Ostwald process, and a third interfacial tension that controls the relaxation of capillary waves. (See Ref.~\cite{cates-2025} for a discussion of these three different tensions.)
We then address the $\ofour$ theory, first in the simplest case of constant coefficients, and then in other cases. We discuss the regimes in which the Ostwald process is reversed and the capillary waves are unstable, generalizing results already known for AMB+. We also find new generic physics in the $\ofour$ theory: For example, there exists a regime, which we refer to as the forbidden region, where phase coexistence is destabilized at nonlinear level by a mechanism involving the shrinking to zero of the interfacial width. Additionally, our analysis clarifies the assumptions under which a mechanically defined interfacial tension~\cite{kirkwood-1949,bialke-2015,solon-2018-njp,langford-2025} can be directly linked to the (conceptually distinct~\cite{cates-2025}) interfacial tensions that control diffusive fluxes and hence the Ostwald process.

In Sec.~\ref{sec:coarsegraining}, we use systematic coarse-graining to explicitly construct the $\ofour$ theory corresponding to a well-studied particle-based model for MIPS. Specifically, we consider quorum-sensing particles undergoing additional translational diffusion (so-called `thermal QS active particles' or tQSAPs). We first show that a coarse-graining approach widely used in the literature, the diffusion-drift (DD) approximation, maps tQSAPs to a specific $\ofour$ theory. However, this method does not consistently generate {\em all} terms with four gradients at continuum level. We thus introduce a different approach, based on capturing fast-evolving orientational moments via a multiple-scale (MS) analysis, and show that this makes the gradient expansion consistent to $\ofour$. The formal distinction between the DD and MS approximations is explored further by reference both to a field-theoretical toy model and to the motion of a single active Ornstein--Uhlenbeck (AOUP) particle in an external potential. 

Using the MS approach, we then find explicitly for tQSAPs all five coefficient functions that appear in the $\ofour$ theory. 
Furthermore, we show that the MS approach for tQSAPs can be perturbatively justified, using as a small parameter the ratio between the largest microscopic length scale and the interfacial width.
Finally, we show that, at $\ofour$ level, the MS theory reduces to the DD approximation when the range parameter, $\gamma$, of the QS interaction is much larger than all other microscopic scales. To our knowledge, this is the first time that the widely used DD approximation has been analytically justified in any asymptotic limit.
\par  
In Sec.~\ref{sec:qs_application}, we construct explicit predictions at $\ofour$ level for tQSAPs, and compare them to particle simulations. From the two $\ofour$ theories obtained in Sec.~\ref{sec:coarsegraining}, via the DD and MS analyses, respectively, we predict both the binodals and interfacial tensions. We find that the MS-based theory gives quantitatively accurate binodals and interfacial tensions not only within, but well beyond, its domain of perturbative validity. In particular, the MS-$\ofour$ theory accounts well for the $\gamma$ dependence of the binodals, whereas in DD-$\ofour$ theory there is no such dependence. Surprisingly,
MS also predicts a phenomenon previously unseen in QS active particles and missed by other coarse-graining techniques: The liquid binodal is a reentrant function of the translational diffusivity. Our numerical simulations of tQSAPs verify this prediction.

Our various results are summarized and discussed in Sec.~\ref{sec:discussion} where we also identify avenues for future work.

\section{\texorpdfstring{$\ofour$}{O(grad 4)} field theory}
\label{sec:ofour_intro}
In this section, we introduce the generic class of field theories that we will analyze in this work. 
We are interested in capturing the phase separation of active systems that can be described by a single scalar field $\rho(\xv, t)$ (such as single-species systems or two-species systems with an incompressibility condition, which can be reduced to one composite variable). This field is assumed to be conserved, and therefore obeys a continuity equation:
\begin{equation}
    \partial_t\rho(\xv, t) = -\div\Jv,\label{eq:continuity}
\end{equation}
where $\Jv(\xv,t)$ is the current. In principle, $\Jv$ will have a deterministic and a stochastic part, so that Eq.~\eqref{eq:continuity} is a stochastic partial differential equation (PDE). In this work, we restrict study to the deterministic dynamics only, and neglect the noise term entering $\Jv$. This approximation is well-founded unless one wants to investigate phenomena that are directly controlled by noise, as, for example, nucleation~\cite{cates-2023}, the roughening of the liquid-vapor interface~\cite{besse-2023}, or bubbly phase separation~\cite{tjhung-2018}. In most other cases, including bulk phase equilibria away from the critical point, modest amounts of noise should merely renormalize the coefficients of the deterministic theory~\cite{chaikin2000principles,bray-1994}.

In an equilibrium (passive) setting, the current is then
\begin{equation}
    \Jv_\mathrm{pass} = -\mob[\rho]\grad\fdv{\mathcal{F[\rho]}}{\rho}
    \,,\label{eq:J_pass}
\end{equation}
where $\mob[\rho]>0$ is the collective mobility that in general is a functional of the density, and $\mathcal{F}[\rho]$ is the free-energy functional.
The free-energy functional is commonly expanded to second order in gradients (leading to the generic Cahn--Hilliard form for $\Jv$):
\begin{equation}
    \mathcal{F}[\rho] = \int\dd\xv\left[f(\rho) + \frac{K(\rho)}{2}|\grad\rho|^2\right]
    \,,
    \label{eq:fenergy_pass}
\end{equation}
where $d$ is the spatial dimension. Working to this order in gradients, which amounts to including all possible terms containing up to four spatial derivatives on the right-hand side (RHS) of Eq.~\eqref{eq:continuity}, 
is necessary to obtain a finite interface width and a finite surface tension, and it is thus the lowest order possible to address phase separation. Note that Eq.~\eqref{eq:fenergy_pass}
is insufficient if $K(\rho)$ becomes negative; a higher-order gradient term is then needed to stabilize microphase-separated patterns such as lamellar phases~\cite{gompper1994phase}.
\par
The dynamics of an active system does not have to obey a free-energy minimization principle. Thus, in an active field theory, generic terms are allowed in the current that break the gradient structure imposed in Eq.~\eqref{eq:J_pass}. We thus expand in gradients up to the order necessary to again capture the physics of interfaces, \emph{i.e.}, fourth order in spatial derivatives, and perform the expansion directly in the current rather than at the level of a free energy. The most general current is then
\begin{equation}
    \begin{aligned}
        \Jv ={}&{} -\grad[\mub(\rho) - K(\rho)\laplacian\rho + \lambda(\rho)|\grad\rho|^2]\\
        &{}+ \zeta(\rho)\laplacian\rho\grad\rho + \nu(\rho)|\grad\rho|^2\grad\rho
        \,,
        \label{eq:o4}
    \end{aligned}
\end{equation}
which, once inserted in the continuity equation \eqref{eq:continuity}, leads to a dynamics for the density field that contains all possible terms with four derivatives. Equations~\eqref{eq:continuity} and~\eqref{eq:o4} will be referred to as the $\ofour$ theory. We will refer in the following to $f(\rho)$ as the effective free-energy density (although it has no thermodynamic interpretation in general active systems). This is defined up to an irrelevant constant via $f'(\rho)= \mub(\rho)$, where here and below the prime $(')$ denotes the derivative with respect to $\rho$.  As in the passive case, we assume $K(\rho)>0$, since Eq.~\eqref{eq:continuity} otherwise has an unchecked instability at small wavelengths that requires higher-order gradient terms for the theory to be regularized.

Note that it is generally possible to factor out a mobility function $\mob(\rho)$ from the current by writing $\Jv = \mob(\rho)\tilde\Jv$, where $\tilde\Jv$ is still of order $\ofour$ and of the form of Eq.~\eqref{eq:o4}. However, in the absence of noise, this does not give any additional freedom in defining the current, which is completely characterized by the five `coefficient functions' $\mub, K, \lambda, \zeta, \nu$ appearing in the canonical form of Eq.~\eqref{eq:o4}. The results presented in this work, including all definitions of interfacial tensions, use the canonical form and do not factor out such a mobility function unless explicitly stated. However, the effect of factoring out a density-dependent mobility on the tensions is considered in Sec.~\ref{sec:ofour_mob}.

The $\ofour$ theory includes cases already known in the literature. First, the passive case with unit mobility $(\mob=1)$ and free energy as in Eq.~\eqref{eq:fenergy_pass} is recovered by setting $\lambda(\rho) = -K'(\rho)/2$, as well as $\zeta(\rho)=0=\nu(\rho)$. Moreover, if $K(\rho)$ is constant, we then recover the original Cahn--Hilliard equation~\cite{cahn-1958}.  
\par
A second important case is found by expanding for small departures from the critical density~$\rhoc$, setting $\phi \coloneqq \rho-\rhoc$. Assuming $\phi$ to be small leads in equilibrium to Model B, whose local free energy $f$ is quartic in $\phi$, with $K$ a constant and $\lambda=\zeta=\nu=0$~\cite{hohenberg-1977}. The lowest order at which active terms arise is $\mathcal{O}(\phi^2)$; expanding all terms (except for $\mub$) up to this order sets $K$ to be at most linear in $\phi$, and $\lambda$ and $\zeta$ to be constants, with $\nu=0$. This procedure recovers AMB+~\cite{tjhung-2018}. We later study the distinct case where $\nu$ is nonzero even though all coefficient functions are constant (see Sec.~\ref{sec:ofour_constant_coeffs}).
\section{Phase separation in \texorpdfstring{$\ofour$}{O(grad 4)} theory}
\label{sec:ofour_framework}
We now derive the general framework for describing phase separation in any $\ofour$ theory defined by Eqs.~\eqref{eq:continuity} and \eqref{eq:o4} with generic coefficient functions $\mub, K, \lambda, \zeta, \nu$. We show how to obtain the binodal densities in Sec.~\ref{sec:ofour_coexistence} and the nonequilibrium interfacial tensions driving the Ostwald process and the capillary-wave dynamics in Sec.~\ref{sec:ofour_tensions}. We then apply our insights to two simple cases: in Sec.~\ref{sec:ofour_constant_coeffs} to
the $\ofour$ theory with constant coefficients and, in Sec.~\ref{sec:ofour_mob}, to an extension of AMB+ that includes a $\rho$-dependent mobility. Crucially, the results found here will then be applied in Sec.~\ref{sec:qs_application} to quantitatively address phase separation in tQSAPs, after deriving the associated coefficient functions by explicit coarse-graining in Sec.~\ref{sec:coarsegraining}.
\subsection{Binodal densities: anomalous phase coexistence}
\label{sec:ofour_coexistence}
We now show how to obtain the binodal densities in $\ofour$ theory and the spatial profile of the interface. This is achieved by generalizing a change of variables (into so-called `pseudovariables') previously introduced for simpler models, including, but not limited to, AMB+~\cite{aifantis1983mechanical,aifantis1983equilibrium,solon-2018-njp,tjhung-2018}. This generalization is quite nontrivial---in particular, when $\nu(\rho)\neq \zeta'(\rho)/2$ in Eq.~\eqref{eq:o4}, it requires us to abandon the `uncommon tangent construction' that has proved a powerful tool for AMB+ and related models~\cite{wittkowski-2014,solon-2018,tjhung-2018}.

\subsubsection{Coexistence conditions}
To compute the binodals, we can assume the system to have a flat stationary interface connecting the vapor and liquid densities $\rho_\mathrm{V,L}$ at $x=\mp\infty$, with the $x$ axis perpendicular to the interface, so that the system is effectively one-dimensional.
Then, the current is given by $\Jv = J\vu{e}_x$. We can now define a nonlocal chemical potential $\mu(x)$ such that $J = -\partial_x \mu(x)$; it is given by
\begin{equation}
\begin{aligned}
    \mu(x) ={}&{} \mub(\rho) - K(\rho)\partial^2_x\rho + \bar\lambda(\rho)(\partial_x\rho)^2\\
    &{}+ \int_x^\infty\bar\nu(\rho(\tilde x))[\partial_{\tilde{x}}\rho(\tilde{x})]^3\dd\tilde{x}\,,\label{eq:non-loc-mu}
\end{aligned}
\end{equation}
where we have introduced the notations 
\begin{equation}
    \bar\lambda = \lambda - \zeta/2 \, , \qquad \qquad \bar\nu = \nu - \zeta'/2\,.
    \label{eq:shorthands}
\end{equation}
For a stationary profile, $J$ must be uniform. We assume that it vanishes at infinity, which then implies that $J = 0$ everywhere. As a consequence, $\mu(x)=\mathrm{const.} = \mub(\rhoL)$, since all other terms in Eq.~\eqref{eq:non-loc-mu} vanish at $x=\infty$.
Interestingly, when $\bar\nu \neq 0$, the bulk chemical potential $\mub(\rho)$ is unequal in the two phases (this feature is absent in AMB+~\cite{tjhung-2018}), with the jump in $\mub$ given by
\begin{equation}
    \label{eq:mu0-jump}
    \Delta\mub = \int_{-\infty}^\infty\bar\nu(\rho(x))[\partial_{{x}}\rho({x})]^3\dd{x}.
\end{equation}
Here and below, we define $\Delta X \coloneqq X(\rhoL) - X(\rhoV)$.
\par
To proceed, let us introduce two functions of the density $\psi^{(i)}(\rho)$ with $i = 1,2$, referred to as the \emph{pseudodensities}, chosen to be the two linearly independent solutions of the second-order ordinary differential equation (as shown below in Eq.~\eqref{eq:psiSv}, this choice allows to rewrite $\Jv$ as the divergence of a local tensor):
\begin{equation}
    K(\rho)\psi''(\rho) + [2\bar\lambda(\rho)+K'(\rho)]\psi'(\rho) + 2\bar\nu(\rho) \psi(\rho) = 0\,.\label{eq:psi}
\end{equation}
We multiply Eq.~\eqref{eq:non-loc-mu} by $\partial_x\psi^{(1,2)}(\rho)$ and integrate through the interface. Integration by parts yields the two conditions:
\begin{equation}
    \Delta\Pi^{(1,2)} = \int^{\rhoL}_{\rhoV} \mub'(u) \psi^{(1,2)}(u)\dd{u}=0\,,\label{eq:binodals}
\end{equation}
where we have introduced two distinct pseudopressures:
\begin{equation}\label{eq:pseudopressure}
\Pi^{(i)}(\rho) \coloneqq \int^\rho \mub'(u) \psi^{(i)}(u)\dd{u}.
\end{equation}
The two equalities in Eq.~\eqref{eq:binodals} reduce to known phase coexistence conditions in certain cases. First, note that whenever $\bar\nu = 0$, one solution to Eq.~\eqref{eq:psi} can be chosen as $\psi^{(1)} = 1$. The corresponding pseudopressure condition then restores the equality of the bulk chemical potential: $\Delta\Pi^{(1)} = \Delta\mub = 0$. This holds for all equilibrium theories, and also for AMB+.
\par
The second independent solution of Eq.~\eqref{eq:psi}, on the other hand, is nontrivial even when $\bar\nu = 0$. In the equilibrium case, where $\bar\lambda = -K'/2$, it can be chosen to be $\psi^{(2)} = \rho$; this recovers the equality of the thermodynamic pressure $P = \mub\rho - f(\rho)$ at coexistence, since then $\Delta\Pi^{(2)} = \Delta P = 0$~\cite{solon-2018-njp}.
In AMB+, the second, nontrivial solution for the pseudodensity $\psi^{(2)}(\rho)$ results instead in the equality of the pseudopressure that was previously defined in Refs.~\cite{solon-2018-njp,tjhung-2018} (corresponding to $\Delta\Pi^{(2)}=0$). 
\par
In contrast, for the generality of $\ofour$ theories, which have $\bar\nu \neq 0$, {\em both} solutions of Eq.~\eqref{eq:psi} are nontrivial. The two conditions of Eq.~\eqref{eq:binodals} then allow us to find the binodals without knowledge of $\Delta\mub$, which is nonvanishing in general, and calculable {\em a posteriori} using Eq.~\eqref{eq:mu0-jump}.
\par
\begin{figure}[tbp]
    \centering
    \includegraphics[width=\columnwidth]{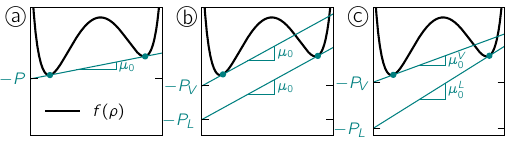}
    \caption{Stepwise abolition of common tangent construction in $\ofour$ theory, for a bulk (effective) free-energy density defined via $f'(\rho)=\mub(\rho)$. a) Passive case (common tangent construction): $\Delta P = \Delta\mub = 0$; b) AMB+ (uncommon tangent construction): $\Delta P \neq 0$, $\Delta\mub = 0$; c) $\ofour$ theory (no tangent construction): $\Delta P \neq 0$, $\Delta\mub \neq 0$.
    \label{fig:tangents}}
\end{figure}

In summary, while in equilibrium systems the knowledge of $f(\rho)$ is sufficient to obtain the binodals, this is not true for AMB+~\cite{cates-2025}, and even less true for $\ofour$ theory. More specifically, a dependence of the binodals on $K(\rho)$, $\bar\lambda(\rho)$, and $\bar\nu(\rho)$ follows from the fact that the $\psi^{(i)}$ entering Eq.~\eqref{eq:binodals} depend on all these functions. Furthermore,
$\ofour$ theory completes the stepwise abolition in active systems of the common tangent construction. In equilibrium models, the tangents of $f(\rho)$ at the binodals have equal slopes (arising from $\Delta\mub = 0$) and intercepts (due to $\Delta P = 0$). This statement is equivalent to the Maxwell construction on $P(\rho)=\mub\rho-f$~\cite{chaikin2000principles}.
In AMB+, the equality of chemical potential survives while that of pressure is superseded, leading to parallel tangents with differing intercepts; this is the \emph{uncommon tangent} construction discussed in Refs.~\cite{wittkowski-2014,solon-2018-njp}. In general, $\ofour$ theory gives instead $\Delta\mub \neq 0$: thus, the equality of slopes is abolished as well, and the tangents determined by Eq.~\eqref{eq:binodals} are geometrically unrelated (Fig.~\fref{fig:tangents}). 

To explicitly solve the pseudodensity equation~\eqref{eq:psi} and to obtain the binodal densities, specific choices of coefficient functions of the $\ofour$ theory must be made. The task will be completed for the case of constant coefficients in Sec.~\ref{sec:ofour_constant_coeffs}, and for two distinct $\ofour$ theories found by coarse-graining tQSAPs in Sec.~\ref{sec:coarsegraining_qsap}.
\subsubsection{Interface profiles}\label{sec:interfacial-profile}
In the following, we will need to know the density profile $\rho(x)$ of a flat interface at stationarity. To obtain this, we start from the requirement $J(x) = -\partial_x\mu = 0$. Changing variables from $x$ to $\rho$, we define the function $w(\rho)\coloneqq(\partial_x\rho)^2$, which from differentiation of Eq.~\eqref{eq:non-loc-mu} obeys:
\begin{equation}
    Kw'' - (2\bar\lambda-K')w' + 2(\bar\nu-\bar\lambda')w = 2\mub'\,.\label{eq:w}
\end{equation}
This \emph{interface equation} is a linear, second-order, inhomogeneous ordinary differential equation for $w(\rho)$. It can be solved by providing two initial conditions, for example, Dirichlet conditions at the binodal densities, $w(\rhoV)=w(\rhoL)=0$. For a general $\ofour$ theory, the solution must be sought numerically; however, in certain cases such as that with constant coefficients, the solution can be found analytically (see Eq.~\eqref{eq:w_constcoeff} below).
We further show in App.~\ref{app:ofour} that interfaces in $\ofour$ theory are always monotonic.

\subsection{Nonequilibrium interfacial tensions}
\label{sec:ofour_tensions}
We now derive the various nonequilibrium interfacial tensions determining, first, the departures of the coexisting densities from the binodals in the presence of a curved interface (these control the Ostwald process as we also show), and second, the dynamics of capillary waves on a flat interface. This analysis generalizes to the case of $\ofour$ theory the results previously obtained for AMB+ in Refs.~\cite{tjhung-2018,fausti-2021} and reviewed in Ref.~\cite{cates-2025}.
\subsubsection{Pseudostress tensor}
We start by establishing an identity that will be useful later, generalizing a strategy introduced in Ref.~\cite{solon-2018-njp}. For any solution $\psi^{(i)}(\rho)$ of the ODE \eqref{eq:psi}, the $\ofour$ current of Eq.~\eqref{eq:o4} can be rewritten as \footnote{Here, we assume that $\psi^{(i)}(\rho) \neq 0$ between the binodals.}:
\begin{equation}
    \Jv = [\psi^{(i)}(\rho)]^{-1}\div\Sv^{(i)}\,,\label{eq:psiSv}
\end{equation}
with $\Sv^{(i)}$ a second-rank pseudostress tensor~\footnote{Note that a term $\grad\grad\rho$ could in principle appear in such a pseudostress tensor; however, its coefficient can be set to zero by reabsorbing it into the other terms.}:
\begin{equation}
\begin{aligned}
    \Sv^{(i)} ={}&{} -\big[\Pi^{(i)} -\psi^{(i)} K\laplacian\rho
    \\
    &{}\qquad+ \left(\left(\bar\lambda+\zeta\right)\psi^{(i)}-K\psi^{(i)\prime}\right)|\grad\rho|^2\big]\id
    \\
    &{}+ (\zeta\psi^{(i)}-K\psi^{(i)\prime} )\grad\rho\grad\rho\,.\label{eq:S_psi}
\end{aligned}
\end{equation}
Note that setting $\Jv = \vb{0}$ in Eq.~\eqref{eq:psiSv} results in $\Sv = \mathrm{const.}$ for a flat interface, which immediately recovers the condition $\Delta\Pi^{(i)} = 0$ because the gradient terms in $\Sv$ vanish inside the bulk phases. Finally, whenever $\bar\nu = 0$, $\psi^{(1)} = 1$; in that case (which includes AMB+) the current is \emph{already} the divergence of a pseudostress tensor. 
\par 
In passing, we note that a current in the form of Eq.~\eqref{eq:psiSv} was used in Ref.~\cite{deluca-2024} to show that an extension of AMB+ to include nonconstant mobility exhibits hyperuniformity (HU) in the coarsening regime of phase ordering, with an HU exponent $\varsigma = 4$ as defined there. The fact that even for $\bar\nu \neq 0$ the current can be brought into the form of Eq.~\eqref{eq:psiSv} shows that, during coarsening, the $\ofour$ theory is likewise HU with exponent $\varsigma = 4$, with the factor $1/\psi^{(i)}(\rho)$ then taking the role of the density-dependent mobility. 
\subsubsection{Effect of curvature on coexistence}
We next find the coexisting densities in the presence of a curved interface. We consider a spherically symmetric state, with a liquid droplet or vapor bubble of radius $R$ and density $\rho_\mathrm{in}$, immersed in the majority phase at density $\rho_\mathrm{out}$, at stationarity.
\par
It is convenient to introduce special solutions of the pseudodensity equation \eqref{eq:psi}, denoted as $\psiVL$, by imposing Dirichlet boundary conditions at the vapor or liquid densities, respectively:
\begin{align}
    \psiV(\rhoV) = 0\,, &&\psiL(\rhoL) = 0\,.\label{eq:psi_anch}
\end{align}
We refer to the resulting solutions, each of which is a function of $\rho$, as \emph{phase-anchored} pseudodensities. The corresponding pseudopressures, defined via Eq.~\eqref{eq:pseudopressure}, are denoted as $\Pi_\mathrm{V,L}(\rho)$.

Let us further denote $\Pi_\mathrm{V,L}^\mathrm{in,out} = \Pi_\mathrm{V,L}(\rho_\mathrm{in,out})$. Setting $\Jv = \vb{0}$ in Eq.~\eqref{eq:psiSv}, we obtain $(\div\Sv_\mathrm{V,L})\vdot\vu{e}_r = 0$. Expressing the divergence in spherical coordinates and integrating the resulting equation once, we find
\begin{equation}\label{eq:Laplace_pseudopressure_deriv}
\begin{aligned}
    \Pi_\mathrm{V,L}^\mathrm{in} ={}&{} \Pi_\mathrm{V,L}(\rho) -K\psiVL\partial_r^2\rho+ \bar\lambda\psiVL(\partial_r\rho)^2\\
    &{}-(d-1)\frac{\partial_r\rho}{r}K\psiVL\\
    &{}+ (d-1)\int_0^r\frac{K\psi'_\mathrm{V,L}-\zeta\psiVL}{\tilde{r}}(\partial_{\tilde{r}}\rho)^2\dd{\tilde{r}}\!.
\end{aligned}
\end{equation}
Evaluating this at $r=\infty$ and assuming a narrow interface width ($\xi\ll R$) inside the integral, we obtain a curvature-dependent pseudopressure difference across the interface:
\begin{equation}
\begin{aligned}
    \Pi_\mathrm{V,L}^\mathrm{in} - \Pi_\mathrm{V,L}^\mathrm{out} ={}&{} \frac{d-1}{R}\frac{\Delta\psiVL}{\Delta\rho}\sigmaVL + \mathcal{O}(R^{-2})\,.\label{eq:Laplace_pseudopressure}
\end{aligned}
\end{equation}
These differences are controlled by the \emph{nonequilibrium interfacial tensions}:
\begin{equation}
\begin{aligned}
    \sigmaVL = \frac{\Delta\rho}{\Delta\psiVL}\int \left[K \dv{\psiVL}{\rho}- \zeta \psiVL\right](\partial_x\rho)^2\dd{x},
\end{aligned}
\label{eq:sigma}
\end{equation}
where the integral is through the interface, which can be taken to be flat to this order in $1/R$. Note that our definitions for $\sigmaVL$ are consistent with those adopted in Ref.~\cite{cates-2025}. (For reasons explained there, these differ by the factor $\Delta\rho/\Delta\psiVL$ from those used previously in, \emph{e.g.}, Refs.~\cite{tjhung-2018,fausti-2021}.)
\par
Computing the interfacial tensions given by Eq.~\eqref{eq:sigma} in practice requires knowledge of the interface profile $w(\rho)$. Indeed, with a change of variables from $x$ to $\rho$, we obtain
\begin{equation}
    \sigmaVL = \frac{\Delta\rho}{\Delta\psiVL}\int_{\rhoV}^{\rhoL} \left[K\dv{\psiVL}{\rho}- \zeta\psiVL\right]\sqrt{w(\rho)}\dd{\rho},\label{eq:sigma_rho}
\end{equation}
where $w(\rho)$ is the solution of Eq.~\eqref{eq:w}.

We are now ready to compute the shift in the coexisting densities due to curvature from Eq.~\eqref{eq:Laplace_pseudopressure}. For definiteness, we consider the case of a liquid droplet in a vapor background and set $\rho_\mathrm{in, out} = \rhoLV + \delta_\mathrm{in, out}$. Because $\delta_\mathrm{in, out}$ are of order $1/R$, we now expand Eq.~\eqref{eq:binodals} around the binodals, finding
\begin{subequations}
\label{eq:Pi_OW}
\begin{align}
    \Pi_\mathrm{V}^\mathrm{in} - \Pi_\mathrm{V}^\mathrm{out} &= \mub'(\rhoL)\psiV(\rhoL)\delta_\mathrm{in} + \mathcal{O}(\delta^2)\,,\\
    \Pi_\mathrm{L}^\mathrm{in} - \Pi_\mathrm{L}^\mathrm{out} &=-\mub'(\rhoV)\psiL(\rhoV)\delta_\mathrm{out} + \mathcal{O}(\delta^2)\,.
\end{align}
\end{subequations}
Equating this result with Eq.~\eqref{eq:Laplace_pseudopressure} and using Eq.~\eqref{eq:psi_anch}, we obtain the binodal density shifts outside and inside the liquid droplet:
\begin{equation}
\begin{aligned}
    \delta_\mathrm{in,out} = \frac{d-1}{R}\frac{\sigmaVL}{\mub'(\rhoLV)\Delta\rho} + \mathcal{O}(R^{-2})\,.
\end{aligned}\label{eq:curvature_shifts}
\end{equation}
Turning now to the case of a vapor bubble immersed in liquid, an identical calculation shows that Eq.~\eqref{eq:curvature_shifts} acquires an additional minus sign (on top of exchanging $\mathrm{L}\leftrightarrow \mathrm{V}$), resulting in
 \begin{equation}
\begin{aligned}
    \delta_\mathrm{in,out} = -\frac{d-1}{R}\frac{\sigma_\mathrm{L,V}}{\mub'(\rho_\mathrm{V,L})\Delta\rho} + \mathcal{O}(R^{-2})\,.
\end{aligned}\label{eq:curvature_shifts-v}
\end{equation}

In Sec.~\ref{reversalO} below, we will show that the interfacial tensions $\sigmaVL$ obtained with the phase-anchored pseudodensities govern the Ostwald process, and their signs determine whether the process is forward or reversed. In the latter case, based on what is known for AMB+, the system is expected to undergo bubbly phase separation or microphase separation (rather than bulk phase separation), which involve noise in an essential way~\cite{cates-2025,fausti-2024,tjhung-2018}. We work without noise and so focus on positive tensions, and bulk phase separations only, in this work.
\par

\subsubsection{Mechanical interfacial tension in systems where the pressure is a state function}
\label{sec:sigma_mech}
In generic active systems, including QS particles, pressure is not a state function \cite{solon-2015-np}. However, in the special case of single-species, active particle systems with isotropic, torque-free, conservative pairwise interactions and constant particle mobility (here set to unity), it reacquires this role \cite{solon-2015-prl, takatori-2014, takatori-2015}. In such cases, the particle current takes the form $\Jv = \div\Sv$, with $\Sv$ the \emph{mechanical} stress tensor \cite{cates-2025,solon-2018-njp,langford-2025}. It is generally not clear whether $\Sv$ is local in $\rho$ and its gradients, as assumed in Refs.~\cite{omar-2023,langford-2024-interface, langford-2025}. But if so, given its tensorial character, one can expand it as $\Sv = -\mathcal{P}(\rho)\vb{1} + \mathcal{O}(\nabla^2)$, where $\mathcal{P}$ is now the mechanical pressure.

Comparing this with the general expansion of the $\ofour$ current given in Eq.~\eqref{eq:o4}, we see that for this specific case the mechanical pressure in fact corresponds to the bulk chemical potential in the terminology used here: $\mub(\rho) = \mathcal{P}(\rho) $. (For a discussion of this mismatch in nomenclature, see Refs.~\cite{cates-2025,langford-2025}.) Furthermore, assuming that the current is the divergence of a {\em local} second-rank tensor, it follows that, while $\nu,\zeta\neq0$ in general, $\bar\nu = 0$. For a flat interface, the mechanical pressure is constant in the two phases; in this context, this corresponds to the statement we made earlier that for $\bar\nu = 0$ there is no jump in $\mub$ (Eq.~\eqref{eq:mu0-jump}). 

In the presence of a curved interface, a pressure difference arises between the two phases (the so-called Laplace pressure). Due to the correspondence above, in our theory this corresponds to a jump in the bulk chemical potential. Using $\mub(\rho_\mathrm{in, out}) = \mub(\rho_\mathrm{L, V}) + \mub'(\rho_\mathrm{L, V})\delta_\mathrm{in, out}$, the jump in $\mathcal P=\mub$ follows directly from Eq.~\eqref{eq:curvature_shifts} as
\begin{equation}
\begin{aligned}
    \mathcal P^\mathrm{in} - \mathcal P^\mathrm{out} &= \frac{d-1}{R}\sigma_\mathrm{mech} + \mathcal{O}(R^{-2})\,.
\end{aligned}
\end{equation}
Here,
\begin{equation}\label{eq:sigma_mech}
    \sigma_{\mathrm{mech}} = \frac{\sigmaV - \sigmaL}{\Delta\rho}\,
\end{equation}
represents the \emph{mechanical tension} as defined previously in the literature \cite{kirkwood-1949,bialke-2015,solon-2018-njp,langford-2025}. 

Summarizing the above, we have shown that, in systems where the pressure is a state function, the mechanical tension controlling the mechanical Laplace pressure is a linear combination of the Ostwald tensions obeying Eq.~\eqref{eq:sigma}, under the assumption that the mechanical stress tensor is expandable locally in $\rho$ and its derivatives. We emphasize that $\sigma_{\mathrm{mech}}$ is only well-defined for systems where the pressure indeed has a mechanical interpretation. Hence, the above results should not be applied in other cases, such as, for example, in AMB+ with $\zeta=0$, where $\sigmaV=\sigmaL$ and Eq.~\eqref{eq:sigma_mech} gives a vanishing result~\cite{tjhung-2018}.

For an {\em equilibrium} system undergoing Brownian dynamics with unit particle mobility, the collective mobility obeys $\mathcal{M}(\rho) = \rho$ and $\Jv = -\rho\grad\mu = \div\Sv$ with $\mu$ the thermodynamic chemical potential and $\Sv$ the mechanical stress. Here, $\Sv$ is necessarily local for any equilibrium system with short-range interactions~\cite{chaikin2000principles}.
Also, with $\mathcal{M}(\rho) = \rho$ we have $\sigmaVL = \rhoLV\sigma_\mathrm{eq}$ as proven below in Eq.~\eqref{eq:sigma_AMBp_mob}. One then finds from Eq.~\eqref{eq:sigma_mech} that, as expected, $\sigma_{\mathrm{mech}}=\sigma_\mathrm{eq}$, the equilibrium interfacial tension. 

The above reasoning exemplifies how quite general arguments applied at field level can explain some of the more surprising outcomes observed in particle-level simulations of active systems. Specifically, and in contrast to equilibrium, for active systems satisfying the mechanical assumptions detailed above, Eq.~\eqref{eq:sigma_mech} explains why the mechanical tension can be negative even when capillary waves are stable~\cite{bialke-2015,solon-2018-njp} (see also~Eq.~\eqref{eq:sigma_rel} below).
\subsubsection{Ostwald process and its reversal}\label{reversalO}
We now return to the general $\ofour$ case and investigate the Ostwald process, 
\emph{i.e.}, the dynamics of a liquid droplet in a supersaturated majority vapor phase, at density $\rho_\infty = \rhoV + \epsilon$. The supersaturation is taken to represent the mean shift in the binodals resulting from (say) a distant population of droplets~\cite{bray-1994}. For $\epsilon \neq \delta_\mathrm{out}$, there will be a diffusive current flowing between the supersaturated vapor at infinity and the surface of the chosen droplet, driving the Ostwald process, which we now calculate.

Assuming radial symmetry, we can write $\Jv = J\vu{e}_r$ with $J = -\partial_r\mu(r)$ (so that $\div \Jv = -\laplacian\mu$) with a nonlocal chemical potential $\mu(r)$ obeying
\begin{align}
    \mu(r) ={}&{} \mub(\rho) - K(\rho)\partial^2_r\rho - \frac{d-1}{r}K(\rho)\partial_r\rho  + \bar\lambda(\rho)(\partial_r\rho)^2\nonumber\\
    &{}+ \int_r^\infty\zeta(\rho(\tilde r))\frac{[\partial_{\tilde{r}}\rho(\tilde{r})]^2}{\tilde r}\dd\tilde{r}\nonumber\\
    &{}+ \int_r^\infty\bar\nu(\rho(\tilde r))[\partial_{\tilde{r}}\rho(\tilde{r})]^3\dd\tilde{r}\,.\label{eq:non-loc-mu-r}
\end{align}
As is standard in discussing the Ostwald process, we assume the current to be quasi-stationary. We thus solve the Laplace equation $\laplacian\mu = 0$ outside of the droplet, with boundary conditions set by the densities at infinity and close to the droplet, $\rho_\infty$ and $\rho_\mathrm{out}$, respectively~\cite{bray-1994}. This results in 
\begin{equation}
    \mu(r) = \frac{R^{d-2}}{r^{d-2}}(\mub(\rho_\mathrm{out})-\mub(\rho_\infty)) + \mub(\rho_\infty)
\end{equation}
for $d>2$, and
\begin{equation}
    \mu(r) = \frac{\mub(\rho_\infty)\ln(r/R)-\mub(\rho_\mathrm{out})\ln(r/\Lambda)}{\ln(\Lambda/R)}
\end{equation}
in $d=2$, with $\Lambda$ an upper cutoff for the coordinate $r$.

The inward current is given by evaluating the gradient of this solution at $r=R$, $-J_\mathrm{in} = -\partial_r\mu|_{r=R}$, giving
\begin{equation}
    J_\mathrm{in} = C_d\frac{\mub(\rho_\mathrm{out})-\mub(\rho_\infty)}{R}\,,
\end{equation}
with $C_{d>2} = d-2$ and $C_2 = (\ln \Lambda/R)^{{-1}}$. Conservation of mass yields the condition
\begin{equation}
    \dv{t}[V_R(\rho_\mathrm{out}-\rho_\mathrm{in})] = A_R J_\mathrm{in}\,,
\end{equation}
where $A_R$ denotes the surface area of a droplet of volume $V_R$.
Using $\mub(\rho_\mathrm{out})-\mub(\rho_\infty) = (\delta_\mathrm{out}-\epsilon)\mub'(\rhoV) + \mathcal{O}(\delta^2,\epsilon^2)$, we obtain
\begin{equation}
    \dot R = \frac{(d-1)C_d\sigmaL}{(\Delta\rho)^2 R}\left(\frac{1}{R_\mathrm{c}}-\frac{1}{R}\right) + \mathcal{O}(R^{-2})\,,\label{eq:Rdot}
\end{equation}
where we have introduced the critical radius
\begin{equation}
    R_\mathrm{c} = \frac{(d-1)\sigmaL}{\epsilon \mub'(\rhoV)\Delta\rho}\,.\label{eq:Rcrit}
\end{equation}
For a vapor bubble, Eqs.~\eqref{eq:Rdot} and \eqref{eq:Rcrit} are still valid upon interchanging the labels L and V.

The sign of $\sigmaL$ determines the stability of the fixed point $R_\mathrm{c}$ in Eq.~\eqref{eq:Rdot}. For $\sigmaL > 0$, the fixed point is unstable and the Ostwald ripening process for liquid droplets proceeds as usual: Small droplets shrink and large ones grow. For $\sigmaL < 0$, the fixed point is stable and the Ostwald process is reversed~\cite{cates-2025,tjhung-2018}. Similarly, negative $\sigmaV$ causes reversal of the Ostwald process for vapor bubbles.

Note that for the specific case of pairwise repulsive self-propelled particles, the corrections to the coexisting densities $\delta_\mathrm{in,out}$ found in Ref.~\cite{solon-2018-njp}, along with Eq.~\eqref{eq:curvature_shifts}, imply that the Ostwald process is reversed for vapor bubbles ($\sigmaV<0$) and forward for liquid droplets ($\sigmaL>0$). This is consistent with the numerical observation of bubbly phase separation in which finite bubbles pervade a continuous liquid phase~\cite{tjhung-2018}. Moreover, under the assumptions of Sec.~\ref{sec:sigma_mech} above, these results imply that $\sigma_\mathrm{mech}$ of Eq.~\eqref{eq:sigma_mech} is negative, which is indeed consistent with the numerical findings of Ref.~\cite{solon-2018-njp} discussed at the end of that section.

\subsubsection{Capillary waves}
We now turn to capillary-wave dynamics. We consider an interface in $d$ dimensions, whose fluctuations around a planar ground state are described by a height function $\hat h(\xvpa, t)$, where $\xvpa$ denotes the $d-1$ in-plane coordinates and $\xpe$ the normal coordinate. Following Ref.~\cite{fausti-2021}, we make an ansatz for the field of the form
\begin{equation}\label{eq:cw-capillary}
    \rho(\xv,t) = \varrho(\xpe-\hat h(\xvpa, t))\,,
\end{equation}
where $\varrho$ describes the interface profile. In principle, curvature-dependent corrections to $\varrho$ in Eq.~\eqref{eq:cw-capillary} appear; they, however, turn out to be negligible at leading order in $\hat h$ and $q=|\qv|$ (where $\qv$ is the wavevector along the $\xvpa$ direction) and thus we do not consider them here.
As shown in Appendix~\ref{app:cw}, the dynamics of the height function is most conveniently written after Fourier transforming along the $\xvpa$ direction and then obeys
\begin{equation}\label{eq:cw-pt-h}
    \partial_t h(\qv,t) = -\frac{2q^3\sigmacw}{(\Delta\rho)^2}h(\qv,t) + \mathcal{O}(q^4h, h^2)\,,
\end{equation}
where we have introduced the capillary-wave interfacial tension
\begin{equation}\label{eq:sigma_cw}
    \sigmacw = \frac{\Delta\rho}{\Delta\psicw}\int_{\rhoV}^{\rhoL} \left[K\dv{\psicw}{\rho}- \zeta\psicw\right]\sqrt{w(\rho)}\dd{\rho}.
\end{equation}
The pseudodensity $\psicw$ appearing in $\sigmacw$ solves Eq.~\eqref{eq:psi} with boundary condition $\psicw(\rhoV) = -\psicw(\rhoL)$, and is thus given by
\begin{equation}\label{eq:psi_cw}
    \psicw(\rho) = \frac{\psiV(\rho)}{\psiV(\rhoL)}-\frac{\psiL(\rho)}{\psiL(\rhoV)}\,.
\end{equation}
From Eq.~\eqref{eq:sigma_cw}, using the facts that $\Delta\psiL = -\psiL(\rhoV)$, $\Delta\psiV = \psiV(\rhoL)$, and $\Delta\psicw = 2$, we find
\begin{equation}
    \sigmacw = \frac{\sigmaV + \sigmaL}{2}\,.\label{eq:sigma_rel}
\end{equation}
Thus, the relation \eqref{eq:sigma_rel} previously found for AMB+ \cite{fausti-2021} is generalized to arbitrary $\ofour$ theories.

Capillary waves are stable at long wavelengths as long as $\sigmacw > 0$; conversely, a negative capillary-wave interfacial tension indicates unstable interfaces~\cite{fausti-2021}. An important general consequence of Eq.~\eqref{eq:sigma_rel} is that at $\ofour$ level capillary waves cannot be unstable without the reversal of the Ostwald process for at least one phase. (This was previously known for AMB+ only~\cite{fausti-2021}.)
Note, however, that Eq.~\eqref{eq:sigma_rel} relies on the diffusive structure of the $\ofour$ theory: in particular, $\sigmacw$ is not determined in terms of $\sigmaVL$ for {\em wet} active systems in which $\rho$ is coupled to a conserved fluid momentum~\cite{tiribocchi2015active,singh2019hydrodynamically,cates-2025}.

The result \eqref{eq:sigma_rel} may appear surprising for the following reason. In AMB+, one can explain it by considering that $\sigmaV$ and $\sigmaL$ control the curvature-induced fluxes on either side of a symmetric, sinusoidally perturbed flat interface. The liquid and vapor phases thus contribute equally to the diffusive relaxation rate, which then involves the average of the two tensions. However, the $\ofour$ theory includes an extension of AMB+ with density-dependent mobility (see Sec.~\ref{sec:ofour_mob} below) for which one might expect $\sigmaVL$ in Eq.~\eqref{eq:sigma_rel} to be weighted by the corresponding mobilities $\mob(\rhoVL)$. This is resolved by noting that the canonical form, Eq.~\eqref{eq:o4}, of $\ofour$ theory effectively selects a `constant $\mathcal{M}$ gauge'. A different gauge choice would indeed lead to mobility factors in Eq.~\eqref{eq:sigma_rel}, but this gauge absorbs them into the interfacial tensions. This is shown explicitly in Sec.~\ref{sec:ofour_mob} below (see Eq.~\eqref{eq:variantCW}). 
\subsection{\texorpdfstring{$\ofour$}{O(grad 4)} theory with constant coefficients}
\label{sec:ofour_constant_coeffs}
We now apply the above framework to derive binodals and interfacial tensions for the simplest $\ofour$ theory, obtained by setting the coefficient functions $K>0$, $\lambda$, $\zeta$, and $\nu$ to be constants.
For simplicity, we further choose a quartic bulk free energy of the form $f(\rho) = -a\rho^2/2 + b\rho^4/4$, with $a, b> 0$, which in equilibrium has binodals $\rho_\pm = \pm \sqrt{a/b}$. (The effect of considering instead a Flory--Huggins form for the local free energy $f(\rho)$ is discussed in App.~\ref{app:flory}.)
Due to the symmetry of $f(\rho)$, the theory is symmetric under the transformation $\rho \to -\rho$ so long as the parameters transform as $K, \lambda, \zeta, \nu \to K, -\lambda, -\zeta, \nu$. This symmetry will be visible in both the binodals and the interfacial tensions found below. 

With these choices, we arrive at a minimal extension of AMB+ whose only addition is a term $\nu|\grad\rho|^2\grad\rho$ in the current. Besides obtaining explicit binodals and tensions for this case, we will find that it exemplifies an unexpected feature of the $\ofour$ theory. This is the presence of a forbidden region of parameters where binodals do not exist, and whose emergence is associated with interfaces becoming infinitesimal in width.

\subsubsection{Binodals and the forbidden region}\label{sec:binforbid}
In the case of constant coefficients, Eq.~\eqref{eq:psi} for the pseudodensity $\psi(\rho)$ is the equation for a damped harmonic oscillator with likewise constant coefficients. We choose two explicit, linearly independent solutions as
\begin{align}
\psi^{(1)}(\rho) = e^{-\bar\lambda\rho/K}a_\omega(\rho)\,,&&\psi^{(2)}(\rho) = e^{-\bar\lambda\rho/K}b_\omega(\rho)\,,\label{eq:psi_const}
\end{align}
where we have introduced
\begin{subequations}
\begin{align}
    \omega &\coloneqq 2\nu/K - (\bar\lambda/K)^2\,,\label{eq:omega}\\
    a_\omega(\rho)&\coloneqq \begin{cases}
        \cosh(\sqrt{|\omega|}\rho) &\text{for } \omega \leq 0\,,\\
        \cos(\sqrt{\omega}\rho) &\text{for } \omega > 0\,,
    \end{cases}\\
    b_\omega(\rho)&\coloneqq \begin{cases}
        \sinh(\sqrt{|\omega|}\rho) &\text{for } \omega < 0\,,\\
        \rho &\text{for } \omega = 0\,,\\
        \sin(\sqrt{\omega}\rho) &\text{for } \omega > 0
        \,.
    \end{cases}
\end{align}
\end{subequations}
The sign of $\omega$ thus determines whether the pseudodensities $\psi^{(1,2)}(\rho)$ are oscillatory functions or not (in the latter case, a one-to-one mapping between the pseudodensities $\psi^{(i)}$ and physical density $\rho$ ceases to exist).

\begin{figure}[tbp]
    \centering
    \includegraphics[width=\columnwidth]{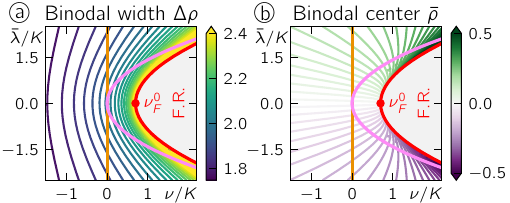}
    \caption{Binodals in the constant-coefficient $\ofour$ theory (with free-energy parameters $a=b=1$). a) Binodal width $\Delta\rho=\rhoL-\rhoV$ as a function of $\nu$ and $\bar\lambda$. b) Binodal center $\bar\rho=(\rhoL+\rhoV)/2$ as a function of $\nu$ and $\bar\lambda$. The gray area is the forbidden region (F.R.) where no binodal solutions exist. Orange, red, and pink lines as in Fig.~\fref[b]{fig:constantcoeff_forbiddenregion}.
    \label{fig:constantcoeff_binodals}}
\end{figure}

We are now ready to compute the binodals. As detailed in App.~\ref{app:bincalc}, these can be easily obtained numerically by minimizing the square of $\Delta\Pi^{(1,2)}$ in Eq.~\eqref{eq:binodals}. The resulting phase diagram, as a function of $\nu/K$ and $\bar{\lambda}/K$,
is reported in Fig.~\fref{fig:constantcoeff_binodals}, where we show the mean of the binodals $\bar\rho$ and the width of the miscibility gap $\Delta\rho$, defined so that $\rhoVL = \bar\rho \mp \Delta\rho/2$.

As shown in Fig.~\fref[a]{fig:constantcoeff_binodals}, the miscibility gap (of width $\Delta\rho$) narrows upon increasing $|\bar\lambda|$ throughout the phase diagram. Further, the binodal mean $\bar\rho$ monotonically increases with $\bar\lambda$ for any $\nu<\nuF^0$ (the value of this $\nuF^0$ will be computed below). These two properties are already present in AMB+~\cite{tjhung-2018}. In contrast, in $\ofour$ theory $\bar\rho$ becomes nonmonotonic in $\bar\lambda$ when $\nu>\nuF^0$. Even more surprisingly, in the $\ofour$ theory we find a region of the phase diagram (gray background in Fig.~\fref{fig:constantcoeff_binodals}) where no solution for the binodals exists, appearing at $\nu > \nuF^0$, which we refer to as the \emph{forbidden region}.
\par
The existence of the forbidden region can be understood by the following physical argument. 
The term $\nu|\grad\rho|^2\grad\rho$ in the current acts as a pump across interfaces (see Fig.~\fref[a]{fig:constantcoeff_forbiddenregion}), which for $\nu>0$ brings material from the dilute phase to the dense phase. Thus, increasing $\nu$ leads to a wider miscibility gap $\Delta\rho$ everywhere (see Fig.~\fref[a]{fig:constantcoeff_binodals}). In addition, increasing $\nu$ also leads to an increase in $|\bar\rho|$, unless $\bar\lambda=0$. Now, the pump created by $\nu\neq0$ is stronger when the interface is steeper, due to its dependence on $|\grad\rho|^2$. This creates positive feedback for $\nu>0$: The stronger the $\nu$ term, the stronger the phase separation, the steeper the interface, which in turn strengthens the $\nu$ pump. If this feedback is not compensated by other terms in the current, the interface will form a shock where its steepness diverges, leading to a pathology in the $\ofour$ theory. This results in the forbidden region where no binodals can be found.
\par
A more quantitative version of this argument is the following. Treating the $\nu$ term as a correction to the antidiffusive part of the current, we can introduce $a_\mathrm{eff} = a+\nu|\grad\rho|^2 \simeq a+\nu \Delta\rho^2/\xi^2$, with $\xi$ the interfacial width. This results in $\xi^2 = K/(2a_\mathrm{eff})$, which can be solved for $\xi^2$ to give
\begin{equation}\label{eq:xi_forbidden_approximation}
    \xi^2 \simeq \frac{K}{2a} - \frac{\nu\Delta\rho^2}{2}\,.
\end{equation}
Hence, for sufficiently large $\nu$ we expect $\xi\to 0$. We thus interpret the forbidden region as being due to interfacial width falling to zero beyond which there is no solution involving conventional phase separation~\footnote{For $\bar\lambda=0$, Eq.~\eqref{eq:xi_forbidden_approximation} gives that $\xi\to0$ as $\nu\to\nuF^0 \approx 0.23K$ for $a=b=1$; this rough estimate should be compared to the exact result from Eq.~\eqref{eq:nuF0}, $\nuF^0 \approx 0.70K$. Thus, Eq.~\eqref{eq:xi_forbidden_approximation} correctly captures the mechanism leading to the forbidden region, but since it uses a small $\nu$ expansion, should not be relied upon for quantitative predictions.}.
\par
The boundary of the forbidden region $\nuF(\bar\lambda)$ can be computed exactly, as detailed in App.~\ref{app:forbidden_region}. For $\bar\lambda = 0$, the maximal value of $\nu$ for which the theory is well-defined reads (red dot in Fig.~\fref[b]{fig:constantcoeff_forbiddenregion}):
\begin{equation}\label{eq:nuF0}
   \nuF^0\coloneqq\nuF(\bar\lambda = 0) = \frac{Kb}{a}\left(\frac{3\pi^2}{8}-3\right).
\end{equation}
Moreover, $\nuF(\bar\lambda)$ can be found analytically when $\omega\leq 0$:
\begin{align}\label{eq:nuF_omeganeg}
    \nuF(\bar\lambda) = \frac{3Kb}{2a}\left(\sqrt{1+\frac{4a\bar\lambda^2}{3bK^2}}-1\right) &&\text{for }\omega\leq 0\,.
\end{align}
which corresponds to the dotted red line in Fig.~\fref[b]{fig:constantcoeff_forbiddenregion}. For $\omega>0$ and $\bar\lambda \neq 0$, finding $\nuF(\bar\lambda)$ requires solving a transcendental equation (Eq.~\eqref{eq:forbidden_omegapos}), which is easy to do numerically (solid red line in Fig.~\fref[b]{fig:constantcoeff_forbiddenregion}).

\begin{figure}[tbp]
    \centering
    \includegraphics[width=\columnwidth]{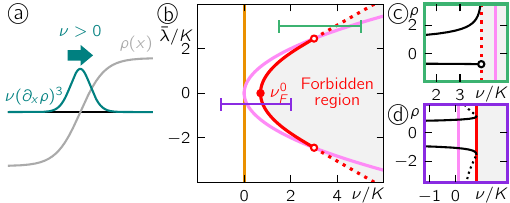}
    \caption{Forbidden region in the constant-coefficient $\ofour$ theory (with free-energy parameters $a=b=1$). a) The $\nu$ term acts as a density pump across an interface. The arrow shows the direction of the current for positive $\nu$. b) Characterization of the forbidden region. The orange vertical line here corresponds to AMB+ ($\nu = 0$). The red line indicates the boundary $\nuF$ of the forbidden region; it crosses the pink line, to the right of which the pseudodensity becomes oscillatory, at $\nu = 3Kb/a$. As $\nu\uparrow\nuF$, depending on whether the pseudodensities are oscillatory or not, either one of the binodals diverges (dotted red line, panel c) or both end at finite values, annihilating with another, unphysical solution of Eq.~\eqref{eq:binodals} (solid red line, panel d). The green and purple line segments in panel b) indicate the range of values of $\nu$ for which binodals are shown in panels c,d).\label{fig:constantcoeff_forbiddenregion}}
\end{figure}

\par
The behavior of the binodals as one approaches the forbidden region is determined by the sign of $\omega$ at $\nuF$. For $\omega \leq 0$, one binodal diverges while the other remains finite (Fig.~\fref[c]{fig:constantcoeff_forbiddenregion}). On the other hand, for oscillatory pseudodensities ($\omega>0$), there is a second, unphysical solution of the binodal conditions, which annihilates with the physical solution at $\nuF$, so that the binodal densities at the boundary of the forbidden region remain finite (Fig.~\fref[d]{fig:constantcoeff_forbiddenregion}).
\par
The instability in the forbidden region is \emph{nonlinear} in character (since the $\nu$ term is nonlinear in $\rho$): If a homogeneous state is prepared for $\nu>\nuF$ and perturbed, transiently one observes the formation of phase-separated domains of finite size. As interfaces form, they steepen beyond control, eventually leading to numerical instabilities as they reach the discretization scale.

The mechanism just outlined differs from equilibrium systems, where values of $K<0$ result in the growth of perturbations of any wavelength around the homogeneous equilibrium state. In contrast to ours, the latter is a \emph{linear} instability emerging immediately when that state is perturbed, without interfaces first needing to form. 

We expect that both instabilities can be cured by introducing higher-order gradient terms, \emph{e.g.}, by going to $\osix$. In the equilibrium case, this results in finite wavelength patterning. It is an intriguing possibility that, rather than this outcome or the restoration of normal phase separation, qualitatively new phenomena might arise in the forbidden region once these higher-order gradient terms are included. We defer the study of any such outcomes to future work. However, one possibility, suggested by the diverging binodal density seen at $\ofour$ level, is that some active systems (without hard-core repulsions) might undergo a condensation transition, rather than standard phase separation, in part or all of this region~\footnote{Condensation involves the formation of regions of unbounded density. This is not prevented by the chosen quartic $f(\rho)$ in the $\ofour$ theory addressed in Sec.~\ref{sec:binforbid}, though it would be by a Flory--Huggins $f(\rho)$ (see App.~\ref{app:flory}, where we show that the forbidden region survives but the binodal densities remain finite) or another similar choice that diverges at a saturating density to reflect the presence of hard-core repulsions.}.

\begin{figure}[tbp]
    \centering\includegraphics[width=\columnwidth]{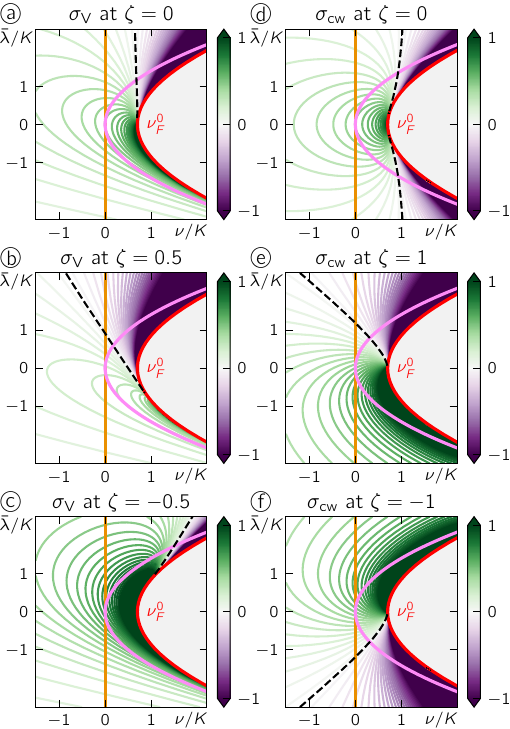}
    \caption{Interfacial tensions in the constant-coefficient $\ofour$ theory (with free-energy parameters $a=b=1$), as functions of $\nu/K$ and $\bar\lambda/K$ for different $\zeta$. Orange, pink, and red curves as in Figs.~\fref{fig:constantcoeff_binodals} and \fref{fig:constantcoeff_forbiddenregion}; the dashed black line indicates a sign change in $\sigma$. a--c) Ostwald interfacial tension for vapor bubbles. The liquid droplet interfacial tension can be found by setting $\bar\lambda, \zeta, \nu \to -\bar\lambda, -\zeta, \nu$. d--f) Capillary-wave interfacial tension.\label{fig:constantcoeff_tensions}}
\end{figure}
\subsubsection{Interfacial tensions}
We now turn to the calculation of the interfacial tensions for the constant-coefficient model using the expression in Eq.~\eqref{eq:sigma_rho}. To compute these, we introduce the phase-anchored pseudodensities $\psiVL(\rho) = \psi^{(2)}(\rho-\rhoVL)$, where $\psi^{(2)}$ is given in Eq.~\eqref{eq:psi_const}. We further need the interfacial profile $w(\rho)$. This is obtained by solving Eq.~\eqref{eq:w} with boundary conditions $w(\rhoVL) = 0$, giving
\begin{equation}\label{eq:w_constcoeff}
    w(\rho) = \frac{W(\rhoL)e^{2\bar\lambda\rho}\psiV(\rho)}{e^{2\bar\lambda\rhoL}\psiV(\rhoL)} + \frac{W(\rhoV)e^{2\bar\lambda\rho}\psiL(\rho)}{e^{2\bar\lambda\rhoV}\psiL(\rhoV)}- W(\rho)\,,
\end{equation}
where 
\begin{equation}W(\rho) \coloneqq \frac{3bK}{\nu^2}-\frac{6b{\bar\lambda} (\bar\lambda + \nu\rho)}{\nu^3} - \frac{3b\rho^2-a}{\nu}\,.
\end{equation}
Using this expression with the binodals $\rhoVL$ computed previously, we obtain the two tensions $\sigmaVL$ numerically from 
the integral in Eq.~\eqref{eq:sigma_rho}. The capillary-wave interfacial tension $\sigmacw$ is then found as the average of these, via Eq.~\eqref{eq:sigma_rel}. The resulting tensions are shown in Fig.~\fref{fig:constantcoeff_tensions} for selected values of $\zeta$.
\par
We first note that for sufficiently large $\nu$, the interfacial tensions can become negative even at $\zeta = 0$. This is in contrast to AMB+ ($\nu = 0$), where $\zeta\neq 0$ is required for this outcome. In particular, for $\zeta = 0$ the Ostwald process is reversed for vapor (liquid) bubbles whenever $\nu > \nuF^0$ and $\bar\lambda > 0$ ($\bar\lambda < 0$). Capillary waves become unstable for even larger $\nu$, for both signs of $\bar\lambda$.
\par 
Once $\zeta\neq 0$, the loci of sign reversal for all interfacial tensions are shifted, intersecting the $\nu=0$ axis at values of $\bar\lambda$ derived previously for the AMB+ case~\cite{tjhung-2018}. As there, and regardless of $\nu$, positive values of $\zeta$ facilitate the reversal of the Ostwald process for vapor bubbles while negative values shrink the parameter region where the reversal occurs, and {\em vice versa} for liquid droplets. Likewise, for $\bar\lambda\zeta>0$ (resp.~$<0$), the region of unstable capillary waves grows (resp.~shrinks) on increasing $\zeta$.

\par
Notably, no parameter region is found where the tensions $\sigmaVL$ are simultaneously negative. The absence of such a regime implies that one can expect to observe bubbly phase separation (that is, noisy microphase separation~\cite{tjhung-2018}) on one side of the liquid-vapor phase diagram, or on the other, or of course on neither (as in equilibrium), but not on both. This is a feature of the constant-coefficient theory, and does not mean that such a simultaneous reversal of the Ostwald process for both phases is not possible in a general $\ofour$ theory.
\subsection{AMB+ with density-dependent mobility}
\label{sec:ofour_mob}
Another interesting special case of the $\ofour$ theory arises when AMB+ is generalized to allow an arbitrary positive density-dependent mobility $\mob(\rho)$:
\begin{equation}
\begin{aligned}
    \Jv &= -\mob(\rho)\bigl[\grad(\mub^0(\rho)+\lambda_0|\grad\rho|^2-K_0\laplacian\rho)\\
    &\phantom{= -\mob(\rho)[}+\zeta_0\laplacian\rho\grad\rho\,\bigr]\,.\label{eq:JAMB+_mob}
\end{aligned}
\end{equation}
While this mobility will not affect any static properties resulting from $\Jv=\vb{0}$, such as the binodals, it will impact dynamical properties, such as the nonequilibrium interfacial tensions, as we now show.
Expanding out the current, we can read off the $\ofour$ coefficient functions written in the canonical form, Eq.~\eqref{eq:o4}:
\begin{equation}\label{eq:AMB+Mneq0}
\begin{aligned}
\mub' &= \mob\mub^{0\prime}\,,&&\\
K&=\mob K_0\,, & \lambda&=\mob\lambda_0\,,\\
\zeta&=\mob\zeta_0-\mob'K_0\,,& \nu&=\mob'\lambda_0\,.
\end{aligned}
\end{equation}
Using these coefficient functions in Eq.~\eqref{eq:psi}, we find two independent solutions for the pseudodensities:
\begin{equation}
    \psi^{(1)}(\rho) = \frac{1}{\mob(\rho)}\,,\quad \psi^{(2)}(\rho) = \frac{\psi^0(\rho)}{\mob(\rho)}\,.\label{eq:psi_mob}
\end{equation}
Here, $\psi^0(\rho)$ denotes the AMB+ pseudodensity introduced in Ref.~\cite{tjhung-2018}, which reads
\begin{equation}\label{eq:psi_AMBp}
    \psi^0(\rho) = \frac{K_0}{2\bar\lambda_0}\left(1 - \mathrm{e}^{-\frac{2\bar\lambda_0}{K_0}\rho}\right).
\end{equation}
As expected, the constant-mobility binodals are unchanged, since the factor $\mob(\rho)$ in $\mub'$ cancels out when multiplied with the pseudodensities in Eq.~\eqref{eq:binodals}.
\par 
To calculate the interfacial tensions for the model of Eq.~\eqref{eq:JAMB+_mob}, we take a linear combination of the two pseudodensity solutions of Eq.~\eqref{eq:psi_mob} to obtain the phase-anchored pseudodensities $\psiVL(\rho)$:
\begin{equation}
    \psiVL(\rho) = \frac{\psi^0(\rho)-\psi^0(\rhoVL)}{\mob(\rho)}\,.\label{eq:psiVL_ambp}
\end{equation}
The interfacial tensions then read
\begin{equation}\label{eq:sigma_AMBp_mob}
    \sigmaVL = \mob(\rhoLV)\sigmaVL^0\,,
\end{equation}
with the AMB+ interfacial tensions $\sigmaVL^0$ \cite{tjhung-2018,cates-2025}:
\begin{equation}\label{eq:sigma_AMBp}
    \sigmaVL^0 =\!\frac{\Delta\rho}{\Delta\psi^0}\!\int_{\rhoV}^{\rhoL}[K_0\psi^{0\prime} - \zeta_0(\psi^0-\psi^0(\rhoVL))]\sqrt{w(\rho)}\dd{\rho}.
\end{equation}
The prefactor $\mob(\rhoVL)$ appearing in Eq.~\eqref{eq:sigma_AMBp_mob} reflects the modulation of the strength of the diffusive current in the bulk between droplets (bubbles) due to the mobility, resulting in a change in the growth rate of a given droplet (bubble) in Eq.~\eqref{eq:Rdot}.
Using Eqs.~\eqref{eq:AMB+Mneq0} and \eqref{eq:sigma_AMBp_mob} in Eq.~\eqref{eq:curvature_shifts}, we note that the shifts in binodal densities $\delta_\mathrm{in,out}$ due to interfacial curvature are independent of $\mob$, as expected for any static property.

As reported in Eq.~\eqref{eq:sigma_rel}, in canonical gauge the capillary-wave tension is given by the average of the two Ostwald tensions $\sigmaVL$. For AMB+ with density-dependent mobility, we thus find from Eq.~\eqref{eq:sigma_AMBp_mob}:
\begin{equation}   
\label{eq:variantCW}
 \sigmacw = \frac{\mob(\rhoV)\sigma^0_\mathrm{L} + \mob(\rhoL)\sigma^0_\mathrm{V}}{2}\,.
\end{equation}
This reflects the physical competition between currents on the vapor and liquid side that results in the relaxation (or amplification) of capillary waves~\cite{fausti-2021}: The strength of these currents is proportional to the mobilities in the two phases, affecting that competition. This was previously discussed after Eq.~\eqref{eq:sigma_rel} above, where use of the canonical gauge absorbs factors of $\mob(\rhoVL)$ into the tensions $\sigmaVL$. As anticipated there, these factors now emerge explicitly in Eq.~\eqref{eq:sigma_AMBp_mob} and hence Eq.~\eqref{eq:variantCW}.

All the results in this section can also be obtained with a different method, presented in Appendix~\ref{app:mob}. As explained there, this allows a further generalization to cover the case in which the mobility $\mob[\rho]$ in AMB+ depends not only on $\rho$ but also on its spatial derivatives.
\section{Systematic coarse-graining of the microscopic \MakeLowercase{t}QSAP model}
\label{sec:coarsegraining}
We now turn to a widely used microscopic model of active phase-separating systems, tQSAPs, and find the corresponding $\ofour$ theory by explicit coarse-graining. 
Since the positions of self-propelled particles are coupled to their orientations, the coarse-graining procedure leads to an infinite orientational-moment hierarchy. 
As we shall discuss, even the standard DD approximation~\cite{cates-2013, cates-2015, solon-2015, solon-2018, solon-2018-njp, dinelli-2024}, based on adiabatic elimination of nematic and polar order, leads to a nontrivial $\ofour$ theory. However, we will see that this theory neglects some terms with four spatial derivatives that should be present in a fully consistent coarse-graining of tQSAPs. 

To encompass these terms by going beyond DD poses technical challenges, which we solve by employing an alternative MS analysis to fully capture the effects of fast-relaxing orientational moments. Formally, the MS approach is perturbative in a small parameter $\epslvap$, the ratio of the largest microscopic length to the smallest length of interest in phase separation. The latter is the interfacial width, upon which interfacial tensions---and in active systems also the binodals---crucially depend. Our MS analysis allows us to consistently derive all terms with four orders in gradients in the dynamics for the density field, because counting powers of $\epslvap$ and the number of gradients turns out to be equivalent.

The key results of this section will be given in Eqs.~\eqref{eq:o4-coeffs_dimful}, where we report the coefficient functions of the $\ofour$ theory associated with tQSAPs as obtained from both the DD and MS approaches. Furthermore, we find a regime where DD approximation is valid: This corresponds, in addition to small $\epslvap$, to assuming that the persistence length and the length scale associated with translational diffusion over the persistence time are both small with respect to the length scale on which particles perform quorum sensing. This is the first time, to the best of our knowledge, in which a specific parameter regime for interacting active particles is identified wherein the DD approximation is formally justifiable.

We start in Sec.~\ref{sec:fail_mini} by discussing, within the context of a field-level toy model that includes only density and polarization, why a consistent technique for eliminating orientational moments (the polarization, in this case) is needed at all. We do so by showing how the standard adiabatic elimination used in DD does not capture all terms at $\ofour$ level (which equates to $\mathcal{O}(\epslvap^2)$). Furthermore, we show that an iterative extension of this procedure that was previously proposed~\cite{bickmann2020predictive, vrugt-2023, kalz-2024}, while appealingly simple, fails in this setting because it leads to an unphysical linear instability of the homogeneous state. 

Within this toy model, we then discuss in Sec.~\ref{sec:multiscale_mini} how to perturbatively capture fast-relaxing orientational moments by employing a temporal MS analysis for small $\epslvap$. As a second example, we also use MS analysis to obtain the effective dynamics for the position of a single AOUP in an external potential in the limit of small persistence time.

We then introduce and coarse-grain the full tQSAP system in Sec.~\ref{sec:coarsegraining_qsap}.  
We derive the associated hierarchy for the orientational moments in Sec.~\ref{sec:fpe_qsap}, apply the MS analysis
to the associated moment equations,
and obtain the associated $\ofour$ theory for the density field in Sec.~\ref{sec:ofour_qsap}. In Appendix \ref{app:abps}, we repeat the analysis with active Brownian dynamics replacing the Ornstein--Uhlenbeck dynamics of our tQSAP model (keeping other aspects of its specification unaltered). There, we obtain a somewhat different $\ofour$ theory for the active Brownian version, see Eqs.~\eqref{eq:abp-o4-coeffs_dimful}.
\subsection{Limitations of diffusion-drift theory and iterative adiabatic elimination}
\label{sec:fail_mini}
We start from the following `toy' model, describing the coupled dynamics of two fields corresponding to the density~$\rho$ and polarization~$\pol$:
\begin{subequations}\label{eq:minimal model dimless}
\begin{align}
    \epsfac\partial_t \rho &= -\frac{1}{ \epslvap}\div \pol + \etaTsq \laplacian \rho
    \,,
    \label{eq:minimal model density dimensionless}
    \\
    \epsfac\partial_t \pol &= - \frac{1}{\epslvap^2} \pol 
    - \frac{1}{\epslvap}\grad \rho + \etaTsq\laplacian \pol 
    \label{eq:minimal model polarization dimensionless}
    \,.
\end{align}
\end{subequations}
As discussed in more detail after Eqs.~\eqref{eq:dt_moments} below, this model can be interpreted as a minimal continuum description of {\em noninteracting} active particles with constant self-propulsion and thermal diffusion~$\etaTsq$; there, the physical role of $\epslvap$ as a smallness parameter will become evident.
\par 
Inspecting Eqs.~\eqref{eq:minimal model dimless} shows that the polarization field~$\pol$ is fast compared with the density~$\rho$ when $\epslvap\downarrow0$. It is then natural to look for a closed equation for the density by setting $\partial_t\pol = \vb{0}$. To leading order, this gives $\pol = - \epslvap\grad \rho + \mathcal{O}(\epslvap^3 \nabla^3)$ which, re-injected into Eq.~\eqref{eq:minimal model density dimensionless}, gives:
\begin{align}\label{eq:p_trho-minimal-1}
    \epsfac\partial_t \rho &= \left(1 + \etaTsq\right) \laplacian{\rho} + 
    \mathcal{O}(\epslvap^2 \nabla^4)
    \,.
\end{align}
This is standard adiabatic elimination, and it is correct to order $\mathcal{O}(\epslvap^0\nabla^2)$. 
\par 
Many approaches in the literature, as for example the DD approximations of Refs.~\cite{cates-2013, cates-2015, solon-2015, solon-2018, solon-2018-njp, dinelli-2024}, employ adiabatic elimination only to this leading order to obtain a closed equation for the density field. As Eq.~\eqref{eq:p_trho-minimal-1} demonstrates, however, this omits some terms of fourth order in gradients, even if others might be supplied elsewhere in the theory (such as QS interactions with a finite interaction kernel~\cite{cates-2013}). The DD approach is thus incomplete at $\ofour$.

Other authors have tried to address this issue by extending the adiabatic elimination iteratively to the next order in a natural-looking if somewhat informal manner~\cite{bickmann2020predictive, vrugt-2023, kalz-2024}. 
To follow this path, we exploit the fact that $\pol$ is fast and again set $\partial_t\pol = \vb{0}$. From Eq.~\eqref{eq:minimal model polarization dimensionless}, this gives
\begin{align}
    \pol  =- \epslvap\grad \rho + \epslvap^2 \etaTsq\laplacian \pol 
    \, ,    
    \label{eq:constitutive eq polarization minimal model}
\end{align}
and, now solving iteratively for $\pol$, 
\begin{align}
     \pol = - \epslvap\grad \rho - \epslvap^3 \etaTsq\grad \laplacian \marginaldistro + \mathcal{O}(\epslvap^5\nabla^5)
     \,.
     \label{eq:mini_pol-naive}
\end{align}
Re-injecting this result into Eq.~\eqref{eq:minimal model density dimensionless}, we finally get
\begin{align}\label{eq:p_trho-minimal}
    \epsfac\partial_t \rho & = \left(1 + \etaTsq\right) \laplacian \rho + \epslvap^2\etaTsq \nabla^4 \rho + \mathcal{O}(\epslvap^4\nabla^6)
    \, .
\end{align}
As expected, and in contrast with Eq.~\eqref{eq:p_trho-minimal-1}, Eq.~\eqref{eq:p_trho-minimal} now contains terms with four derivatives. 

Yet Eq.~\eqref{eq:p_trho-minimal} leads to unphysical results. To see this, we perform a linear stability analysis around the homogeneous state $\rho(\xv, t) \to \rho_0 + \delta \rho(\xv, t)$. Fourier transforming in space
(with $\qv$ denoting the wave-vector and $q\coloneqq|\qv|$ the wavenumber), 
we obtain
\begin{align}
    \epsfac \partial_t \delta \rho_{\qv} 
    =  - (1 + \etaTsq) q^2 \delta\rho_{\qv}
    + \epslvap^2\etaTsq q^4 \delta\rho_{\qv}
    \,.
    \label{eq:minimodel_unstable}
\end{align}
Because $\epslvap^2\etaTsq >0$, there is a linear instability at large wavenumbers $q$. This contradicts Eqs.~\eqref{eq:minimal model dimless}, which are linearly stable at all wavevectors. Indeed, linearization and Fourier transforming those equations in space gives
\begin{align}
    \epsfac\partial_{t} 
    \begin{pmatrix}
        \rho_{\qv} 
        \\
        \pol_{\qv}
    \end{pmatrix}
    = 
    \begin{pmatrix}
        -\etaTsq q^2& - i \qv^\mathsf{T}/\epslvap
        \\
        - i \qv/\epslvap& -\left(\epslvap^{-2} +\etaTsq q^2\right) \id
    \end{pmatrix}
    \begin{pmatrix}
        \rho_{\qv} 
        \\
        \pol_{\qv}
    \end{pmatrix}
    \,.\label{eq:minimodel_stable}
\end{align}
The eigenvalues of this matrix have a negative real part for all $\qv$, implying linear stability. We thus conclude that the instability of Eq.~\eqref{eq:p_trho-minimal} is an artefact of the iterative approach, which is thus unfaithful to Eqs.~\eqref{eq:minimal model dimless}.

\subsection{Multiple-scale expansion}
\label{sec:multiscale_mini}\label{sec:multiscale_toy}
We next show how to derive a closed $\ofour$ equation for the density $\rho$ that remains faithful to the original model, Eqs.~\eqref{eq:minimal model dimless}. In Sec.~\ref{sec:multiscale_qsap-toy-model} below, we do so systematically using MS analysis in time~\cite{kevorkian-1996,bocquet-1997, pavliotis-2008}.
Then, before applying this technique to the full moments equation hierarchy associated with tQSAPs in Sec.~\ref{sec:coarsegraining_qsap}, we briefly summarize another example of how the MS analysis allows us to reliably obtain the effective time evolution for a different active particle system: a single Active Ornstein--Uhlenbeck particle in an external potential.
\subsubsection{MS analysis of the toy model}
\label{sec:multiscale_qsap-toy-model}
Following the standard MS procedure~\cite{kevorkian-1996,bocquet-1997, pavliotis-2008}, we approximate the solution of Eqs.~\eqref{eq:minimal model dimless} by the perturbative expansion
\begin{subequations}\label{eq:multiple scale fields ansatz minimal model}
\begin{align}
    \rho &=\rho^{(0)}+\epslvap \rho^{(1)}+\epslvap^2 \rho^{(2)} + \epslvap^3 \rho^{(3)} +\mathcal{O}(\epslvap^4) \,,\label{eq:rho-perturbative-def}
     \\[0.4em]
    \pol &=\pol^{(0)} + \epslvap\pol^{(1)} + \epslvap^2 \pol^{(2)} + \epslvap^3 \pol^{(3)} + \mathcal{O}(\epslvap^4)\label{eq:p-perturbative-def-c}
    \,, 
\end{align}
\end{subequations}
and by introducing the following times: 
\begin{align}
    t_0 = t\,, \quad t_1 = \epslvap t\,,\quad t_2 = \epslvap^2 t
    \,,\quad t_3 = \epslvap^3 t
    \,, 
    \label{eq:qs-multi_time-scales}
\end{align}
which implies that the time derivatives are replaced by
\begin{align}
    \frac{\partial}{\partial t}
    \to \frac{\partial}{\partial t_0}
    + \epslvap \frac{\partial}{\partial t_1}
    + \epslvap^2 \frac{\partial}{\partial t_2}
    + \epslvap^3 \frac{\partial}{\partial t_3}
    + \cdots
    \, . 
    \label{eq:multiple scale time derivative ansatz minimal model}
\end{align}
As can be seen from Eqs.~\eqref{eq:minimal model density dimensionless} and~\eqref{eq:p-perturbative-def-c}, in order to obtain all terms of order $\mathcal{O}(\epslvap^2\nabla^4)$ in the density equation, we only need $\pol^{(i)}$ for $i\leq3$.
Hence, we insert Eqs.~\eqref{eq:multiple scale fields ansatz minimal model} and \eqref{eq:multiple scale time derivative ansatz minimal model} into Eqs.~\eqref{eq:minimal model dimless}, and solve for $\pol^{(i)}$ order by order. At leading order, Eq.~\eqref{eq:minimal model polarization dimensionless} gives 
\begin{align}
    \pol^{(0)} = \vb{0}
    \, .
    \label{eq:pol-eps0-toy}
\end{align}
Using Eq.~\eqref{eq:pol-eps0-toy} in Eqs.~\eqref{eq:minimal model dimless} leads, at order $1/\epslvap$, to the trivial conclusion that $0 = 0$, and 
\begin{align}
    \pol^{(1)} = - \grad \rho^{(0)}
    \,. 
    \label{eq:pol-eps1-toy}
\end{align}
We can now find the next-order terms, $\mathcal{O}(\epslvap^0)$, {\em i.e.} $\mathcal{O}(1)$. Inserting Eq.~\eqref{eq:pol-eps1-toy} into Eqs.~\eqref{eq:minimal model dimless}, we get
\begin{subequations}
\begin{align}
    \epsfac \frac{\partial}{\partial t_0} \rho^{(0)}
    &= \left(1 + \etaTsq\right) \nabla^2 \rho^{(0)}
    \,,
    \label{eq:rho-eps0-toy}
    \\
    \pol^{(2)} &= - \grad \rho^{(1)}
    \,. 
    \label{eq:pol-eps2-toy}
\end{align}
\end{subequations}
We next iterate to order $\mathcal{O}\left(\epslvap\right)$ by the same procedure, obtaining
\begin{equation}
        \pol^{(3)} = - \grad \rho^{(2)} + \grad \nabla^2  \rho^{(0)}
    \,. 
    \label{eq:pol-eps3-toy}
\end{equation}
Injecting Eqs.~\eqref{eq:pol-eps0-toy}, \eqref{eq:pol-eps1-toy}, \eqref{eq:pol-eps2-toy} and \eqref{eq:pol-eps3-toy} into Eq.~\eqref{eq:minimal model density dimensionless} and using Eq.~\eqref{eq:rho-perturbative-def} we get: 
\begin{align}
    \epsfac\partial_t \rho = \left(1 + \etaTsq\right) \laplacian \rho  
    -\epslvap^2 \nabla^4 \rho  + \mathcal{O}(\epslvap^4\nabla^6)
    \,.
    \label{eq:minimal-model-multiple-scale-result}
\end{align}
We argue, on both physical and mathematical grounds, that Eq.~\eqref{eq:minimal-model-multiple-scale-result} must replace the result \eqref{eq:p_trho-minimal} of the less formal iterative approach described above. The crucial difference is that our MS analysis does not neglect $\partial_t$ in Eq.~\eqref{eq:minimal model polarization dimensionless}.
At variance with Eq.~\eqref{eq:p_trho-minimal}, but in agreement with the full model in Eqs.~\eqref{eq:minimal model dimless}, this coarse-grained theory is linearly stable for all wavenumbers $q$. 
\par 
It should be noted that our derivation implicitly assumed that the initial condition for $\pol^{(1)}$ is already compatible with Eq.~\eqref{eq:pol-eps1-toy}. If this is not the case, (temporal) boundary-layer terms initially arise. Because they decay rapidly in time and we are interested only in the long-time evolution, we do not consider them further here.
\subsubsection{MS analysis of a single active particle in a potential}
\label{sec:multiscale_aoup-1}
To further exemplify the MS approach, we now consider a single AOUP in an external potential and show that multiple-scale analysis allows us to recover the effective time evolution for its positional density, as an expansion in small persistence time~$\tau$.
Several schemes have been proposed to approximate the dynamics, but going beyond the leading order in $\tau$ has remained a challenge~\cite{grigolini-1986, peacock-lopez-1988, wittmann-2025, fox-1986a, fox-1986b, jung-1987, haenggi-1995, maggi-2015}. For example, the widely employed unified colored noise approximation (UCNA) is inconsistent with the exactly known stationary measure beyond leading order~\cite{martin-2021}. Here, we report a consistent MS dynamics to order $\tau^2$.

We consider a single AOUP with position $\xv$ and orientation $\uv = (u_1, u_2,..,u_d)^\mathsf{T}$ in spatial dimension~$d$:
\begin{subequations}\label{eq:aoups ext-pot-langevin}
\begin{align}
    \dot{\xv} &= - \aoupmob \grad \externalPotential(\xv) + v_0 \uv 
    + \sqrt{2T} \, \Lvt\,,
     \label{eq:aoups ext pot langevin r}
    \\[0.4em]
    \dot{\uv} &= - \frac{\uv}{\tau} + \sqrt{\frac{2}{\tau}} \, \Lv
    \,, 
    \label{eq:aoups ext pot langevin u}
\end{align}
\end{subequations}
where $T$ is the thermal diffusivity, $\externalPotential(\xv)$ an external confining potential, $\aoupmob$ the
mobility, $v_0$ the characteristic self-propulsion speed, and the two vectors $\Lvt$ and $\Lv$ have components which are independent Gaussian white noises with zero mean and unit variance.

The Fokker--Planck equation corresponding to Eqs.~\eqref{eq:aoups ext-pot-langevin} for the PDF $\jointpdf = \jointpdf(\xv, \uv, t)$ reads
\begin{equation}
\begin{aligned}
    \partial_t \jointpdf &= \tau^{-1}
    \grad_{\mathbf{u}} \vdot \left[ \left(\mathbf{u} + \grad_{\mathbf{u}} \right) \jointpdf
    \right]  -v_0\mathbf{u}   \vdot \grad_{\xv} \jointpdf 
    \\
    &\phantom{=} + \aoupmob \grad_{\xv} \vdot \left[\left(\grad_{\xv} \externalPotential  \right) \jointpdf \right] + T \nabla_{\xv}^2 \jointpdf
    \, .
    \label{eq:fpe-aoup}
\end{aligned}
\end{equation}
From this, we can easily obtain the coupled evolution for the positional density $\marginaldistro(\xv,t) = \int \dd \uv \, \jointpdf(\xv,\uv,t) $, polarization, and higher-order orientational moments. In Appendix~\ref{app:aoup}, we then apply an MS expansion as in the previous subsection, in the limit of small $\tau$, and get the closed evolution for $\rho$ to order $\tau^2$: 
\begin{equation}
\begin{aligned}
    \partial_t \marginaldistro
    &=  \left(v_0^2 \tau + T \right) \laplacian \marginaldistro
    + 
     \aoupmob \div \left[\left(\grad \externalPotential  \right) \marginaldistro \right]
     \\[0.4em]
     &\phantom{=}- v_0^2 \tau^2 \aoupmob \grad\grad\boldsymbol{:} \left[\left(\grad\grad \externalPotential  \right)     \marginaldistro 
     \right]
     \,,
     \label{eq:result aoups external potential}
\end{aligned}
\end{equation}
where $\vb{A} \boldsymbol{:} \vb{B} = \sum_{\alpha\beta} A^{\alpha\beta} B^{\alpha\beta}$ (where Greek indices refer to spatial coordinates).
This result confirms the one obtained via an extension of the Fox theory in Ref.~\cite{baek-2023} upon the mapping  $v_0^2 \tau \to D$ to uniformize our notations with those of Ref.~\cite{baek-2023}, and considering only terms up to order $\mathcal{O}(D \tau)$. 
For the one-dimensional case with $T=0$, the same result has also been derived in Ref.~\cite{van-kampen-1989} using homogenization theory and the projection operator formalism. 
On the other hand, following Refs. \cite{jung-1987,martin-2021,baek-2023}, we show in Appendix~\ref{app:aoup} that the UCNA result is, as expected, incorrect beyond order $\mathcal{O}(\tau^0)$ at dynamical level.
\subsection{Thermal quorum-sensing active particles}
\label{sec:coarsegraining_qsap}
In this section, we finally use our consistent MS approach to coarse-grain the dynamics of tQSAPs. We first introduce in Sec.~\ref{sec:particle-model_qsap} the particle model involved, and in Sec.~\ref{sec:fpe_qsap} derive the associated hierarchy of orientational moment evolutions, then in Sec.~\ref{sec:multiscales_qsap} apply MS analysis to this problem to obtain a closed evolution equation for the density field alone. In Sec.~\ref{sec:mean-field}, we discuss and apply a mean-field approximation valid at high densities. This leads to Eq.~\eqref{eq:qs_all-result} below, from which, by further expanding the self-propulsion speed in gradients of the density field (or, equivalently, in $\epslvap$), we finally in Sec.~\ref{sec:ofour_qsap} obtain the fully quantified $\ofour$ theory associated with tQSAPs (Eqs.~\eqref{eq:o4-coeffs_dimful}). After reviewing the regime of validity of the derivation, and addressing a limit in which we recover the result of the DD approximation, we study the linear stability of homogeneous states within our $\ofour$ theory in Sec.~\ref{sec:linear-stability_qsap}.
\subsubsection{tQSAP model}
\label{sec:particle-model_qsap}
We consider quorum-sensing active particles~\cite{cates-2013, cates-2015, solon-2015, solon-2018, solon-2018-njp, martin-2021,dinelli-2024}, undergoing thermal diffusion on top of their self-propulsion. 
We work in two dimensions, although the coarse-graining methods that follow are not restricted to the case $d=2$. 
These particles interact via a dependence of their propulsive speed $\qsspeed$ on the nearby particle density, causing their slowing down from a free particle speed $v_0$ to a lower speed $v_1$ in dense regions. 
For a strongly enough decreasing speed, the combination of persistent motion and density-dependent slowdown triggers MIPS~\cite{tailleur-2008, cates-2015}. 
Originally motivated by bacterial biology \cite{schnitzer-1993, miller-2001,liu-2011,fu-2012,mukherjee-2019}, the tQSAP system is now a paradigmatic model to study MIPS more widely. 
It can be implemented for several different types of microscopic dynamics characterized by a speed parameter. 
The analytically simplest choice, made below, is that of AOUPs, but active Brownian particles (ABPs, see App.~\ref{app:abps}) or run-and-tumble particles~\cite{tailleur-2008} could also be chosen.
\par 
We thus consider a collection of $N$ interacting AOUPs.  
The dynamics of particle $i=1,\ldots,N$ with position $\xv_i$ and orientation $\uv_i$ follows: 
\begin{subequations}\label{eq:qs_langevin}
\begin{align}
    \dot{\xv}_i  &=  \qsspeed[\rho_{\mathrm{d}}](\xv_i) \uv_i 
    + \sqrt{2T} \, \Lvt_i \,,
    \\[0.4em]
    \dot{\uv}_i  &=  -\frac{\uv_i}{\tau} + \sqrt{\frac{2}{\tau}} \, 
   \Lv_i
   \,, \label{eq:qs_langevin_u}
\end{align}
\end{subequations}
where $\tau$ denotes the persistence time, $T$ the thermal diffusivity, and  $\Lv$,~$\Lvt$ are vectors whose elements are independent Gaussian white noises with zero mean and unit variance.
In Eqs.~\eqref{eq:qs_langevin}, the thermal diffusivity $T$ is an independent parameter, unrelated to the velocity autocorrelation time $\tau$. The diffusivity of an isolated particle, including the active contribution, is then  $D_\mathrm{part} = T+ v_0^2\tau$. (Note that, as often done in AOUPs literature, we omit the $d$ dependence in Eq.~\eqref{eq:qs_langevin_u}, resulting in $D_\mathrm{part}$ independent of $d$; to reinstate the $d$ dependence of the translational diffusivity, it is sufficient to take $\qsspeed \to \,\qsspeed/\sqrt{d}$, so that $D_\mathrm{part} = T + v_0^2\tau/d$.)
\par
Although our results up to and including Sec.~\ref{sec:mean-field} do not depend on it, we now assume that the QS speed functional $\qsspeed[\rho_{\mathrm{d}}]$ obeys
\begin{subequations}\label{eq:vqs}
\begin{align}
    \qsspeed[\rho_{\mathrm{d}}](\xv,t)   &= 
    \vloc(\rho_{\mathrm{d},\kernel}(\xv,t))\,,
    \\
    \rho_{\mathrm{d},\kernel}(\xv,t) 
    &= \int \dd  \xv^\prime  \, \rho_{\mathrm{d}}(\xv^\prime,t) \kernel(\xv - \xv^\prime) 
    \,,\label{eq:rho_kernel}
\end{align}
\end{subequations}
where $\rho_{\mathrm{d}}$ is the empirical density,
\begin{align}
     \rho_{\mathrm{d}}(\xv,t) = \sum_{i=1}^N \delta(\xv - \xv_i(t))
     \,,
     \label{eq:rho_empirical}
\end{align}
$\vloc(\rho)$ is some purely local function of its argument and $\kernel(\xv)$ is the QS kernel (see Sec.~\ref{sec:qs_application} for an explicit choice of $\vloc$ and $\kernel$).
Equation~\eqref{eq:vqs} expresses that the particles adapt their speed by instantaneously measuring the density in their vicinity, weighted by the kernel.
\subsubsection{Dynamical hierarchy for the orientational moments}
\label{sec:fpe_qsap}
We next consider the $N$-body Fokker--Planck equation corresponding to tQSAPs obeying Eqs.~\eqref{eq:qs_langevin}. We use the shorthand notations $\allxs = (\xv_1,\hdots,\xv_N)$, $\allus = (\mathbf{u}_1,\hdots,\mathbf{u}_N)$, and $v_i(\allxs) \coloneqq \qsspeed[\rho_{\mathrm{d}}](\xv_i)$ and obtain
\begin{align}
    \partial_t \jointpdf(\allxs,\allus, t) &= \tau^{-1} \grad_{\uv_i} \vdot \left[\left(\uv_i + \grad_{\uv_i} \right) \jointpdf
    \right] + T \laplacian_\allxs \jointpdf\nonumber\\
    &\phantom{{}={}}- \grad_i \vdot(v_i(\allxs) \uv_i \jointpdf)
    \,,
    \label{eq:Nbody_FPE_QS}
\end{align}
where $\grad_i = \grad_{\xv_i}$ and $\laplacian_\allxs \coloneqq \grad_i\vdot\grad_i$; here and in the following, summation over particle indices is implied in all terms where the index appears more than once.
\par 
We make the Fokker--Planck equation dimensionless by the following rescaling: $\xv \to {\xv}/ \ellr$,~$t \to t/\tir$, where $\ellr$ and $\tir$ are, respectively, a macroscopic length and time scale. In general $\ellr$ and $\tir$ are the length and time scales over which $\rho$ varies, so that in a system subject to bulk phase separation $\ellr$ is of order the interfacial width $\xi$. (Different choices may be appropriate to study other types of states.) It is useful to introduce the dimensionless quantities
\begin{align}
    \epslvap &\coloneqq \frac{v_0\tau}{\ellr} \, ,
    & \epst &\coloneqq \frac{\tau}{\tir} \,,
    \label{eq:dimless_nos}
\end{align}
which compare the basic microscopic length scale (persistence length in the dilute limit $v_0\tau$) and time scale (persistence time $\tau$) to $\ellr$ and $\tir$, respectively.
We further introduce
\begin{align}
    A &\coloneqq \frac{v_0}{v_1}\,,& \etaT &\coloneqq \frac{\sqrt{T\tau}}{v_1\tau} \,.
    \label{eq:dimless_nos_2}
\end{align}
Here, $A$ can be viewed as a ratio of phase-specific persistence lengths $v_0\tau$ and $v_1\tau$, and $\etaT$ compares the diffusive length scale $\sqrt{T\tau}$, with the high-density persistence length.

Our target in the following is to describe the large-scale phenomenology in the diffusive limit: $\epst =\epsfac \epslvap^2$, where $\epsfac = \mathcal{O}(1)$ is positive and constant, and $\epslvap \downarrow 0$. (Our physical conclusions will not depend on the arbitrary scale $\epsfac$.) As we will show, this limit corresponds to a gradient expansion where $\ellr$, the length scale at which we are interested in describing the system, is taken to be larger than all microscopic length scales~\footnote{We will see that terms with $n$ derivatives in the final equation of the density field will correspond to terms of order $\epslvap^{n-2}$: expanding in gradients and in powers of $\epslvap$ are thus equivalent.}. In a phase-separated state, this becomes formally true in the vicinity of the critical point, where $\ellr\sim\xi$ diverges. 

With these rescalings, Eq.~\eqref{eq:Nbody_FPE_QS} takes the form
\begin{equation}
\begin{aligned}
    \epsfac\partial_t \jointpdf &= \frac{1}{\epslvap^2}\grad_{\uv_i} \vdot \left[\left(\uv_i + \grad_{\uv_i} \right) \jointpdf
    \right] + \frac{\etaTsq}{A^2}\laplacian_\allxs \jointpdf\\
    &\phantom{{}={}}- \frac{1}{\epslvap}\grad_i \vdot\left(v_i(\allxs) \uv_i \jointpdf\right)
    \,,
    \label{eq:Nbody_FPE_QS_dimless}
\end{aligned}
\end{equation}
where, after rescaling and with a slight abuse of notation, we restore $v_i/v_0\to v_i$. 
We now consider the moments of the $N$-body probability distribution $\jointpdf$, obtained by integrating out the orientational variables:
\begin{subequations}\label{eq:moments}
\begin{align}
    \hat\rho(\allxs,t) &= \int  \dd{\allus} \jointpdf(\allxs, \allus,t) 
    \,, 
    \\
    \hat\pol_i(\allxs,t) &= \int \dd{\allus} \jointpdf(\allxs, \allus,t)\,\uv_i
    \,,
    \\
    \hat\Qv_{ij}(\allxs,t) &= \int \dd{\allus} \jointpdf(\allxs, \allus,t)\,\uv_i\otimes\uv_j
    \,,
    \\
    \hat{\Tv}^{[n]}_{i_1,\ldots,i_n}(\allxs,t) &= \int \dd{\allus} \jointpdf(\allxs, \allus,t) \bigotimes_{m=1}^n \uv_{i_m}
    \,.
\end{align}
\end{subequations}
The $n$-th moment is a symmetric rank-$n$ tensor (in both spatial and particle indices). Note that we choose not to make the second moment and higher-order ones traceless. From Eqs.~\eqref{eq:Nbody_FPE_QS_dimless} and \eqref{eq:moments}, we obtain in App.~\ref{app:moment-eqs_qs-aoups} the time evolution for the infinite hierarchy of moments as
\begin{subequations}
\label{eq:dt_moments}
\begin{align}
    \epsfac\partial_t\hat\rho&= -\frac{1}{\epslvap}\grad_i\vdot\left(v_i \hat\pol_i \right) + \frac{\etaTsq}{A^2}\laplacian_\allxs \hat\rho\label{eq:dt_density-qs}
    \,,
    \\
    \epsfac\partial_t \hat\pol_i &= -\frac{1}{\epslvap^2}\hat\pol_i -\frac{1}{\epslvap}\grad_j\vdot\left(v_j \hat\Qv_{ij}\right) 
    \nonumber
    \\
    &\phantom{{}={}}+ \frac{\etaTsq}{A^2}\laplacian_\allxs \hat\pol_i 
    \,,\label{eq:dt_pol-qs}
    \\
    \epsfac\partial_t \hat\Qv_{ij} &= -\frac{2}{\epslvap^2}\hat\Qv_{ij} 
    -\frac{1}{\epslvap}\grad_k\vdot\left(v_k \hat\Tv^{[3]}_{ijk}\right)
     \nonumber
    \\
    &\phantom{{}={}}+ \frac{2}{\epslvap^2}\hat\rho\delta_{ij}\id
     + \frac{\etaTsq}{A^2}\laplacian_\allxs \hat\Qv_{ij} \,, \label{eq:dt_Q-qs}
     \\
    \epsfac\partial_t\hat{\vb{T}}^{[n]}_{i_1,\ldots,i_n} &= -\frac{n}{\epslvap^2} \hat{\Tv}^{[n]}_{i_1,\ldots,i_n} - \frac{1}{\epslvap}\grad_{i_0} \vdot \left(v_{i_0}\hat{\vb{T}}^{[n+1]}_{i_0,i_1,\ldots,i_n}\right)\nonumber\\
    &\phantom{{}={}} + \frac{n(n-1)}{\epslvap^2}\left(\delta_{i_1,i_2}\id\otimes\hat{\vb{T}}^{[n-2]}_{i_3,\ldots,i_n}\right)^\symm\nonumber\\
    &\phantom{{}={}} + \frac{\etaTsq}{A^2}\nabla^2_{\allxs}\hat{\vb{T}}^{[n]}_{i_1,\ldots,i_n}
    \,.\label{eq:dt_generalmoment-qs}
\end{align}
\end{subequations}
Here $\symm$ indicates symmetrization with respect to all spatial and particle indices, and the divergence always acts on the last spatial index, so that, \emph{e.g.}, ${\grad_j\vdot\hat\Qv_{ij}=\sum_{j, \beta}\partial{\hat Q^{\alpha\beta}_{ij}}/\partial{x^\beta_j}}$ (where Greek indices refer to spatial coordinates). 
\par 
We now want to describe the large-scale behavior of our system by obtaining a closed equation for the density field, leveraging the fact that every orientational moment relaxes fast compared with the density field on large scales. This, however, poses a significant challenge, as we have seen in Sec.~\ref{sec:fail_mini} for the toy model. 

We note in passing that the toy model can be obtained from Eqs.~\eqref{eq:dt_moments} by setting $v_i(\allxs)$ to be constant (hence, $A=1$) and assuming, as in 
the diffusion-drift approximation, that $\hat{\vb{T}}^{[n]}=0$ for $n > 2$, and $\hat{\Qv} = \hat\rho\id$. Since with constant self-propulsion the particles are noninteracting, one may then replace $\hat{\rho}$ and $\hat{\pol}_i$ with their one-particle versions to arrive at Eqs.~\eqref{eq:minimal model dimless} for the toy model.
\subsubsection{Multiple-scale expansion}
\label{sec:multiscales_qsap}
We start from the infinite hierarchy for the orientational moments in Eqs.~\eqref{eq:dt_moments}, and extend here the MS technique introduced in Sec.~\ref{sec:multiscale_mini} to capture and eliminate all moments but the density field. 
In Eqs.~\eqref{eq:dt_moments}, it is apparent that for $\epslvap \downarrow 0$ all $\hat\Tv^{[n]}$ with $n\geq 1$ are fast compared to the $N$-body density $\hat\marginaldistro(\allxs,t)$, which is slow. From power counting in Eqs.~\eqref{eq:dt_moments}, it transpires that each moment  $\hat\Tv^{[n]}$ contributes terms of order $\epslvap^n$ and higher in the equation for the $N$-body density $\hat\rho$. 
Since every $\epslvap$ arrives with a $\grad$, to obtain all terms of fourth order in gradients in the density dynamics, we only need moments up to $\hat\Tv^{[4]}$; therefore, we can set $\hat\Tv^{[5]} = \vb{0}$, as the corresponding contribution is negligible.
\par 
To this end, we follow the same procedure as in Sec.~\ref{sec:multiscale_mini} with the complication that we now have four moments to eliminate instead of the polarization only. The calculation is detailed in App.~\ref{app:coarsegraining_QS} and gives
\begin{equation}
\begin{aligned}
     \epsfac\partial_t \hat\rho &=  
    \grad_i \vdot\left[v_i \grad_i\left(v_i \hat\rho\right)\right] 
    +\frac{\etaTsq}{A^2} \nabla_i^2 \hat\rho
    \\
    &+\frac{\epslvap^2\etaTsq}{A^2}  \grad_i \vdot\left\{v_i \grad_i\left[\laplacian_\allxs\left(v_i \hat\rho\right) -v_i \laplacian_\allxs \hat\rho\right]\right\} \\
    & -\epslvap^2\grad_i \vdot\left(v_i \grad_i\left\{v_i \grad_j \vdot\left[v_j \grad_j\left(v_j  \hat\rho\right)\right]\right\}\right)
    \\
    &+\epslvap^2 \grad_i \vdot\left(v_i \grad_j \vdot\left\{v_j\left[\grad_i(v_i \grad_j\left(v_j \hat\rho)\right)\right]^\symm\right\}\right)
    \, ,
    \label{eq:qs_many-result}
\end{aligned}
\end{equation}
where we remind that summation over $i,j$ is implied.
In principle, the MS analysis presented here could be extended to derive terms of order $\mathcal{O}(\epslvap^4\nabla^6)$ or higher but we leave this interesting task for future investigations. 
\subsubsection{Mean-field limit}
\label{sec:mean-field}
We have derived the closed time evolution of the many-body distribution $\hat\rho$. This quantity depends on the spatial coordinates $\allxs$ of all $N$ particles. 
We now set out to obtain the dynamics for the one-particle density~$\rho$ (normalized to $N$):
\begin{align}
    \rho(\xv_1,t) \coloneqq N\int \prod_{i=2}^N \dd \xv_i \, \hat\rho(\allxs,t)
    \,.
\end{align}
To derive the time evolution for this density, we integrate Eq.~\eqref{eq:qs_many-result} over the spatial coordinates of the remaining ${N-1}$ particles, which yields an infinite BBGKY hierarchy of particle correlation functions. This is because $v_i = v_i(\allxs)$ in Eq.~\eqref{eq:qs_many-result} depends on the position of other particles due to the QS interaction.

Here, we take the mean-field limit, by assuming that the $N$-body density factorizes into one-body densities,
\begin{align}
    \hat\rho(\allxs, t) \approx \prod_{i=1}^N \rho(\xv_i,t)
    \,, 
\end{align}
and that the empirical density $\rho_{\mathrm{d}}$ in the QS speed can be approximated by the one-body density $\rho$. (From this, it obviously follows that $v_1(\allxs) = \qsspeed[\rho_{\mathrm{d}}](\xv_1) \approx \qsspeed[\rho](\xv)$.)  
The mean-field approximation is justified as long as the QS interaction range of the particles is large enough with respect to the typical interparticle distance such that each particle interacts with many others~\cite{demery2014generalized,bouchet2010thermodynamics,bouchet2016perturbative}. 
In this case, the one-particle density $\rho(\xv,t)$ obeys
\begin{equation}
\begin{aligned}
    \epsfac\partial_t \rho &=  
   \div\left[\qsspeed \grad\left(\qsspeed \rho\right)\right]  
    +\frac{\etaTsq}{A^2} \laplacian \rho
    \\
    &+\frac{\epslvap^2\etaTsq}{A^2}  \div\left\{\qsspeed \grad \left[\laplacian( \qsspeed \rho) -\qsspeed \laplacian \rho\right]\right\} \\
    & -\epslvap^2\div\left(\qsspeed \grad\left\{\qsspeed \div\left[\qsspeed \grad(\qsspeed  \rho)\right]\right\}\right)
    \\
    &+\epslvap^2 \div\left(\qsspeed \div\left\{\qsspeed[\grad (\qsspeed \grad(\qsspeed \rho))]^\symm\right\}\right)
    \,,
    \label{eq:qs_all-result}
\end{aligned}
\end{equation}
where $\qsspeed$ is now a functional of $\rho$, not $\hat\rho$.  
\par
Equation~\eqref{eq:qs_all-result} should be compared to the result of standard (non-iterated) adiabatic elimination in the DD approximation widely used in the literature~\cite{cates-2013, cates-2015, solon-2015, solon-2018, solon-2018-njp, martin-2021,dinelli-2024}:
\begin{equation}
\begin{aligned}
    \epsfac\partial_t \rho &=  
    \div\left[\qsspeed \grad(\qsspeed \rho)\right] 
    +\frac{\etaTsq}{A^2} \laplacian \rho
    \,. 
    \label{eq:qs_diff-drift}
\end{aligned}
\end{equation}
As can be immediately seen, the DD result corresponds to the leading-order contribution in $\epslvap$ in our multiscale formalism. Although DD has proved useful in the past to describe phase-separating systems while also expanding $\qsspeed$ in gradients to obtain {\em some} stabilizing terms of order $\epslvap^2\nabla^4$~\cite{cates-2013, cates-2015, solon-2015, solon-2018, solon-2018-njp, dinelli-2024}, it is clear from Eqs.~\eqref{eq:qs_all-result} and \eqref{eq:qs_diff-drift} that DD does not provide {\em all} terms at this order, and is therefore not a consistent coarse-graining at $\ofour$ level. Therefore, the MS approach is an essential step in the construction of an active Cahn--Hilliard theory whose five coefficient functions are microscopically determined, which is our goal here.
\subsubsection{Mapping to \texorpdfstring{$\ofour$}{O(grad 4)} theory}
\label{sec:ofour_qsap}
Equation~\eqref{eq:qs_all-result} is not yet local in $\rho$ and its gradients due to the presence of the QS speed $\qsspeed[\rho]$, which is nonlocal via Eqs.~\eqref{eq:vqs}. 
For simplicity, and although more general choices could be made,
we assume the QS kernel $\kernel$ to be isotropic (and normalized to unity) while $\vloc(\rho)$ is a local, monotonically decreasing function that captures the slowing down in dense regions (a specific choice for $\vloc(\rho)$ and $\kernel$ will be made in Sec.~\ref{sec:qs_application}). 
\par
To proceed, we expand the convolved density $\rho_\kernel$ in gradients~\cite{cates-2015}:
\begin{equation}
    \rho_\kernel \approx \rho 
    + \frac{\gamma^2}{(v_0\tau)^2}\epslvap^2 \laplacian \rho + \mathcal{O}\left(\epslvap^4\nabla^4\right)
    \,.
    \label{eq:density grad expansion}
\end{equation}
Here, $\gamma$ is a QS interaction length scale, within which particles sense their neighbors:
\begin{equation}
    \gamma  \coloneqq \left(\frac{1}{2}\int \dd{\xv}\, x^2 \kernel(\xv)\right)^{\frac{1}{2}}
    \label{eq:gamma}
    \, .
\end{equation}
Substituting Eq.~\eqref{eq:density grad expansion} into the definition of~$\qsspeed[\rho]$ yields
\begin{equation}
    \qsspeed[\rho] = \vloc(\rho) + \frac{\gamma^2}{(v_0\tau)^2}\epslvap^2 \vloc'(\rho)\laplacian \rho
    \,.
    \label{eq:vqs_grad-expansion_resc}
\end{equation}
%\par
%
Using Eq.~\eqref{eq:vqs_grad-expansion_resc} in Eq.~\eqref{eq:qs_all-result} and keeping all the terms to order $\mathcal{O}(\epslvap^2\nabla^4)$ finally yields an equation for the density in $\ofour$ theory form.
To find its coefficient functions, we now undo the rescaling in time and space in terms of $\ellr$ and $\tir$, reinstating the original variables of the particle model of Eqs.~\eqref{eq:qs_langevin}. This yields Eqs.~\eqref{eq:continuity} and \eqref{eq:o4} with the following microscopically derived coefficient functions:
\begin{subequations}
\label{eq:o4-coeffs_dimful}
    \begin{align}
        \mub(\rho) &= T \rho + \tau\int^\rho \dd{u}
        \left(\vloc \vloc' u + \vloc^2(u)\right)\,,\label{eq:coeffs_dimful_mub}
        \\
        K(\rho) &= - \gamma^2\tau\,\vloc \vloc' \rho
        - \cms T\tau^2 \vloc \vloc' \rho\,,\label{eq:coeffs_dimful_K}
        \\
        \lambda(\rho) &=  \cms T\tau^2 \vloc   
        \left(\vloc'' \rho + 2\vloc'\right) 
        + \frac{1}{2} \cms \tau^3 \tilde g(\rho)
        \,,\\
        \zeta(\rho) &=  - \gamma^2\tau\, \vloc \vloc'  + \cms T\tau^2  \vloc'^2 \rho 
        + \cms \tau^3 \tilde g(\rho)\,,
        \\
        \nu(\rho) &= \cms T\tau^2 \vloc' \left(\vloc'' \rho + 2\vloc'\right) 
        + \frac{1}{2} \cms \tau^3 \tilde g'(\rho)
        \,.
    \end{align}
\end{subequations}
Here, $\tilde g(\rho) \coloneqq  \vloc (\vloc \vloc'^2 \rho + \vloc^2 \vloc')$ is introduced to ease notation. Notice that, as expected, the diffusive scale parameter $\epsfac$ no longer appears in the coarse-grained theory. 
In Eqs.~\eqref{eq:o4-coeffs_dimful}, we have also introduced a bookkeeping parameter $\cms$ that is unity in the MS result but vanishes when employing the DD approximation. (The latter can be checked by inserting Eq.~\eqref{eq:vqs_grad-expansion_resc} into Eq.~\eqref{eq:qs_diff-drift}.) In the following, we will refer to these cases as the MS-$\ofour$ and DD-$\ofour$ theories, respectively, for the tQSAP system. 
\par

It is useful to recall the domain of validity of the MS-$\ofour$ theory as we have derived it for tQSAPs. Our coarse-graining is built on an expansion on $\epslvap$, such that the scale parameter $\ellr$ setting the window of investigation is taken to be much larger than any microscopic length scale. (This is equivalent to a gradient expansion as we have seen.) When studying phase separation, $\ellr$ should be taken of order the interfacial width, $\ellr\sim \xi$. Formally, therefore, the required separation of length scales occurs on approach to the mean-field critical point, where $\xi$ diverges. (Recall there is no noise in our treatment.)
As long as the particle density is sufficiently high to also justify the mean-field treatment of interactions that was used in Sec.~\ref{sec:mean-field}~\footnote{If noise were present, this requirement would exclude a fluctuation dominated region, namely, the Ginzburg interval, that surrounds the critical point itself.}, we can expect
our MS-$\ofour$ theory to approach {\em quantitative} validity in the near-critical region of the mean-field phase diagram. Moreover, as it does in equilibrium, the gradient expansion may offer a good semiquantitative description across much wider regions of parameter space. Both aspects will be confirmed numerically in Sec.~\ref{sec:qs_application}.
\par
Importantly, our MS derivation further allows us to discover a regime in which the DD-$\ofour$ theory is quantitatively justified: namely, when the $\cms$ terms in Eqs.~\eqref{eq:o4-coeffs_dimful} are negligible compared to the $\gamma^2$ terms. Comparing the prefactors of these terms reveals that this is the case when the QS length $\gamma$ is much larger than both the diffusive length,  $ \sqrt{T\tau}$ and the (bare) persistence length, $v_0\tau$. When this happens, although the DD approach omits terms that should formally be included at $\ofour$, these terms are negligible in practice.
In App.~\ref{app:abps}, we derive the $\ofour$ theory of tQSAPs with ABP dynamics, obtaining in Eqs.~\eqref{eq:abp-o4-coeffs_dimful} the coefficient functions replacing Eqs.~\eqref{eq:o4-coeffs_dimful}. While for AOUPs, $K(\rho)>0$ always in Eq.~\eqref{eq:coeffs_dimful_K}, consistently with our assumption in Sec.~\ref{sec:ofour_intro}, for ABPs $K(\rho)$ is no longer strictly positive. There, it can change sign between the spinodals, doing so far from the critical point, where the MS expansion is not guaranteed to be valid. This suggests the need for higher-order gradient terms in that regime for ABPs, a possibility we defer to future work.
\subsubsection{Linear stability}
\label{sec:linear-stability_qsap}
It is informative to compare our MS-$\ofour$ and DD-$\ofour$ theories at linearized level.
To this end, we consider a homogeneous state at density $\rho_0$ and apply a small perturbation to it: $\rho(\xv, t) = \rho_0 + \delta \rho(\xv, t)$. 
In Fourier space, the dynamics of the perturbation then reads
\begin{equation}
\begin{aligned}
    \partial_t \delta \rho_{\qv} 
    &=  - \left[T  + 
        \tau \vloc(\rho_0)\left(\vloc'(\rho_0) \rho_0 + \vloc(\rho_0)\right)\right] q^2 \delta\rho_{\qv}
    \\
    &\phantom{=}
    + \vloc'(\rho_0) \vloc(\rho_0) \rho_0 \left(\gamma^2\tau  
        + \cms T\tau^2   \right) q^4 \delta\rho_{\qv}
    \,.
    \label{eq:qs_linear}
\end{aligned}
\end{equation}
We thus find the spinodal instability triggering MIPS when $\vloc(\rho)$ is sufficiently strongly decreasing or, precisely, when
\begin{align}
    \mub'(\rho_0) = T  + 
    \tau\vloc(\rho_0)\left(\vloc'(\rho_0) \rho_0 + \vloc(\rho_0)\right) 
    <0 
    \,.
    \label{eq:mips_threshold}
\end{align}
We have thus shown that the spinodal instability is predicted identically by the MS-$\ofour$ and DD-$\ofour$ theories. Furthermore, for $\vloc'(\rho_0) < 0$, the $q^4$ term is stabilizing in both theories, regularizing that instability at small length scales, although its amplitude differs for the two theories, once translational diffusion is present ($T> 0$).

\section{Phase coexistence in \MakeLowercase{t}QSAPs}
\label{sec:qs_application}
In this section we derive theoretical predictions for phase coexistence in tQSAPs (with Ornstein--Uhlenbeck orientation dynamics).
We then compare these to direct numerical simulations of the particle model. The theoretical results in this section are obtained by applying the results of Sec.~\ref{sec:ofour_framework} to the MS-$\ofour$ and DD-$\ofour$ theories for tQSAPs, as derived in Eqs.~\eqref{eq:o4-coeffs_dimful}.
\par
We first discuss in Sec.~\ref{sec:qsap_dimensionless} the four nondimensional parameters of microscopic tQSAPs; then, in Secs.~\ref{sec:qsap_theory} Sec.~\ref{sec:qsap_theory_tensions} we show, respectively, how the predictions for binodals and interfacial tensions obtained from DD-$\ofour$ differ from those obtained from MS-$\ofour$. We will show that MS-$\ofour$ theory also predicts distinctive qualitative phenomenology, absent in DD-$\ofour$. This includes the emergence of reentrant phase separation at small translational diffusivity and the dependence of the binodals and Ostwald tensions on the QS length scale $\gamma$ (defined in Eq.~\eqref{eq:gamma} above).  

In Sec.~\ref{sec:qsap_sim}, we compare these theoretical predictions with particle simulations.
We confirm the accuracy of MS-$\ofour$ in and around its formal range of validity for both the binodals and the interfacial tensions, and show that phase separation is indeed reentrant at small translational diffusion, as predicted by MS-$\ofour$ but missed by the DD approximation.
In App.~\ref{app:all_details} we discuss further theoretical predictions obtained from DD-$\ofour$ and MS-$\ofour$ theories. 
\subsection{QS speed response and nondimensional units}
\label{sec:qsap_dimensionless}
As in other studies of QS particles~\cite{cates-2013, cates-2015, solon-2015, solon-2018, solon-2018-njp, dinelli-2024}, we make the following choice for the QS speed $\vloc(\rho)$:
\begin{equation}\label{eq:qs_vloc}
    \vloc(\rho) = v_0 - \frac{v_0-v_1}{2}\left[1 + \tanh(\frac{\rho - \rhom}{\rhow})\right]\,.
\end{equation}
This describes a slowdown in speed from $v_0$ to $v_1$ over a density interval of width $\rhow$ centered around the midpoint density $\rhom$. For the kernel~$\kernel$, we choose~\cite{solon-2015,solon-2018-njp,dinelli-2024}
\begin{equation}\label{eq:kernel}
    \kernel(\mathbf{r}) = \frac{1}{Z} \Theta(\rcut - r)
    \exp\left(-\frac{\rcut^2}{\rcut^2 - r^2}\right)\,,
\end{equation}
where $Z$ ensures normalization. It follows from Eq.~\eqref{eq:gamma} that $\gamma = c_\gamma \rcut$ with proportionality constant $c_\gamma \approx 0.361$ in $d=2$.
Numerical simulations presented in this section were performed with a parallel molecular dynamics code simulating Eqs.~\eqref{eq:qs_langevin} with the choices given by Eqs.~\eqref{eq:qs_vloc} and~\eqref{eq:kernel}. For details, see App.~\ref{app:sims}.

With these choices made, the $\ofour$ coefficient functions of Eq.~\eqref{eq:o4-coeffs_dimful} can now be rewritten in terms of four dimensionless parameters of the microscopic dynamics. Two of these, $A = v_0/v_1$ (as in Eq.~\eqref{eq:dimless_nos_2}) and 
\begin{equation}
    S=\frac{\rhom}{\rhow}
\end{equation}
characterize the QS speed response via Eq.~\eqref{eq:qs_vloc}.
The third, $\etaT$ (defined in Eq.~\eqref{eq:dimless_nos_2}), compares the persistence length in the fluid phase $v_1\tau$ to the thermal diffusive length. The fourth and final parameter,
\begin{equation}
    \alphag = \frac{v_1\tau}{\gamma}
    \,,
   \end{equation}
compares the persistence length to the QS kernel scale.  

We may now express the $\ofour$ coefficient functions in terms of the reduced microscopic parameters $A,S,\etaT,\alphag$ by choosing $v_1\tau$, $\tau$, and $\rhom$ as units of space, time, and density, respectively (then, $v_1$ is the unit for $\vloc(\rho)$). With these rescalings, Eqs.~\eqref{eq:o4-coeffs_dimful} read
\begin{subequations}
\label{eq:coefficients}
    \begin{align}
        \mub(\rho) &= \etaTsq \rho + \int^\rho \dd{u}
        \left(\vloc \vloc' u + \vloc^2(u)\right)\,, \label{eq:coeffs_mub}
        \\
        K(\rho) &= - \alphag^{-2} \vloc \vloc' \rho
        - \cms\etaTsq \vloc \vloc' \rho\,,
        \label{eq:coeffs_K}
        \\
        \lambda(\rho) &=  \cms\etaTsq \vloc   
        \left(\vloc'' \rho + 2\vloc'\right) 
        + \frac{1}{2} \cms g(\rho)\,,\\
        \zeta(\rho) &=  - \alphag^{-2} \vloc \vloc'  + \cms\etaTsq  \vloc'^2 \rho 
        +\cms g(\rho)\,,
        \\
        \nu(\rho) &= \cms\etaTsq \vloc' \left(\vloc'' \rho + 2\vloc'\right) 
        + \frac{1}{2} \cms g'(\rho)
        \,,
        \label{eq:coeffs_nu}
    \end{align}
\end{subequations}
with $g(\rho) = \vloc \left(\vloc \vloc'^2 \rho + \vloc^2 \vloc'\right)$. Note that the parameters $\alphag, \etaT$ enter explicitly, whereas the QS response parameters $A,S$ enter these equations via $\vloc$ and its derivatives, including those in $g(\rho$) and the integral in Eq.~\eqref{eq:coeffs_mub}.
\subsection{Binodals: theoretical predictions}
\label{sec:qsap_theory}
We now apply the results of Sec.~\ref{sec:ofour_framework} to predict the binodal densities of tQSAPs. We will directly compare the MS-$\ofour$ and DD-$\ofour$ theories, which correspond to choosing $\cms = 1,0$, respectively, in Eqs.~\eqref{eq:o4-coeffs_dimful}, or equivalently, Eqs.~\eqref{eq:coefficients}.
\par
The binodal densities are obtained from Eq.~\eqref{eq:binodals}. In the integrand of those conditions, three functions appear: the bulk chemical potential $\mub(\rho)$, which is the same for DD and MS, and the two pseudodensities $\psi^{(1,2)}(\rho)$, encoding the effect on the binodals of the remaining $\ofour$ terms. Notably, therefore, only the combinations of coefficient functions that enter the pseudodensities through Eq.~\eqref{eq:psi} affect the binodals. These are $K(\rho)$ in Eq.~\eqref{eq:coeffs_K}, and
\begin{subequations}\label{eq:coefficients_binodals}
    \begin{align}
        \bar\lambda(\rho) &= \frac{1}{2}\alphag^{-2}\vloc\vloc' + \cms\etaTsq    
        \left(\vloc\vloc'' \rho + 2\vloc\vloc' - \frac{\vloc'^2\rho}{2}\right)
        \,,\\
        \bar\nu(\rho) &= \frac{1}{2}\alphag^{-2} \left(\vloc \vloc''+\vloc'^2\right) + \frac{3}{2}\cms\etaTsq \vloc'^2
        \,.
    \end{align}
\end{subequations}
Note that $g(\rho)$ (as defined after Eq.~\eqref{eq:coefficients}) does not appear in these expressions; hence, the binodals of the MS and DD theories differ only when translational diffusion is present, $\etaT> 0$. (This is, however, not the case for ABPs, see App.~\ref{app:abps}.)

As shown in App.~\ref{app:ofour_qsap_crit}, the MS-$\ofour$ and DD-$\ofour$ theories predict the same mean-field critical point.
Moreover, they yield the same binodal curves whenever the $\cms$ terms in Eqs.~\eqref{eq:coefficients_binodals} are negligible compared to the remaining ones, \emph{i.e.}, whenever $\etaT \ll \alphag^{-1}$, which is equivalent to the diffusive length scale being much smaller than the QS length, $\sqrt{T\tau} \ll \gamma$. Notably, this condition for AOUPs is less stringent than the one discussed in Sec.~\ref{sec:coarsegraining_qsap} under which the entirety of MS-$\ofour$ (not just its binodals) reduces to DD-$\ofour$. The latter additionally requires $v_0\tau\ll\gamma$ so that the $\cms$ terms are negligible throughout Eqs.~\eqref{eq:coefficients} and not just in Eqs.~\eqref{eq:coefficients_binodals}.

Note also that the only $\etaT$ dependence in DD-$\ofour$ theory enters via the bulk chemical potential $\mub(\rho)$ in Eq.~\eqref{eq:coeffs_mub}, reflecting the Fickian diffusive current $\Jv_\mathrm{diff} = -\etaTsq\grad\rho$. In contrast, the MS-$\ofour$ theory additionally has an active, non-Fickian dependence on $\etaT$ in its current, causing the terms in $\cms$ to appear in Eqs.~\eqref{eq:coefficients_binodals}. We discuss the physics of these important contributions in Sec.~\ref{subsec:polatin} below.

\begin{figure}[tbp]
    \centering
    \includegraphics[width=\columnwidth]{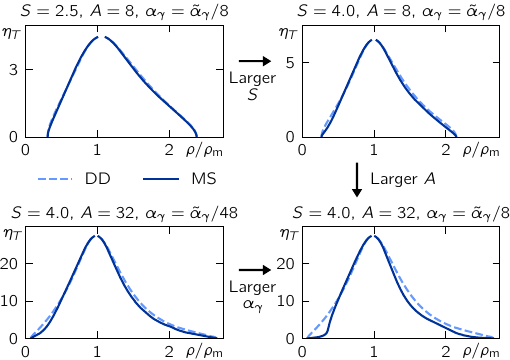}
    \caption{Predictions of binodal curves of DD- and MS-$\ofour$ theories for tQSAPs with various parameter values (see legends), as functions of $\etaT$. The difference between the two theories becomes larger upon increasing $A$, $S$, or $\alpha_\gamma$. The two theories exactly agree on the location of the critical point, and at $\etaT=0$. Here and in subsequent figures, $\alphagt = c_\gamma^{-1} \approx 2.77$ denotes the value of $\alphag$ obtained for $v_1\tau=1$, $\rcut=1$.\label{fig:qsap_theory}}
\end{figure}

To compare the binodals of MS-$\ofour$ and DD-$\ofour$, we solved Eqs.~\eqref{eq:psi} and \eqref{eq:binodals} numerically, with coefficient functions obeying Eqs.~\eqref{eq:coefficients}. (The procedure is discussed in App.~\ref{app:bincalc}.) As shown in Fig.~\fref{fig:qsap_theory}, we find that an extended regime emerges where the binodals of the two theories differ appreciably. This difference is more prominent for larger $\alphag$, $A$, or $S$, as we further explore below.
\begin{figure*}[tbp]
    \centering
    \includegraphics[width=\textwidth]{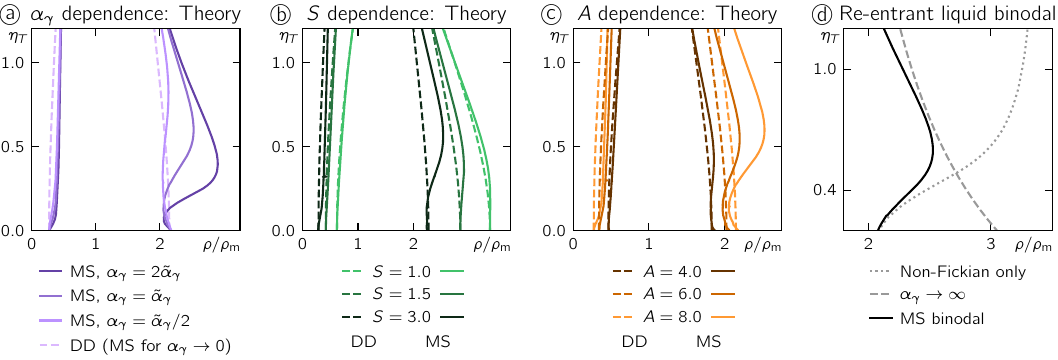}
    \caption{Dependence on QS length ($\gamma$) and reentrant phase separation in $\ofour$ theory of tQSAPs. a) $\alphag$ dependence of binodals ($A=8$, $S=4$). MS-$\ofour$ theory predicts $\gamma$-dependent binodals while DD-$\ofour$ does not. The DD binodals are recovered from MS when $\sqrt{T\tau} \ll \gamma$ or, for fixed $T\tau$ as used here, when $\alphag \to 0$. While the DD binodals are always monotonic as a function of $\etaT$, MS theory shows reentrant phase separation. For larger $\alphag$, the re-entrance is more pronounced and its location moves to smaller values of $\etaT$. b) $S$ dependence of binodals ($A=8$, $\alphag=\alphagt$, with $\alphagt$ given in Fig.~\ref{fig:qsap_theory}). While DD predicts monotonic binodals throughout, MS-$\ofour$ theory predicts more pronounced reentrant phase separation as $S$ is increased. c) Likewise for $A$ dependence of binodals ($S=4$, $\alphag=\alphagt$). d) Re-entrant phase separation in the liquid binodal results from the interplay between the Fickian and non-Fickian dependence on $\etaT$. The non-Fickian dependence on its own (dotted line) shifts the liquid binodal to the right for increasing $\etaT$; the Fickian dependence instead leads to a monotonic decrease of $\rhoL$ as a function of $\etaT$ (cf. DD line in panel a). Together, they give rise to the nonmonotonic behavior seen in MS theory. For large $\etaT$, the MS binodal approaches the $\alphag\to\infty$ limit (dashed line).\label{fig:qsap_RPS_theory}}
\end{figure*}
\subsubsection{Dependence on the QS length scale \texorpdfstring{$\gamma$}{gamma}}
One qualitative difference between DD-$\ofour$ and MS-$\ofour$ theories is that DD binodals do not depend on the QS length parameter $\gamma$, whereas the MS binodals do. 
To see this, note that $\mub(\rho)$ is independent of $\gamma$; the only $\gamma$ dependence lies in the other $\ofour$ coefficients in Eqs.~\eqref{eq:coefficients}. There, for DD, $\alphag^{-2}$ appears as an overall factor in all three functions $K(\rho), \bar\lambda(\rho), \bar\nu(\rho)$: thus, it drops out of the pseudodensity equation \eqref{eq:psi}. This implies that the binodals are independent of $\alphag$ even if other predictions of the DD-$\ofour$ theory are not. In contrast, in the MS-$\ofour$ coefficient functions the presence of the $\cms$ terms means that no such overall factor can be extracted, and the binodals do depend on $\gamma$. In Fig.~\fref[a]{fig:qsap_RPS_theory}, we show the dependence of the MS binodals on $\alphag$. As mentioned already, the DD binodals correspond to the limit $\sqrt{T\tau} \ll \gamma$, or equivalently, to the $\alphag \to 0$ limit of the MS binodals, because the $\cms$ terms are negligible in this limit.
\subsubsection{Re-entrant phase separation}
\label{subsec:reent}
A second qualitative difference between the DD-$\ofour$ and MS-$\ofour$ theories is that, as evident from Fig.~\fref{fig:qsap_RPS_theory}, MS predicts reentrant phase separation (nonmonotonic binodals as a function of $\etaT$) while DD does not.  
We now argue that it is the interplay between the Fickian and non-Fickian $\etaT$-dependent currents, the latter of which are absent in DD, that gives rise to the reentrant phase separation.
To study this interplay, we isolate the two currents by computing the binodals upon separately suppressing the $\etaT$ term in the bulk chemical potential and in the higher-order gradient terms. The Fickian contribution always results in bringing the binodals together as $\etaT$ is increased, reflecting the effect of the diffusive current going from the liquid into the vapor (see dashed line in Fig.~\fref[a]{fig:qsap_RPS_theory}). On the other hand, the non-Fickian contribution on its own shifts both binodals to higher densities, with the effect being more prominent for the liquid binodal (see dotted line in Fig.~\fref[d]{fig:qsap_RPS_theory}). Hence, for the liquid binodal, the non-Fickian and Fickian contributions from $\etaT$ compete with each other, resulting in the nonmonotonic behavior once both effects are taken into account (solid line in Fig.~\fref[d]{fig:qsap_RPS_theory}).
\par
The binodals are, however, monotonic when $\etaT\gg \alphag^{-1}$, approaching the curve given by the $\alphag \to \infty$ limit (see dashed line in Fig.~\fref[d]{fig:qsap_RPS_theory}). Indeed, the non-Fickian shift saturates as $\etaT \gg \alphag^{-1}$, since then the $\cms$ terms in Eqs.~\eqref{eq:coefficients} dominate the DD ones, and $\etaTsq$ becomes an overall factor.
\par
Moreover, the saturation of the non-Fickian shift also sets the location of the binodal re-entrance, which occurs when the diffusive and QS length scales are comparable to each other: This can be expressed either as $\sqrt{T\tau} \sim \gamma$ or $\etaT \sim \alphag^{-1}$.
In contrast, the magnitude (as opposed to location) of the reentrant feature depends on $A,S,\alphag$. As shown in Fig.~\fref[a--c]{fig:qsap_RPS_theory}, the amplitude of the re-entrance grows with larger $\alphag$, $S$, or $A$.
\par
Let us finally notice that re-entrance of the liquid binodal occurs far from the critical point. The formal assumptions of length-scale separation underlying our MS-based coarse-graining procedure are not fulfilled in this regime. However, the hope that the MS-$\ofour$ theory is a reliable guide to the physical behavior beyond its formal validity range is fulfilled. Indeed, we show in Sec.~\ref{sec:qsap_sim} that its prediction of reentrant phase separation is confirmed in particle simulations.

\subsubsection{Polarization at the interface: origin of non-Fickian current}\label{subsec:polatin}
\begin{figure}[tbp]
    \centering
    \includegraphics[width=\columnwidth]{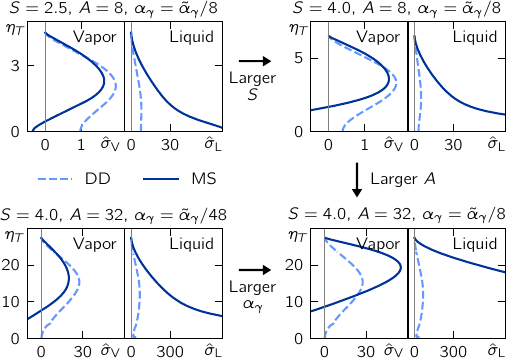}
    \caption{Predictions of interfacial tensions of DD- and MS-$\ofour$ theories for tQSAPs with various parameter values (see legends), as functions of $\etaT$. The rescaled tensions $\hat\sigma_\mathrm{V,L}$ of Eq.~\eqref{eq:sigma_scaling_dimensions} are plotted. As for the binodals in Fig.~\ref{fig:qsap_theory}, the difference between the two theories becomes larger upon increasing $A$, $S$, or $\alpha_\gamma$.\label{fig:qsap_theory_tensions}}
\end{figure}
We next explain in more mechanistic terms why the DD-$\ofour$ theory omits some of the important physics governing phase separation in tQSAPs.
We observe that, in the presence of thermal translational diffusion, QS particles acquire nonvanishing polarization at interfaces.
This is because, at stationarity, the (Fickian) diffusive current from the liquid to the vapor phase has to be balanced by a ballistic current from the vapor to the liquid, which requires particles at the interface to be polarized in this direction.
(This is a different mechanism from self-propelled particles with excluded volume, in which polarization emerges at the interfaces irrespectively of the presence of translational diffusion in the microscopic model~\cite{solon-2018-njp,omar-2023,soto-2024}.) In turn, the interface polarization causes the emergence of non-Fickian $\etaT$-dependent currents that constitute one of the key differences between the DD-$\ofour$ and MS-$\ofour$ theories.

Indeed, if we set $\partial_t\hat\rho=0$ in Eq.~\eqref{eq:dt_density-qs}, then within the mean-field approximation, we get for the one-particle polarization $\pol$:
\begin{equation}
    \pol = \frac{\etaTsq}{\qsspeed[\rho]}\grad\rho
    \,.\label{eq:interface polarization}
\end{equation}
Hence, whenever $\etaT \neq 0$, we do expect a nonvanishing polarization to emerge at the interface. This is confirmed by particle simulations in Fig.~\fref[a]{fig:qsap_RPS_sim}. Due to this localized polarization, the Laplacian term in $\partial_t\hat\pol$ described by Eq.~\eqref{eq:dt_pol-qs} is nonvanishing, leading to extra currents in the proximity of the interface due to $\etaT$. These non-Fickian currents are captured only by MS-$\ofour$ theory and missed by DD-$\ofour$.

Notice that while this is the only reason why the two sets of predictions for binodals differ for AOUPs (because of which, they become identical when $\eta_T=0$), for tQSAPs consisting of ABPs, the binodals differ even when $\eta_T=0$; see App.~\ref{app:abps}.
\subsection{Interfacial tensions: theoretical predictions}
\label{sec:qsap_theory_tensions}
\begin{figure}[tbp]
    \centering
    \includegraphics[width=\columnwidth]{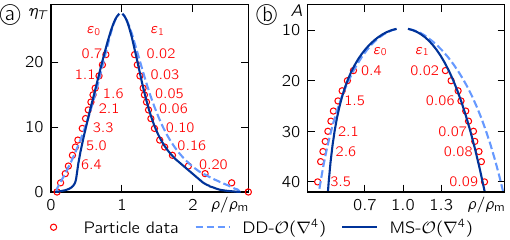}
    \caption{Quantitative comparison of $\ofour$ theory binodals with particle simulations of tQSAPs. a) Phase diagram varying $T$ for $v_0 = 4$, $v_1 = 1/8$, $\rhow = 50$, $\rhom = 200$, $\rcut = 1$ ($\gamma \approx 0.361$), $\tau = 1$  (corresponding to varying $\etaT$ at fixed $A = 32$, $S = 4$, $\alphag=\tilde{\alpha}_\gamma/8$). Note that approaching the critical point more closely makes precise determination of the binodals unfeasible in particle simulations (at constant $\rhom$) due to the increasing noise. 
    We also show estimates for the values of $\epslvap$ and $\epslliq$ defined in Eqs.~\eqref{eq:dimless_nos} and~\eqref{eq:epslliq}. 
    b) Phase diagram varying $v_0$ for $T = 1$, $v_1 = 1/8$, $\rhow = 5$, $\rhom = 20$, $\rcut = 1$, $\tau = 1$ (corresponding to varying $A$ at fixed $S = 4$, $\alphag=\tilde{\alpha}_\gamma/8$, $\etaT = 8$).
    \label{fig:qsap_binodals_quant} } 
\end{figure}
\begin{figure*}[tbp]
    \centering
    \includegraphics[width=\textwidth]{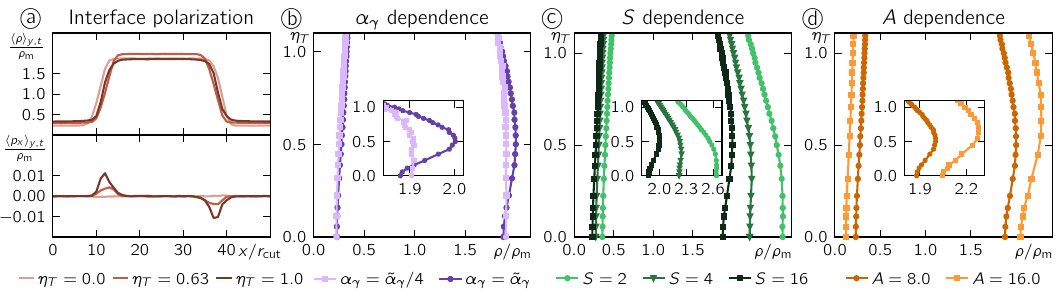}
    \caption{Interface polarization and reentrant phase separation in particle simulations of tQSAPs. a) Density (top) and polarization (bottom) plots (see App.~\ref{app:sims} for the measurement procedure), varying $T$ at fixed $v_0 = 8$, $v_1 = 1$, $\rhow = 1.25$, $\rhom = 20$, $\rcut = 1$, $\tau = 1$ (corresponding to varying $\etaT$ at fixed $A=8$, $S=16$, $\alphag = \alphagt$). For $\etaT>0$, a polarization emerges at the interface, pointing towards the liquid phase, in accordance with Eq.~\eqref{eq:interface polarization}. b) $\alphag$ dependence of binodals (varying $T$ and $v_0\in\{2,8\}$ with $v_1 = v_0/8$ at fixed $\rhow = 1.25$, $\rhom = 20$, $\rcut = 1$, $\tau = 1$, corresponding to varying $\etaT$, $\alphag$ at fixed $A=8$, $S=16$). Re-entrant phase separation is more prominent as $\alphag$ is increased. c) $S$ dependence of binodals (varying $T$ and $\rhow \in \{10,5,1.25\}$ at fixed $v_0=8$, $v_1 = 1$, $\rhom = 20$, $\rcut = 1$, $\tau = 1$, corresponding to varying $\etaT$, $S$ at fixed $A=8$, $\alphag=\alphagt$). Increasing $S$ results in a more prominent binodal re-entrance. d) $A$ dependence of binodals (varying $T$ and $v_0 \in \{8,16\}$ at fixed $v_1 = 1$, $\rhow = 1.25$, $\rhom = 20$, $\rcut = 1$, $\tau = 1$, corresponding to varying $\etaT$, $A$ at fixed $S=16$, $\alphag=\alphagt$). Increasing $A$ results in a more prominent binodal re-entrance. Insets in panels b--d): Close-up views on the liquid binodal.\label{fig:qsap_RPS_sim}}
\end{figure*}
We now report the theoretical predictions of the DD and MS-$\ofour$ theories on the interfacial tensions of tQSAPs. These are obtained by applying Eq.~\eqref{eq:sigma_rho} to the coefficient functions of Eqs.~\eqref{eq:coefficients}. First, let us highlight that unlike the AOUP binodals, the interfacial tensions differ between DD-$\ofour$ and MS-$\ofour$ even in the absence of translational diffusion ($\etaT = 0$), due to the contributions involving $g(\rho)$ in Eqs.~\eqref{eq:coefficients}.

It is convenient to consider the nondimensional version of the Ostwald tensions, defined as 
\begin{equation}
\hat\sigma_\mathrm{V,L}(A,S,\etaT,\alphag)
=
\frac{\tau}{\rhom^2(v_1\tau)^2\gamma}
    \sigmaVL \,.\label{eq:sigma_scaling_dimensions}
\end{equation}
We now show that in DD-$\ofour$ theory, $\hat\sigma_\mathrm{V,L}$ do not depend on $\alphag$. Note first that in DD-$\ofour$ (cf. Eqs.~\eqref{eq:o4-coeffs_dimful}), the gradient terms are proportional to $\gamma^2$, while $\mub$ is $\gamma$-independent. Thus, Eq.~\eqref{eq:w} implies that $w(\rho)\sim \gamma^{-2}$. Using Eq.~\eqref{eq:sigma_rho}, we then find that the tensions are proportional to $\gamma$, $\sigmaVL \sim \gamma$. This, however, implies that there can be no $\gamma$ dependence in the left-hand side of Eq.~\eqref{eq:sigma_scaling_dimensions}, so that in DD-$\ofour$:
\begin{equation}
    \hat\sigma^{(\mathrm{DD})}_\mathrm{V,L}(A,S,\etaT,\alphag) = \hat{\sigma}^{(\mathrm{DD})}_\mathrm{V,L}(A,S,\etaT) \,.
\end{equation}
In contrast, in MS-$\ofour$ theory $\hat\sigma_\mathrm{V,L}$ does
depend on the quorum-sensing length $\gamma$, due to the extra terms involving $\cms$ in Eqs.~\eqref{eq:o4-coeffs_dimful}.

To further compare the dependence of the Ostwald tensions on the parameters in the two $\ofour$ theories, we inspect the interfacial tensions numerically (see App.~\ref{app:bincalc} for details). Using the known binodals from the previous section, we compute the interfacial profile $w(\rho)$ of Eq.~\eqref{eq:w} and the phase-anchored pseudodensities satisfying Eqs.~\eqref{eq:psi} and~\eqref{eq:psi_anch}. We then calculate the tensions $\sigmaVL$ via Eq.~\eqref{eq:sigma_rho}. In Fig.~\fref{fig:qsap_theory_tensions}, we show the two tensions (rescaled according to Eq.~\eqref{eq:sigma_scaling_dimensions}) as a function of $\etaT$, for varying values of $\alphag$, $A$, or $S$. 

As shown in  Fig.~\ref{fig:qsap_theory_tensions}, the magnitude of the liquid tensions predicted from the MS-$\ofour$ and DD-$\ofour$ theories differ significantly. On the other hand, the vapor tension attains comparable, yet different, values within the MS-$\ofour$ and DD-$\ofour$ theories, and it becomes negative within MS-$\ofour$ theory far away from the critical point (where the theory is not expected to be valid).
As with the binodals, we further find that the difference between the tensions predicted by the two theories becomes larger when increasing any of the parameters $\alphag$, $A$, and $S$. Moreover, as expected, Fig.~\ref{fig:qsap_theory_tensions} shows that the rescaled tensions $\hat\sigma_\mathrm{V,L}$ do not depend on $\alphag$ in DD-$\ofour$, while they do in MS-$\ofour$.
We compare these theoretical predictions to data from  particle simulations in the next section.
\subsection{Particle simulations}
\label{sec:qsap_sim}
\begin{figure}[tbp]
    \centering
    \includegraphics[width=\columnwidth]{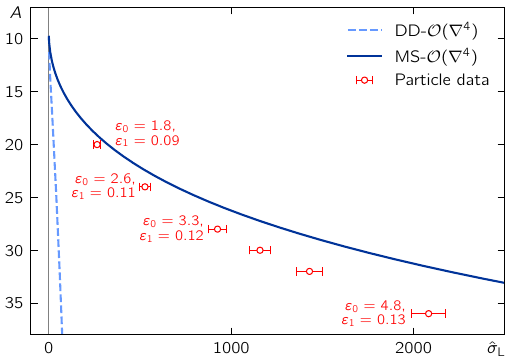}
    \caption{Quantitative comparison of the liquid tension of $\ofour$ theory with particle simulations of tQSAPs. Liquid tension plot varying $v_0$ for $T = 1$, $v_1 = 1/8$, $\rhow = 50$, $\rhom = 200$, $\rcut = 1$, $\tau = 1$ (corresponding to varying $A$ at fixed $S = 4$, $\alphag=\tilde{\alpha}_\gamma/8$, $\etaT = 8$, as in Fig.~\fref[b]{fig:qsap_binodals_quant}). Estimates are indicated for $\epslvap$ and $\epslliq$, obtained by measuring the interface width of the flat binodals. The error bars derive from uncertainties in the droplet radius and binodal densities.
    \label{fig:qsap_tensions_sim}}
\end{figure}
\subsubsection{Binodal densities}
In Fig.~\fref[a]{fig:qsap_binodals_quant}, we report the binodals measured from particle simulations performed at fixed $S$, $A$, and $\alphag$, varying $\etaT$, while Fig.~\fref[b]{fig:qsap_binodals_quant} shows
those obtained by changing $A$ at fixed $S$, $\etaT$ and $\alphag$. Let us recall that the MS-$\ofour$ theory is formally obtained perturbatively in the regime where $\epslvap = v_0 \tau/\ellr$ is small, so that the persistence length {\em at low densities} is small compared to the interfacial width $\xi\simeq \ellr$. However, as shown in Fig.~\fref{fig:qsap_binodals_quant}, it appears that the MS-$\ofour$ binodals are {\em quantitatively} accurate under a much less restrictive condition, namely, that the persistence length {\em at high densities}, $v_1\tau$, is small compared to $\xi$:
\begin{equation}\label{eq:epslliq}
    \epslliq \coloneqq \frac{v_1\tau}{\ellr} \ll 1\,. 
\end{equation}
The values of both $\epslvap$ and $\epslliq$ are annotated on Figs.~\fref[a]{fig:qsap_binodals_quant} and \fref[b]{fig:qsap_binodals_quant}~\footnote{Arguably one should use the length scales $\vloc(\rhoV)\tau$ and $\vloc(\rhoL)\tau$ instead of $v_0\tau$ and $v_1\tau$ but the differences are small in our simulations.}. In each case, we use for $\ellr$ the interfacial width $\xi$ obtained by fitting the interfacial profile to a $\tanh$ function (see App.~\ref{app:sims} for details). 

Given the formal requirement for small $\epslvap$, it is remarkable that our MS-$\ofour$ predictions remain accurate for both binodals up to $\epslvap \lesssim 2$, and for the liquid binodal only, to still higher values of around $\epslvap \approx 4$. However, these correspond to values of $\epslliq$ of around $0.06$ and $0.1$, respectively, so that Eq.~\eqref{eq:epslliq} holds. Throughout this entire regime, at least for the parameters considered in Figs.~\fref{fig:qsap_binodals_quant}, MS gives a significantly more accurate prediction than DD. On the other hand, in Fig.~\fref[b]{fig:app_details} in App.~\ref{app:all_details}
 we show parameters for which $\epslliq\sim \mathcal{O}(1)$, where the liquid binodal is not accurately captured by MS-$\ofour$ theory. We note that DD-$\ofour$ predicts well the vapor binodal (but not the liquid one) throughout Fig.~\fref{fig:qsap_binodals_quant} and also predicts both binodals well in Fig.~\fref[b]{fig:app_details} when $\varepsilon_{0,1}$ are large. Currently, we have no theoretical argument to explain its success in the latter case, where MS-$\ofour$ should, and does, fail. We also notice, however, that this unexpected success of DD is limited to the binodals while, in the same regime, the Ostwald interfacial tensions $\sigmaVL$ are severely misestimated (see Fig.~\fref[c,d]{fig:app_details} in App.~\ref{app:all_details}).
\par
The above findings show that the MS-$\ofour$ theory has a domain of {\em quantitative} accuracy extending well beyond its formal range of validity.
We now turn to verify the two {\em qualitative} predictions made in Sec.~\ref{sec:qsap_theory} above via MS-$\ofour$ theory that are not captured within the DD approximation. These are the dependence of the binodals on the QS length scale~$\gamma$, and the existence of reentrant phase separation at small $\eta_T$. In Fig.~\fref[b]{fig:qsap_RPS_sim}, we show that fixing all other parameters while varying $\alphag$ in particle simulations shows significantly changing binodals, confirming the first prediction. 
\par 
Furthermore, we observe in Fig.~\fref[b--d]{fig:qsap_RPS_sim} parameter regimes where reentrant phase separation appears in the particle simulations at small $\etaT$, confirming the second prediction. In accord with MS-$\ofour$ theory (cf.~Fig.~\ref{fig:qsap_RPS_theory}), the binodal re-entrance is more prominent as $\alphag$, $S$, and $A$ are increased (see Fig.~\fref[b--d]{fig:qsap_RPS_sim}). This suggests that the theory has indeed captured the physical mechanism for re-entrance described in Sec.~\ref{subsec:reent} above.
\par 
Note, however, that the precise parameters at which the re-entrance occurs in particle simulations differ considerably from the theoretical predictions of MS-$\ofour$. This is to be expected insofar as the nonmonotonicity of the liquid binodal emerges at small $\etaT$, far from the critical point and beyond any regime where we expect the MS expansion to be quantitatively valid. This applies even if the criterion used is $\epslliq\ll 1$ of Eq.~\eqref{eq:epslliq}, rather than the formal requirement $\epslvap\ll 1$ of MS-$\ofour$. Indeed, at $\eta_T=0.5$, we find $\epslliq\simeq 1$ and $\epslvap\simeq 5$ for $S=16$.
It is remarkable nonetheless that, even under these conditions, the MS-$\ofour$ theory correctly predicts both the existence of the reentrant phase separation, and its qualitative trends on varying the model parameters, whereas the DD-$\ofour$ theory misses the re-entrance altogether.
\subsubsection{Interfacial tensions}
We now turn to the numerical verification of the predictions of the two $\ofour$ theories on interfacial tensions. To this goal, we make use of Eq.~\eqref{eq:curvature_shifts}: By measuring the differences $\delta_\mathrm{in,out}$ between the binodals of a droplet and those of a flat interface, we can obtain an estimate for $\sigmaVL$. The binodals of the flat interface are found as detailed above (see also App.~\ref{app:sims}); for the curved binodals, we initialize the system with a droplet immersed in vapor. Evaluating $\mub'(\rhoLV)$ from Eq.~\eqref{eq:coeffs_dimful_mub}, we have all quantities required to compute $\sigmaVL$.
\par 
As discussed in Sec.~\ref{sec:qsap_theory_tensions}, the vapor tensions of MS-$\ofour$ and DD-$\ofour$ theories are comparable in magnitude, but MS theory exhibits negative $\sigmaV$ far from the critical point. This would indicate the emergence of reversed Ostwald ripening, which we do not observe in our particle simulations; indeed, the average values for $\sigmaV$ obtained from direct measurement are positive, as we show in Fig.~\fref[a]{fig:app_details} of App.~\ref{app:all_details}. The region where the change of sign of $\sigmaV$ occurs in the predictions of MS-$\ofour$ corresponds to $\epslvap \gtrsim 3$, which is outside of the regime of validity of the theory.

In Fig.~\ref{fig:qsap_tensions_sim}, we compare the liquid tension $\sigmaL$ thus obtained from particle simulations with the theoretical predictions of DD-$\ofour$ and MS-$\ofour$ theories, as a function of $A$. It is evident that DD-$\ofour$ vastly underestimates $\sigmaL$ and misses its qualitative dependence on $A$. In contrast, MS-$\ofour$ correctly captures their magnitude and trend. Notably, as one moves further away from the critical point, leaving the regime of validity of the MS-$\ofour$ theory, the trend in the $A$ dependence of $\sigmaL$ is still correctly captured.
\section{Discussion and Conclusions}
\label{sec:discussion}
To conclude this paper, we review in turn below the general properties of the $\ofour$ theory; its construction for tQSAPs by explicit coarse-graining; and various open questions and implications for future research.
\subsection{General properties of \texorpdfstring{$\ofour$}{O(grad 4)} theory}
We have introduced and studied a generalization of Cahn--Hilliard theory, which we call $\ofour$ theory, to address phase separation in active matter (and other nonequilibrium) systems whose dynamics involves a single order parameter comprising a conserved scalar density. The theory contains all terms with up to four gradients at dynamical level; this requires five separate `coefficient functions' (each a function of the density but not its derivatives).
Unlike field-theoretical models previously studied in the literature~\cite{cates-2025}, $\ofour$ theory does not involve a Taylor expansion in density and so is not restricted to the regime of weak phase separation.
\par 
A major technical achievement of this work is to provide explicit expressions for some of the key observables that govern stationary phase-separated states (and the approach to these states) for $\ofour$ theory in its full generality. These observables are the two coexisting densities (binodals); two interfacial tensions for the Ostwald processes of liquid droplets and vapor bubbles, respectively; and a third tension determining the relaxation of capillary waves. All can be calculated using the general results presented here once the coefficient functions of the $\ofour$ theory are provided, for example by coarse-graining a chosen microscopic model. Accordingly, we expect these results will prove crucial to the future quantitative study of phase separation in a wide range of active matter systems and other systems far from equilibrium. Our results fully generalize earlier works that either assumed weak phase separation~\cite{wittkowski-2014,solon-2018-njp,tjhung-2018,fausti-2021} or addressed phase equilibria solely from the viewpoint of mechanical force balance~\cite{langford-2024-interface,langford-2025}.  

Although we have directly addressed only bulk phase separation, many of our results for $\ofour$ theory are also relevant to other types of structure formation (including microphase separation) far from equilibrium. In particular, we have delineated regimes of negative Ostwald and capillary tensions, which are known to cause bubbly phase separation and active foam formation, respectively~\cite{cates-2025}. Also, we have identified a parameter regime where bulk phase separation is precluded by a previously unknown interfacial instability (reviewed further below) with possibly profound implications for structure formation in that regime. These remain to be explored.
 
While our work considers only a single scalar order parameter, the general techniques developed here may also open the way to analogously general results (avoiding Taylor expansion in conserved densities) for more complex systems that undergo phase separation. These could include multiple-component systems with nonreciprocal interactions~\cite{saha-2020,you2020nonreciprocity,frohoff2023non,frohoff2021suppression,saha-2024,chiu-2025,sahoo-2025,greve-2025}; phase-separating active systems in the presence of fluid flows~\cite{tiribocchi2015active, saintillan-2018, cates-2018}; and active systems with nonconserved dynamics (such as chemical reactions) as well as diffusive relaxation~\cite{brauns2020phase,hyman2014liquid,Weber2019review,li2020non}.

Some of our results are surprising. For example, we discovered that the uncommon tangent construction, which governs binodals in the weak phase separation limit~\cite{wittkowski-2014}, is abolished in the strong separation regime. Within the $\ofour$ theory, it is replaced by a more general procedure that specifies mismatches in not only the intercepts but also the slopes of the tangents to $f(\rho)$ at the binodal densities. (Both mismatches disappear in equilibrium, as is well known~\cite{cates-2018}.)
While a similar phenomenon has been anticipated numerically in traveling states created by anisotropically tumbling particles~\cite{spinney-2025}, we have shown not only how the previous construction fails even for fully isotropic systems at Cahn--Hilliard level, but also what replaces it. Similarly, in nonreciprocal quorum-sensing, the chemical potentials are unequal in the presence of chaotic bands~\cite{duan-2025}; our construction may pave the way to understanding such phenomena.

Surprisingly, we also found a `forbidden region' of parameter space where a nonlinear instability drives the interfacial width to zero at $\ofour$ level. This we precisely delineated in the special case of constant coefficients, for quartic $f(\rho)$ and Flory--Huggins theory. In the former case, depending on the parameters, either one of the binodals grows unbounded on approach to the forbidden region, or both stay finite. Flory--Huggins theory on the other hand enforces a saturating density, but the forbidden region survives. Because the instability stems from runaway sharpening of the interfacial profile caused by active pumping, not by a runaway density, higher derivative terms 
({\em e.g.}, $\osix$) are presumably needed to tame it. We have speculated that in some cases, the end result could be condensation rather than phase separation.

Another unexpected result is that, at $\ofour$ level, there remains a strict relationship between interfacial tensions such that the capillary tension is a specified linear combination of the two Ostwald tensions. This means that the capillary tension cannot be negative without at least one Ostwald tension being so. 

Our results also shed light on claims made in the recent literature concerning an important class of active systems where the diffusive particle flux is set by the divergence of the mechanical stress tensor. Here, the mechanical interfacial tension defined via that stress tensor was recently suggested to be {\em independent} of the Ostwald tensions~\cite{langford-2025}. However, we have shown at $\ofour$ level that in these systems, the mechanical tension is once again a specific linear combination of the two Ostwald tensions, akin to, but distinct from, the capillary tension. 
The latter distinction explains why the mechanical tension could be measured in simulations of pairwise interacting active particles to be negative, even when capillary waves are stable~\cite{solon-2018-njp,bialke-2015}. Any similar relation between the four tensions discussed in this paragraph and yet another interfacial tension, defined as the work needed to create an interface, remains to be investigated~\cite{li2025surface,sun-2025-arxiv}.

\subsection{Construction of \texorpdfstring{$\ofour$}{O(grad 4)} theory by coarse-graining}
As just described, many of the results for $\ofour$ theory have a physical significance and interpretation that does not require one to specify its five coefficient functions. Nonetheless, viewed as a quantitative predictor of phase equilibria, the $\ofour$ theory is necessarily of limited value until those functions are found for any specific system of interest.
Thus, a key part of our work has been to exemplify that all five $\ofour$ coefficient functions can indeed be computed by explicit coarse-graining of a microscopic model comprising tQSAPs.

To do this, we introduced a coarse-graining procedure based on a perturbative expansion and MS analysis, which leverages the fast relaxation of orientational moments and goes beyond approaches previously employed in the literature, such as the DD approximation. The latter uses a one-shot adiabatic elimination of orientational degrees of freedom, which captures some but not all of the contributions needed for a consistent theory at $\ofour$ level. We carefully compared the predictions of the two resulting theories, MS-$\ofour$ and DD-$\ofour$.

Formally, MS-$\ofour$ is precisely justified in the limit in which the interfacial width $\xi$ is much larger than any microscopic length scale. However, we found (by comparing with particle simulations) that it remains quantitatively accurate beyond this regime: Empirically, only the liquid-phase persistence length, which is much smaller than its counterpart in the vapor phase, needs to be small compared to $\xi$. Even when the two lengths are comparable, the theory retains semiquantitative or qualitative accuracy across wide regions of parameter space.

We found substantive differences between the MS- and DD-$\ofour$ theories, such as the unexpected prediction in MS of a reentrant phase boundary on the liquid side. This occurs at relatively low values of the thermal diffusivity where the theory should not be quantitative, despite which re-entrance is confirmed by our particle simulations. More generally, MS-$\ofour$ gives a good qualitative account of this part of the tQSAP phase diagram, where DD-$\ofour$ misses the re-entrance entirely for reasons we have mechanistically identified in Sec.~\ref{subsec:polatin}. MS-$\ofour$ also captures the magnitude and trend of the nonequilibrium interfacial tensions governing the Ostwald process in a regime where DD-$\ofour$ fails.
Surprisingly, we have shown that the DD-$\ofour$ successfully predicts the binodals in a parameter region where $\xi$ is smaller than microscopic length-scales; while we have no theoretical argument for explaining why this happens, we also notice that the Ostwald tensions are substantially mis-estimated by the DD-$\ofour$ theory in this regime.

While they differ strongly in general, we have shown that the MS- and DD-$\ofour$ theories exactly match when the quorum-sensing length $\gamma$ is much larger than other microscopic length scales. Thus, our approach justifies, for the first time, use of the DD approximation in a specific parameter regime.

Our multiple-scale coarse-graining technique for tQSAPs may prove relevant much more broadly. 
With suitable extensions, it could apply to QS models with multiple species that were used to model bacterial experiments~\cite{curatolo-2020}; QS particles with non-reciprocity~\cite{duan-2023, dinelli-2023,duan-2025}; genetic expression models~\cite{ridgway-2023}; or biased tumbling and traffic congestions~\cite{bertin-2024,spinney-2025}.
The approach might also be combined with density functional theory and the interaction-expansion method~\cite{wittkowski-2011,vrugt-2023} or with kinetic theories~\cite{soto-2024, pinto-goldberg-2025} to cover steric interactions between particles. Even more broadly, we have shown that MS analysis is mandated whenever one needs to eliminate orientational moments while retaining orders in gradients beyond those obtained from single-shot adiabatic elimination (in our setting, these are gradients beyond second order in the equation of motion).

Hence, besides phase separation, several other phenomena arising in active systems likely require an MS-based approach to coarse-graining. Examples might include active systems with orientational order~\cite{baskaran2010nonequilibrium,baskaran2009statistical,ihle-2011, ihle-2016,bertin-2006, bertin-2009,peshkov-2014,bertin2015comparison,chate2020dry}. Here, a spurious linear instability was observed at low noise strengths when truncating the hierarchy of orientational moments~\cite{peshkov-2014,mahault2018outstanding}, which might be cured by MS techniques (as it was in our toy-model, Sec.~\ref{sec:multiscale_qsap-toy-model}). Wet active systems undergoing active turbulence~\cite{saintillan2008instabilities,saintillan2014theory} might likewise require MS analysis.

Finally, we performed extensive simulations of tQSAPs with Ornstein--Uhlenbeck dynamics and, as already mentioned, found excellent quantitative agreement with MS-$\ofour$ across wider regions of parameter space than expected, and good semiquantitative or qualitative agreement in larger regions still. Beyond the several specific points of agreement noted already above, a further main one was the confirmation that, unless the QS length $\gamma$ is the largest microscopic length, the observed phase boundaries depend upon it, as MS-$\ofour$ theory predicts, and DD-$\ofour$ theory overlooks. 
\subsection{Outlook}
Our work raises several open questions. First, it is natural to wonder whether phase separation is reentrant at small translational diffusivity also in models of active particles interacting with two-body forces. In the latter models, reentrant phase separation has sometimes been found, but seemingly involves different mechanisms, such as competing attractive interactions and repulsion-induced MIPS~\cite{redner-2013b}, or freezing of particles at zero translational diffusivity~\cite{hawthorne-2025} (kinetic arrest at low diffusivity is also the cause of reentrance in the QS model of Ref.~\cite{desouza-2025}). Intriguingly, it has also been observed for the \emph{vapor} binodal of ABPs interacting via a soft pairwise potential~\cite{su-2023}. Any connections of these systems to our QS model remain to be established.

It also remains open whether the region of re-entrance is likely to be accessible in laboratory experiments rather than numerical ones. For this, one requires a diffusive length scale $\sqrt{T\tau}$ comparable to the QS length scale $\gamma$. Taking $\tau\sim 1\,\mathrm{s}$ as in most synthetic realizations of self-propelled colloids, room temperature, a solvent such as water and $\gamma$ comparable to the particle size, this requires $\gamma$ on the micron scale, which is likely achievable. However, re-entrance also requires at least one of $\alpha_\gamma, A, S$ to be large; to achieve this simultaneously might be more difficult. 

Throughout this paper, we developed the $\ofour$ theory at deterministic level only. This leaves major avenues open for future study. So far, noise-induced phenomena in  nonequilibrium phase separation, such as nucleation, interfacial fluctuations, bubbly phase separation, and active foam phases, were addressed only in the weak phase separation limit governed by AMB+~\cite{cates-2025,fausti-2021,besse-2023,caballero2024interface,cates-2023}. Extending these analyses to the $\ofour$ theory appears reasonably straightforward, so long as the noise is additive Gaussian and temporally white. However, to use $\ofour$ theory to address such issues quantitatively in specific systems, the noise, like the coefficient functions, ought to be microscopically derived. (It is then not guaranteed to be additive, Gaussian, or white.)
We can expect the noise to be derivable by employing homogenization techniques~\cite{gardiner-2004, pavliotis-2008} rather than multiple-scale analysis.

Noise aside, obtaining a theory consistent to the next order in gradients expansion ($\nabla^6$) might enlarge the regime of quantitative accuracy of the MS-$\ofour$ predictions and perhaps also explain why the DD-$\ofour$ theory is sometimes found to be more accurate than MS-$\ofour$ for predicting the binodals (but not the interfacial tensions) in certain parameter regions far outside the formal regime of validity of either theory. It could also help understand the physics of the forbidden zone in $\ofour$ theory and its interfacial instability, and to study the continuum theory of tQSAPs with ABP dynamics, where we have found that $K(\rho)$ changes sign far from the critical point.

As we have emphasized throughout the paper, to make full use of our results for the $\ofour$ theory in any specific context requires its five coefficient functions to be determined. We have achieved this by explicit coarse-graining of one specific model, namely, tQSAPs. However, this and other particle models generally used for active matter are very idealized; by focusing on the key physical principles, such models deliberately omit many details that would quantitatively change the required coefficient functions. One possible avenue is to constrain these functions directly by measurements: Accessing the coexisting densities and the spectra of capillary waves, for example, should allow inference of the capillary tension, as was done in equilibrium~\cite{aarts2004direct}. 
However, one cannot expect to determine all five functions of density in this way. 

A more promising avenue is then to employ data-driven approaches such as sparse regression~\cite{rudy2017data,brunton2016discovering}. These have recently been successful in learning from data the PDE-level description of several homogeneous active systems~\cite{supekar2023learning,joshi2022data}. Whether they are sufficiently robust to handle the high-order PDEs needed for phase-separating systems is unclear.
However, if the coefficient functions can be learned in this way, this would enable the use of our active Cahn--Hilliard approach as a predictive tool for experimental outcomes, bypassing particle models altogether. 
\begin{acknowledgments}
Discussions with Ananyo Maitra on the coarse-graining developed in this work are kindly acknowledged. FDL acknowledges the support of the University of Cambridge Harding Distinguished Postgraduate Scholars Programme. CN acknowledges the
support of the ANR grant PSAM. CN acknowledges the support of the INP-IRP grant IFAM. The authors would like to thank the Isaac Newton Institute for Mathematical Sciences, Cambridge, for support and hospitality during the programme ``Anti-diffusive dynamics: from sub-cellular to astrophysical scales'' where work on this paper was undertaken. This work was supported by EPSRC Grants No. EP/Z000580/1 and No. EP/Z534766/1.
\end{acknowledgments}
\appendix
\section{Details on phase coexistence properties of \texorpdfstring{$\ofour$}{O(grad 4)} theory}
\label{app:ofour}
In this Appendix, we discuss some details on phase coexistence in $\ofour$ theory, expanding on the results presented in Sec.~\ref{sec:ofour_framework}.
\subsection{Monotonicity of interfaces}
\label{app:interface}
The structure of the interface equation \eqref{eq:w} can be used to prove that interfaces in $\ofour$ theory are monotonic.

We proceed by contradiction: We assume a smooth, nonmonotonic interface between two bulk phases $\rhoVL$ exists. Then, there are at least two local extrema where $\partial_x\rho = 0$, and thus $w(\rho) = 0$. Let us denote the density at the first such point encountered coming from the vapor phase as $\rho_\ast$. We now show that computing $w(\rho)$ results in a contradiction.

In deriving Eq.~\eqref{eq:w}, we assumed $\partial_x\rho \neq 0$; hence, it is only valid piecewise on segments that exclude local extrema. We integrate Eq.~\eqref{eq:w} starting from one phase at $\rhoV$, with $w(\rhoV) = w'(\rhoV) = 0$ (note that $w'(\rho) = 2\partial_x^2\rho$), until we reach $\rho_\ast$. Since we assumed nonmonotonicity, after this point the density decreases again; to proceed, we have to stitch the solution obtained so far together with the next segment, where the equation is now integrated using the initial conditions $w(\rho_\ast)=0$ and $w'(\rho_\ast)$ set to the final value of the previous segment (this corresponds to assuming that the first two derivatives of $\rho$ are continuous). Due to the uniqueness of the solution of Eq.~\eqref{eq:w} given two boundary conditions, however, we would necessarily be retracing the first segment exactly, ending up at $\rhoV$ again, thus never reaching the other binodal $\rhoL$. We conclude that nonmonotonic interfaces between two distinct bulk phases cannot exist in the $\ofour$ theory: Nonmonotonic interfaces might exist in a theory that includes terms with six gradients. In this case, indeed, Eq.~\eqref{eq:w} is augmented by terms quadratic in $w$, which invalidate the argument presented here.
\subsection{Calculation of the capillary-wave dynamics}
\label{app:cw}
We now find the capillary-wave interfacial tension for $\ofour$ theory, generalizing the calculation of Ref.~\cite{fausti-2021}. We assume the existence of an interface described by a height function $\xpe = \hat h(\xvpa, t)$. For specificity, we assume that the interface connects a vapor phase of density $\rhoV$ at $\xpe = -\infty$ with a liquid phase of density $\rhoL$ at $\xpe=+\infty$. The in-plane coordinates $\xvpa$ are assumed to be periodic. We denote the Fourier transform along these coordinates as
\begin{equation}
    \mathcal{F}(\hat h)(\qv, t) \coloneqq h(\qv, t) = \int\dd{\xvpa} \mathrm{e}^{-i \qv\vdot\xvpa} \hat h(\xvpa, t)\,.
    \label{eq:xfourier-transform_def}
\end{equation}

Our goal is to obtain the dynamics of $h(\qv, t)$ from that of the field $\rho$ of Eq.~\eqref{eq:continuity}. As a first step, we introduce a nonlocal chemical potential $\mu$ such that $\laplacian\mu = -\div\Jv$. We write $\mu = \mu_\lambda + \mu_{\zeta,\nu} - \bar\mu$, with a free constant $\bar\mu$ that will be specified below and
\begin{subequations}
\begin{align}
    \mu_\lambda &= \mub(\rho) + \lambda|\grad\rho|^2 - K\laplacian\rho\,,\\
    \laplacian\mu_{\zeta,\nu} &= -\div[\zeta\laplacian\rho\grad\rho+\nu|\grad\rho|^2\grad\rho]\,.
\end{align}
\end{subequations}
We now use the ansatz of Eq.~\eqref{eq:cw-capillary}, obtaining
\begin{subequations}
\begin{align}
    \mu_\lambda ={}& \mub(\varrho) + \lambda(\varrho)\varrho'^2-K(\varrho)\varrho''\nonumber\\
    &+K(\varrho)\varrho'\lappa\hat h+\mathcal \mathcal{O}(\hat h^2)\,,\label{eq:mu_lambda}\\
    \laplacian\mu_{\zeta,\nu} ={}& -\frac{1}{2}\laplacian[\zeta(\varrho)\varrho'^2 + 2I_{\bar\nu}]\nonumber\\
&+\gradpe[\zeta(\varrho)\varrho'^2]\lappa\hat h+\mathcal \mathcal{O}(\hat h^2)\,.
\end{align}
\end{subequations}
Here, $\gradpa = \grad_\xvpa$ and $\gradpe = \partial_\xpe$, and the prime denotes a derivative with respect to the argument. Both $\varrho$ and $I_{\bar\nu}$ in these expressions have the argument $\xpe-\hat h(\xvpa,t)$. The quantity $I_{\bar\nu}$ is defined as
\begin{align}
I_{\bar\nu}(\xpe) = \frac{1}{2}\int_{-\infty}^\xpe\bar\nu(\varrho)\varrho'^3(u)\dd u - \frac{1}{2}\int^{\infty}_\xpe\bar\nu(\varrho)\varrho'^3(u)\dd{u}.
\end{align}
Inserting the ansatz of Eq.~\eqref{eq:cw-capillary} into Eq.~\eqref{eq:continuity}, we obtain
\begin{equation}
\label{eq:cw-continuity}
\nabla^{-2}\partial_t\rho = -\nabla^{-2}(\varrho'\partial_t\hat{h})=\mu\,.
\end{equation}
Here, we have introduced the inverse Laplace operator $\nabla^{-2}$, which is the inverse of the Laplacian such that $\hat \ell = \nabla^{-2}\hat s \Leftrightarrow \laplacian \hat \ell = \hat s$ and $\hat\ell(\xvpa, \xpe=\pm \infty) = 0$. The constant $\bar\mu$ is fixed by this latter condition to be
\begin{equation}
    \bar\mu = \mub(\rhoL) - I_{\bar\nu}(+\infty)\,.\label{eq:cw-barmu}
\end{equation}

We now multiply Eq.~\eqref{eq:cw-continuity} by $\dv{\psicw}{u}$ and integrate over the interface. Upon integrating by parts, the right-hand side then gives
\begin{align}\label{eq:cw-RHS-noFT}
    &-\Delta\Pi_\mathrm{cw}+\Delta(\mub\psicw)-\bar\mu\Delta\psicw\nonumber\\
    &+\lappa\hat h\,\sigma_\lambda+C_0\nonumber\\
    &+\int_{-\infty}^\infty\dd{u}\dv{\psicw}{u}\nabla^{-2}(\gradpe[\zeta(\varrho)\varrho'^2]\lappa\hat h)\nonumber\\
    &+\mathcal \mathcal{O}(\hat h^2)\,.
\end{align}
Here, we have introduced
\begin{subequations}
\begin{align}
\sigma_\lambda ={}& \int\dd{u}\psicw'(\varrho) K(\varrho)\varrho'(u)^2,\\
C_0 ={}& \frac{1}{2}\int\dd{u}\left[K\psicw''+(2\bar\lambda+K')\psicw'(\varrho)+2\bar\nu \psicw'(\varrho)\right]\varrho'^3\nonumber\\
& -\frac{\psicw(\rhoL)+\psicw(\rhoV)}{2}\int\dd{u}\bar\nu(\varrho)\varrho'^3\,.
\end{align}
\end{subequations}
We now use that $\psicw(\rho)$ obeys Eq.~\eqref{eq:psi} with boundary conditions $\psicw(\rhoV)=-\psicw(\rhoL)$. Then, $C_0=0$, while the first line of Eq.~\eqref{eq:cw-RHS-noFT} vanishes using Eqs.~\eqref{eq:mu0-jump}, \eqref{eq:binodals}, and \eqref{eq:cw-barmu}.
We now take the Fourier transform of Eq.~\eqref{eq:cw-continuity} with respect to the $\xvpa$ directions. To deal with the line involving $\nabla^{-2}$ in Eq.~\eqref{eq:cw-RHS-noFT}, we use that the following identity holds for the transforms of $\hat\ell$ and $\hat s$ obeying $\hat\ell = \nabla^{-2}\hat s$:
\begin{equation}
\label{eq:invlapy}
\ell(\qv,\xpe) = -\frac{1}{2q}\int_{-\infty}^\infty\dd{u} e^{-q|\xpe-u|}s(\qv, u)\,,
\end{equation}
where $q = |\qv|$. Using partial integration and $\partial_ue^{-q|u|} = q\,\mathrm{sgn}(u)e^{-q|u|} \eqqcolon q c_q(u)$, we obtain
\begin{align}\label{eq:cw-RHS-withFT}
    &- q^2 h(\qv,t) (\sigma_\lambda+\sigma_\zeta) + \mathcal{O}(h^2)\,,
\end{align}
where
\begin{align}
\sigma_\zeta ={}& \frac{1}{2}\int\dd{u}\int\dd{\tilde u}\dv{\psicw}{u}(u)c_q(u-\tilde u)\zeta(\varrho(\tilde u))\varrho'(\tilde u)^2\,.
\end{align}

We now turn to the left-hand side of Eq.~\eqref{eq:cw-continuity}. Multiplying by $\dv{\psicw}{u}$, integrating over the interface, and Fourier transforming gives
\begin{equation}
\partial_t h(\qv, t)\frac{A(q)}{2q}\,,
\end{equation}
with
\begin{equation}
A(q) ={} \int\dd{u}\int\dd{\tilde u}\dv{\psicw}{u}(u)e^{-q|u-\tilde u|}\varrho'(\tilde u)\,.
\end{equation}

It is easy to show that $\sigma_\zeta = -\int\dd{u}\zeta(\varrho)\psicw(\varrho)\varrho'^2 + \mathcal{O}(q)$, and $A(q) = \Delta\rho\Delta\psicw + \mathcal{O}(q)$. Combining the two sides, we thus finally obtain
\begin{equation}
    \partial_t h(\qv, t) = -\frac{2q^3\sigmacw}{(\Delta\rho)^2}h(\qv,t) + \mathcal{O}(q^4h, h^2)\,,
\end{equation}
where $\sigmacw = \frac{\Delta\psicw}{\Delta\rho} (\sigma_\lambda + \lim_{q\to 0} \sigma_\zeta)$ is given by Eq.~\eqref{eq:sigma_cw} in the main text.
\subsection{AMB+ tensions in case of general mobility functional}
\label{app:mob}
Here, we generalize the results obtained in Sec.~\ref{sec:ofour_mob}, considering AMB+ with a density-dependent mobility \emph{functional} $\mob[\rho]>0$ (meaning that $\mob[\rho](\xv,t)$ is now allowed to depend not only on $\rho(\xv,t)$ but also on its spatial derivatives):
\begin{equation}
\begin{aligned}
    \Jv &= \mob[\rho]\left(-\grad \mu_\lambda+\zeta_0\div(\laplacian\rho\grad\rho)\right) \,,\label{eq:JAMB+_mob-functional}
\end{aligned}
\end{equation}
with
\begin{equation}
    \mu_\lambda = \mub(\rho) - K_0 \nabla^2\rho +\lambda_0 \left(\grad\rho\right)^2
    \,,
\end{equation}
and $K_0, \lambda_0, \zeta_0$ constant. 
In the following, we will write $\mob(\vb{r}) = \mob[\rho](\vb{r},t)$. It is useful for the calculations in this section to define a linear operator $\Gamma$ acting on a function $m(\mathbf{r})$:
\begin{align}\label{eq:Gamma}
    \Gamma m(\mathbf{r})=\grad\vdot \left[\mob(\mathbf{r}) \grad m(\mathbf{r}) \right]
    \,.
\end{align}
\subsubsection{Ostwald ripening}
We consider a spherically symmetric droplet of density $\rho_\mathrm{in} = \rhoL + \delta_\mathrm{in}$ in a supersaturated environment ($\rho_\infty = \rhoV + \epsilon$).
We present the calculation, which generalizes the one presented in Ref.~\cite{cates-2023} to the case of nonconstant $\mob$, in $d=3$. The result can be easily generalized to any $d\geq 2$. We make the ansatz
\begin{align}\label{eq:rho_ansatz_R}
    \rho(\xv,t) =\varrho(r-R)\,,
\end{align}
where $r$ denotes the radial coordinate, $R = R(t)$ is the time-dependent droplet radius, and $\varrho$ denotes the profile of a flat interface (corrections to the ansatz in Eq.~\eqref{eq:rho_ansatz_R} given by the curvature dependence of the interface give negligible corrections in the large $R$ expansion performed below). Our goal is to obtain the dynamics of $R(t)$ from that of the field $\rho$ of Eq.~\eqref{eq:continuity}.
\par 
For spherically symmetric functions $m(r)$ and $n(r)$, we define an inverse of the operator of Eq.~\eqref{eq:Gamma} $\Gamma^{-1}$, such that $m = \Gamma^{-1} n \Leftrightarrow \Gamma m = n$ with the requirements $m(r\to \infty) = 0$ and that $m(r)$ be regular at $r=0$. A closed-form expression for $m(r)$ can be obtained by writing Eq.~\eqref{eq:Gamma} in spherical coordinates:
\begin{align}
    \frac{1}{r^{2}}\partial_r(r^{2}\mob m') = n(r) \,. 
    \label{eq:nomob-dgl}
\end{align}
Integrating Eq.~\eqref{eq:nomob-dgl} twice, we obtain
\begin{align}
    \Gamma^{-1} n(r)
    = - \int_r^\infty \frac{\mathrm{d} r_1}{\mob(r_1) r_1^{2}}\int_0^{r_1} \mathrm{d} r_2 \,r_2^{2} n(r_2)
    \,.
    \label{eq:inverse-Gamma}
\end{align}
We now insert the ansatz of Eq.~\eqref{eq:rho_ansatz_R} into Eq.~\eqref{eq:continuity} with the current~$\Jv$ of Eq.~\eqref{eq:JAMB+_mob-functional} and invert $\Gamma$. We find
\begin{equation}
\begin{aligned}
    \dot{R}  \Gamma^{-1}\left[\varrho'(r-R)  \right] &= \mub(\rho_\infty)-\mu_\lambda
    \\
    &\phantom{=}+ \zeta_0 \Gamma^{-1} \grad \vdot \left(\mob \laplacian\rho \grad \rho\right)
    \,,
    \label{eq:amb+_OW}
\end{aligned}
\end{equation}
where $\mu_\lambda$ is given by Eq.~\eqref{eq:mu_lambda}. We now multiply Eq.~\eqref{eq:amb+_OW} by $\dv{r}\psi^{0}(r-R)$ (where $\psi^0$ is the AMB+ pseudodensity of Eq.~\eqref{eq:psi_AMBp}) and integrate across the interface. Upon shifting the integration variable as $r\to r-R$ and integrating by parts, the right-hand side gives
\begin{align}\label{eq:amb+_OW_RHS}
    \mathrm{RHS}&{}={} [\mub(\rho_\mathrm{in})-\mub(\rho_\infty)]\psi^0(\rho_\mathrm{in})+\int_{\rho_\mathrm{in}}^{\rho_\infty}\dd{\rho}\psi^0(\rho)\mub'(\rho)\nonumber\\
    &\phantom{=}- \int_0^{\infty} \dd{r}\, \varrho'^3\left(\frac{K_0}{2} \psi^{0\prime\prime}(\varrho) + \lambda_0 \psi^{0\prime}(\varrho)\right)\nonumber\\
    &\phantom{=}+2\int_0^\infty\dd{r}\dv{\psi^{0}}{r}\frac{K_0}{r}\varrho'(r)\nonumber\\
    &\phantom{=}+ \zeta_0 \int_0^{\infty} \dd{r} \dv{\psi^{0}}{r}\;  \Gamma^{-1} \div(\mob \laplacian\rho \grad \rho)
    \,.
\end{align}
Expanding the first line in small $\epsilon$ and $\delta_\mathrm{in}$ gives to first order $-\epsilon\mub(\rhoV)\Delta\psi^0$. Upon using Eq.~\eqref{eq:inverse-Gamma} and integrating by parts, the last line of Eq.~\eqref{eq:amb+_OW_RHS} reads
\begin{equation}\label{eq:amb+_OW_RHS_zeta}
\begin{aligned}
    &\zeta_0 \int_0^{\infty} \dd{r}\psi^{0\prime}(\varrho)\left[\frac{\varrho'^3}{2}-2\varrho' 
    \int_r^{\infty} \mathrm{d} r_1 \frac{\varrho'(r_1)^2}{r_1}\right]
     \,.
\end{aligned}
\end{equation}
Inserting Eq.~\eqref{eq:amb+_OW_RHS_zeta} into Eq.~\eqref{eq:amb+_OW_RHS}, using that $\psi^0$ fulfils Eq.~\eqref{eq:psi} and that $\varrho'$ is peaked at $r = R$, we then find
\begin{align}\label{eq:amb+_OW_RHS_bis}
    \mathrm{RHS}=& -\epsilon\mub(\rhoV)\Delta\psi^0 +\frac{2}{R}\frac{\Delta\psi^0}{\Delta\rho}\sigmaL^0 + \mathcal{O}(R^{-2})
    \,,
\end{align}
where $\sigmaL^0$ is given by Eq.~\eqref{eq:sigma_AMBp}.
On the other hand, with Eq.~\eqref{eq:inverse-Gamma}, upon multiplication by $\dv{r}\psi^{0}(r-R)$ and integration, the left-hand side of Eq.~\eqref{eq:amb+_OW} yields, using a change of variables $r \to r - R$ and integrating by parts:
\begin{align}
    &\dot{R}  \int_0^\infty \mathrm{d} r  \frac{\psi^0(r)-\psi^0(\rho_\mathrm{in})}{\mob(r)}\partial_r\left(\frac{1}{r}\right)\int_0^r \mathrm{d}r_1\,r_1^2\varrho'(r_1)\nonumber
    \\[0.4em]
    ={}&-\dot{R}\int_0^\infty \mathrm{d} r  \,\frac{1}{r}  \partial_r
    \left[
    \frac{\psi^0(r)-\psi^0(\rho_\mathrm{in})}{\mob(r)}
    \right]\int_0^r \mathrm{d} r_1\, r_1^2 \varrho'(r_1)\nonumber
    \\
    &- \dot{R}
    \int_0^\infty \mathrm{d}r\, r \frac{\psi^0(r)-\psi^0(\rho_\mathrm{in})}{\mob(r)} \varrho'(r)
    \,.\label{eq:ow-Mpeaked}
\end{align}
We now notice that $\partial_r\left(\frac{\psi^0(r)-\psi^0(\rho_\mathrm{in})}{\mob(r)}\right)$ 
is significantly different from zero only close to $r=R$ (as is $\varrho'$); we can thus expand the last term in Eq.~\eqref{eq:ow-Mpeaked} in $R^{-1}$ for large $R$ and integrate by parts again over $r$ in the first integral. 
This leads to:
\begin{align}
    \mathrm{LHS}=-\dot{R} R \frac{\Delta\psi^0 \Delta\rho}{\mob_\mathrm{V}} 
    +\mathcal{O}(\dot{R} R^0)
    \,,
    \label{eq:amb+_OW_LHS_bis}
\end{align}
where $\mob_\mathrm{V} = \mob(r \to \infty)$. 
Equating Eqs.~\eqref{eq:amb+_OW_RHS_bis} and \eqref{eq:amb+_OW_LHS_bis}, we obtain
\begin{align}
    \dot{R} = 
    \frac{2\mob_\mathrm{V} \sigmaL^0}{(\Delta\rho)^2 R}
    \left[\frac{1}{R_{\mathrm{c}}}-\frac{1}{R}\right]
    + \mathcal{O}(R^{-2})
    \,, 
    \label{eq:Rdot_OR}
\end{align}
where the critical radius reads
\begin{align}
    R_{\mathrm{c}} = \frac{2\sigmaL^0}{\epsilon \mub'(\rhoV)\Delta\rho}\,.
    \label{eq:Rcrit_OR}
\end{align}
Identifying $\sigmaL=\mob_\mathrm{V}\sigmaL^0$ brings Eq.~\eqref{eq:Rdot_OR} into the form of Eq.~\eqref{eq:Rdot}. Thus, we have generalized the result of Sec.~\ref{sec:ofour_mob} (cfr.~Eq.~\eqref{eq:sigma_AMBp_mob}) to mobilities that are functionals of the density. The result can be easily obtained also for a vapor bubble (this only results in replacing in all the expressions above, $\mathrm{L}\leftrightarrow\mathrm{V}$) as well as for any spatial dimension $d \geq 2$. In this latter case, we obtain the same $d$ dependence as in Eq.~\eqref{eq:Rdot}.
\subsubsection{Capillary waves}
We now discuss the relaxation of capillary waves. The calculation is similar as in Sec.~\ref{app:cw}, with the difference that we now have to invert $\Gamma$ rather than $\laplacian$. The inverse operator $\Gamma^{-1}$ is defined as $\hat \ell = \Gamma^{-1}\hat s \Leftrightarrow \Gamma \hat \ell = \hat s$ and $\hat\ell(\xvpa, \xpe=\pm \infty) = 0$.

We start again from the ansatz of Eq.~\eqref{eq:cw-capillary}, assuming an interface that connects a vapor phase $\rhoV$ at $\xpe=-\infty$ with a liquid phase $\rhoL$ at $\xpe=\infty$. Substituting it in Eq.~\eqref{eq:continuity} with the current~$\Jv$ from Eq.~\eqref{eq:JAMB+_mob-functional}, we obtain
\begin{equation}\label{eq:cw-continuity-mob}
    -\Gamma^{-1}(\varrho'\partial_t\hat{h})=\mu_\lambda - \bar\mu -\zeta_0\Gamma^{-1}\div\vb{J}_{\zeta_0}
    \,,
\end{equation}
where $\mu_\lambda$ is defined as in Eq.~\eqref{eq:mu_lambda} and $\bar\mu$ sets the boundary condition required by $\Gamma^{-1}$ and $\vb{J}_{\zeta_0} = \zeta_0\mob \nabla^2 \rho \grad \rho$.

We now multiply Eq.~\eqref{eq:cw-continuity-mob} by $\gradpe\psi^{0}(\varrho[\xpe-\hat h(\xvpa, t)])$ given in Eq.~\eqref{eq:psi_AMBp} and integrate across the interface. We then Fourier transform along  $\xvpa$ (cf.~Eq.~\eqref{eq:xfourier-transform_def}). Integrating by parts, the left-hand side then reads to linear order in $\hat{h}$:
\begin{equation}
\begin{aligned}
    - \int \dd \xvpa \int \dd{\xpe} 
    \varrho'\,\partial_t \hat h\,\Gamma^{-1}
    \left[\psi^{0\prime}(\xpe) e^{-i \qv \cdot \xvpa} \right]
    \, . 
    \label{eq:nomob_cw_lhs-1-sh}
\end{aligned}
\end{equation}
To find the inverse of $\Gamma$ here, we solve
\begin{align}
     \Gamma(m_\qv(\xpe)e^{-i \qv \cdot \xvpa}) = \gradpe\psi^0(\xpe) e^{-i \qv \cdot \xvpa} 
     \,.\label{eq:helperfct-def}
\end{align}
Here, $\mob(\xv)=\mob[\varrho(\xpe-\hat h(\xvpa))]$ appears in $\Gamma$; we aim at obtaining a solution to zeroth order in $\hat{h}$, so we can neglect the dependence of $\mob$ on $\xvpa$. We then have
\begin{align}\label{app:eq:CW-gq-ODE-sh}
    &m_\qv'' + (\partial_{\xpe} \ln \mob) m_\qv' -q^2 m_\qv = \frac{\gradpe\psi^0(\xpe)}{\mob(\xpe)}
    \,, 
\end{align}
which should be solved for $m_\qv$. We solve Eq.~\eqref{app:eq:CW-gq-ODE-sh} iteratively. Defining $\tilde m_\qv(\xpe)= (\partial_{\xpe} \ln \mob) m_\qv'(\xpe)$ and taking it as given, the solution to Eq.~\eqref{app:eq:CW-gq-ODE-sh} with the appropriate boundary condition is
\begin{equation}
\begin{aligned}
\label{app:eq-CW-iterative_2}
    m_\qv(\xpe) &= -\frac{1}{2q} \int \dd{u} e^{-q|\xpe -u|} \, \times \\ 
    &\phantom{{}={}} \times \left[\frac{\gradpe\psi^0(\xpe)}{\mob(u)}-\tilde m_\qv(u)\right]
    \,,
\end{aligned}
\end{equation}
where $q = |\qv|$. We are interested in a small $q$ expansion; we find
\begin{subequations}
\begin{align}
    m_\qv(\xpe)  &= -\frac{1}{2q} \int \dd{u}  
    \left[\frac{\gradpe\psi^0(\xpe)
    (u)}{\mob(u)}-\tilde m_\qv(u)\right]
    \nonumber\\
    &\phantom{{}={}} +\mathcal{O}(q^0)\,, \label{eq:helperfct} \\[0.4em]
    m_\qv'(\xpe)&=  \frac{1}{2}\int \dd{u} \, \frac{\mathrm{sgn}(\xpe-u)}{\mob(u)} \, \times 
    \nonumber\\ 
    &\phantom{{}={}} \times \left[
    \gradpe\psi^0(\xpe)-\mob'(u)m_\qv'(u)
    \right]\nonumber\\
    &\phantom{{}={}}+\mathcal{O}(q^1)\label{eq:helperfctprime}
    \,.
\end{align}
\end{subequations}
Equation~\eqref{eq:helperfctprime} is a closed equation for $m_\qv'$. At this order, it is solved by
\begin{equation}
\begin{aligned}
    m_\qv'(\xpe) &= 
    \frac{\psi^0(\xpe)}{\mob(\xpe)} - \frac{\mob_\mathrm{V} \psi^0(\rhoL) + \mob_\mathrm{L} \psi^0(\rhoV)}{(\mob_\mathrm{V} + \mob_\mathrm{L})\mob(\xpe)}
    \, .
      \label{eq:nomob_cap-gprime}
\end{aligned}
\end{equation} 
Here, $\mob_\mathrm{V} = \mob(\xpe\to-\infty)$ and $\mob_\mathrm{L} = \mob(\xpe \to+\infty)$. Inserting Eq.~\eqref{eq:nomob_cap-gprime} into Eq.~\eqref{eq:helperfct} gives
\begin{align}
    &m_\qv(\xpe) =  -\frac{\Delta \psi^0}{2q \mob_{\mathrm{av}}} +\mathcal{O}(q^0,h^1)
    \,, 
\end{align}
where $\mob_{\mathrm{av}} = (\mob_\mathrm{V} + \mob_\mathrm{L})/2$.
Using this result and Eq.~\eqref{eq:helperfct-def}, Eq.~\eqref{eq:nomob_cw_lhs-1-sh} becomes
\begin{equation}
\begin{aligned}
     \text{LHS} &= \frac{\Delta \psi^0\Delta\rho}{2q \mob_{\mathrm{av}} } \partial_t h_{\qv}
     \, .
    \label{eq:nomob-cap_lhs-final}
\end{aligned}
\end{equation}
On the other hand, upon multiplying by $\gradpe\psi^0$, integrating across the interface and Fourier-transforming, the right-hand side of Eq.~\eqref{eq:cw-continuity-mob} reads
\begin{equation}
\begin{aligned}
    &\int \dd \xvpa \int \dd \xpe  e^{-i \qv \cdot \xvpa}
   \gradpe\psi^{0}(\xpe)  \, \times \\ 
   &\phantom{{}={}} \times 
   \left[\mu_\lambda - \bar\mu - \Gamma^{-1}\div\vb{J}_{\zeta_0}\right] \,.
     \label{eq:nomob_cw_rhs-1}   
\end{aligned}
\end{equation}
The $\mu_\lambda - \bar\mu$ part of this equation is manipulated as in Sec.~\ref{app:cw}. On the other hand, in the last term of Eq.~\eqref{eq:nomob_cw_rhs-1}, integrating by parts gives
\begin{equation}
    -\int \dd \xvpa \int \dd \xpe \div\vb{J}_{\zeta_0}\Gamma^{-1}[\gradpe\psi^0(\xpe)e^{-i \qv \cdot \xvpa}]
    \,,
\end{equation}
which using Eq.~\eqref{eq:helperfct-def} and integrating by parts yields
\begin{align}
     q^2 h_{\qv} \zeta_0 
    \int \dd{u}  \,   \varrho'(u)^2 \mob(u) m_\qv'(u)
    \, . 
    \label{eq:nomob-cap_rhs-2}
\end{align}
Using Eq.~\eqref{eq:nomob_cap-gprime} here and combining it with the result from the $\mu_\lambda - \bar\mu$ part of Eq.~\eqref{eq:nomob_cw_rhs-1}, we find
\begin{equation}
    \mathrm{RHS} = -\frac{\Delta\psi^0\sigma_{\mathrm{cw}} q^2}{\mob_\mathrm{av}\Delta\rho}h_{\qv}
    \,, 
    \label{eq:nomob-cap_rhs-final}
\end{equation}
where the capillary tension is a weighted average of the liquid and vapor tension, which generalizes the result from Sec.~\ref{sec:ofour_mob}:
\begin{equation}
\begin{aligned}
    \sigma_{\mathrm{cw}}
    &= \frac{\mob_\mathrm{V} \sigma_{\mathrm{L}}^0 + \mob_\mathrm{L} \sigma_{\mathrm{V}}^0}{2}
    \,. 
\end{aligned}
\end{equation}
Equating Eq.~\eqref{eq:nomob-cap_rhs-final} with Eq.~\eqref{eq:nomob-cap_lhs-final}, the time evolution of capillary waves at leading order in $h, q$ finally reads
\begin{align}
    \partial_t h_{\qv} = -\frac{2 \sigma_{\mathrm{cw}} q^3}{\left(\Delta\rho\right)^2}h_{\qv}
    +\mathcal{O}(q^4 h, h^2)
    \,.
    \label{eq:height-fourier}
\end{align}
\section{Details on \texorpdfstring{$\ofour$}{O(grad 4)} theory with constant coefficients}
\subsection{Derivation of the forbidden region}
\label{app:forbidden_region}
\begin{figure}[tbp]
    \centering
    \includegraphics[width=\columnwidth]{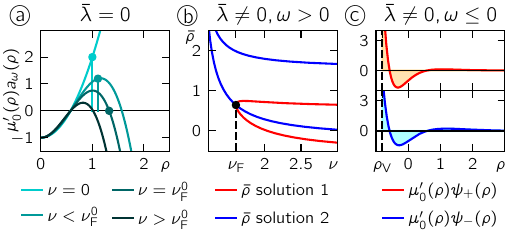}
    \caption{Obtaining the boundary of the forbidden region. a) Case $\bar\lambda=0$: Eq.~\eqref{eq:lambda0_binodal} loses its first solution at $\nu = \nuF^0$, beyond which no binodals exist. b) Case $\bar\lambda\neq 0, \omega > 0$: at $\nuF$, $\Delta\rho = \pi/\sqrt{\omega}$ holds. The forbidden region boundary is found by solving Eq.~\eqref{eq:forbidden_omegapos_cond} for $\bar\rho$ (blue corresponds to the solutions coming from the cosine and red to the sine integral), which have to agree: The intersection point gives the value of $\nuF$. c) For $\omega\leq 0$ and $\bar\lambda >0$, the forbidden region is found by requiring $\rhoL=\infty$. The two conditions then have to agree on the value of $\rhoV$. \label{fig:forbidden_app}}
\end{figure}
We now discuss how to obtain the boundary of the forbidden region $\nuF(\bar\lambda)$ for the $\ofour$ theory with constant coefficients discussed in Sec.~\ref{sec:ofour_constant_coeffs}.

Let us start from the case $\bar\lambda = 0$, where the theory is $\rho \to -\rho$ symmetric and obtaining the binodals is particularly simple. To find the binodals, we use the pseudodensities of Eq.~\eqref{eq:psi_const} to solve the conditions of Eq.~\eqref{eq:binodals}. Because $\mub'(\rho)$ is even and $\psi^{(2)}$ odd, the condition $\Delta\Pi^{(2)} = 0$ is trivially fulfilled by $\rhoV = -\rhoL$. On the other hand, the condition $\Delta\Pi^{(1)} = 0$ reduces to
\begin{equation}
    \int_0^{\rhoL}\dd{\rho} \mub'(\rho)a_\omega(\rho) = 0\,.\label{eq:lambda0_binodal}
\end{equation}
For $\nu\leq 0$, $a_\omega(\rho)=\cosh(\sqrt{\omega}\rho)$ increases monotonically for $\rho > 0$, so that this equation has a unique solution $\rhoL$. On the other hand, for $\nu>0$, $a_\omega(\rho)=\cos(\sqrt{\omega}\rho)$ is oscillatory and Eq.~\eqref{eq:lambda0_binodal} formally has a countably infinite set of solutions.

The first such solution in magnitude exists for $\nu \leq \nuF^0$. It disappears when the binodal coincides with the first zero of $a_\omega(\rho)$, $\rho^0_\mathrm{F} = \pi/(2\sqrt{2\nuF^0/K})$ (see Fig.~\fref[a]{fig:forbidden_app}). Explicitly solving the integral of Eq.~\eqref{eq:lambda0_binodal} for $\omega > 0$ gives
\begin{equation}
\left(\rhoL^2-\frac{2}{\omega}-\frac{a}{3b}\right)\sin(\sqrt{\omega}\rhoL)+\frac{2\rhoL}{\sqrt{\omega}} \cos(\sqrt{\omega} \rhoL) = 0\,.
\end{equation}
For $\rhoL = \rho^0_\mathrm{F}$, the last term vanishes and one obtains a linear equation for $\nuF^0$, which is solved to give Eq.~\eqref{eq:nuF0}.

What about the other solutions that appear for $\nu > 0$? These emerge as more semiperiods of $a_\omega(\rho)$ are involved in the cancellation of the integral in Eq.~\eqref{eq:lambda0_binodal}. We find that these solutions are unphysical: Using the corresponding binodals in the solution $w(\rho)$ of the interface equation \eqref{eq:w} with $w(\rhoV) = w(\rhoL) = 0$ reveals that it has additional zeros between the two binodals. While generally this could point to a nonmonotonic interface, we have proven that these are not allowed in $\ofour$ theory in Sec.~\ref{app:interface}. We conclude that only the first solution corresponds to well-defined bulk phase separation, and this solution ceases to exist for $\nu > \nuF^0$, annihilating with the second, unphysical solution at the finite binodal value $\rho^0_\mathrm{F} = \sqrt{a/(3-24/\pi^2)b}$.

A similar argument allows to find the boundary $\nuF(\bar\lambda)$ of the forbidden region for $\bar\lambda\neq 0$ and $\omega>0$. The calculation is complicated by the fact that for $\bar\lambda\neq 0$, the binodals cease to be symmetric, so that $\bar\rho \neq 0$. To proceed, we note that by taking an appropriate linear combination of the pseudodensities of Eq.~\eqref{eq:psi_const} the argument of the trigonometric functions can be shifted by an arbitrary value. We thus find that the binodals must satisfy
\begin{equation}
\int_{-\Delta\rho/2+\bar\rho}^{\Delta\rho/2+\bar\rho}\dd{\rho}\mub'(\rho)\mqty(\cos(\sqrt{\omega}(\rho-\bar\rho))\\\sin(\sqrt{\omega}(\rho-\bar\rho)))e^{-\bar\lambda\rho/K}=0\,.
\end{equation}
Shifting the integration variable as $\rho\to\rho-\bar\rho$ leads to:
\begin{equation}\label{eq:forbidden_omegapos_cond}
    \int_{-\Delta\rho/2}^{\Delta\rho/2}\dd{\rho} \mub'(\rho+\bar\rho)\mqty(\cos(\sqrt{\omega}\rho)\\\sin(\sqrt{\omega}\rho))e^{-\bar\lambda\rho/K}=0\,.
\end{equation}
Just like Eq.~\eqref{eq:lambda0_binodal}, Eq.~\eqref{eq:forbidden_omegapos_cond} has a bifurcation point where the binodal coincides with the zero of the cosine, which corresponds to the boundary of the forbidden region. To find this point, we set $\Delta\rho = \Delta\rho_\mathrm{F} = \pi/\sqrt{\omega}$ and ask when Eq.~\eqref{eq:forbidden_omegapos_cond} admits a solution for $\bar\rho$, which we want to eliminate. Note that because $\mub'$ is quadratic, we find two quadratic equations for $\bar\rho$, one for the cosine and one for the sine integral. These solutions must match; plotting them as a function of $\nu$ at fixed $\bar\lambda$ reveals that they intersect when the discriminant of the equation from the cosine integral vanishes (cfr.~Fig.~\fref[b]{fig:forbidden_app}). This results in the following transcendental equation, which is finally independent of $\bar\rho$:
\begin{equation}
2 c\,\omega \left(\cosh \left(\frac{\pi  \bar\lambda }{\sqrt{\omega} K}\right)+1\right)=3\pi^2\frac{\nu^2}{K^2}\,,\label{eq:forbidden_omegapos}
\end{equation}
with $c = \left(3+ab\nu/K\right)\nu/K-3\bar\lambda^2/K^2$. Solving this equation for $\nu$ gives $\nuF(\bar\lambda)$, which we do numerically to obtain the boundary in Fig.~\fref[b]{fig:constantcoeff_forbiddenregion}.

We now turn to the case $\omega\leq 0$. Note that as $\omega \downarrow 0$, $\Delta\rho_\mathrm{F} \to \infty$, leading to the divergence of one of the binodals (this is what happens at the boundary of the forbidden region for $\omega \leq 0$, see Fig.~\fref[c]{fig:constantcoeff_forbiddenregion}). We use this property to find the boundary $\nuF(\bar\lambda)$ in this regime. 

It is convenient to take linear combinations of the pseudodensities in Eq.~\eqref{eq:psi_const}, denoted as $\psi_\pm(\rho) = e^{-(\bar\lambda/K\pm\sqrt{|\omega|})\rho}$. Now, we impose $\rhoL=+\infty$ in Eq.~\eqref{eq:binodals}, obtaining
\begin{equation}
    \int_{\rhoV}^{\infty}\dd{\rho}\mub'(\rho)\psi_\pm(\rho) = 0\,.
\end{equation}
Computing these integrals explicitly gives two quadratic equations for $\rhoV$. Again, these solutions must match, for $(\rhoV,\rhoL=\infty)$ to be a formal solution of Eq.~\eqref{eq:binodals} (see Fig.~\fref[c]{fig:forbidden_app}); this directly gives $\nuF$ for a fixed $\bar\lambda$, resulting in the function reported in Eq.~\eqref{eq:nuF_omeganeg}. That result is valid also for $\bar\lambda<0$ and for the case $\omega=0$, where the calculation is analogous.

In Appendix~\ref{app:flory} we discuss a free energy that only allows finite density values. There, the forbidden region survives, but its boundary never crosses the line $\omega=0$, so that $\Delta\rho_\mathrm{F}$ is always finite.
\subsection{\texorpdfstring{$\ofour$}
{O(grad 4)} theory with constant coefficients: Flory--Huggins free energy}
\label{app:flory}
We now consider the $\ofour$ theory with constant coefficients with a free-energy density $f(\rho)$ defined on a finite interval $\rho \in (0,1)$. This is the Flory--Huggins free energy:
\begin{equation}
    f(\rho) = \rho\ln\rho + (1-\rho)\ln(1-\rho) + \chi\rho(1-\rho)\,.
\end{equation}
This free energy can be derived from a two-species incompressible lattice gas; the first two terms reflect the entropy of the binary mixture and the last term the interactions, with $\chi$ being the Flory--Huggins parameter. Phase separation can occur for $\chi>2$. Note that the theory is symmetric under the transformation $\rho \to 1-\rho$ (\emph{i.e.}, mirroring the density about $1/2$) when the parameters are changed to $K, \lambda, \zeta, \nu \to K, -\lambda, -\zeta, \nu$. Because the higher-order gradient terms in the current are unaffected by the choice of $f(\rho)$, the pseudodensities of Eq.~\eqref{eq:psi_const} are still valid here.

This free energy still results in a forbidden region. However, unlike the quartic $f(\rho)$ case, for $\omega \leq 0$, the binodals always exist: In that case, we saw in App.~\ref{app:forbidden_region} that for $\omega \leq 0$, $\Delta\rho \to \infty$ as the forbidden region was approached, which here is not allowed. The forbidden region is thus limited to the regime where the pseudodensities are oscillatory ($\omega>0$); there, it depends on the Flory--Huggins parameter $\chi$, as we show below.

Let us consider the case $\bar\lambda=0$ for simplicity. Since $f(\rho)$ is symmetric with respect to $\rho=1/2$, it is convenient to shift the pseudodensities of Eq.~\eqref{eq:psi_const} and consider $\psi^{(1,2)}(\rho-1/2)$, which are still solutions of the pseudodensity equation. Then, as in the quartic $f(\rho)$ case, the condition $\Delta\Pi^{(2)} = 0$ is trivially fulfilled by choosing $\bar\rho = 1/2$. On the other hand, the condition $\Delta\Pi^{(1)} = 0$ gives the equivalent of Eq.~\eqref{eq:lambda0_binodal}:
\begin{equation}
    \int_{1/2}^{1/2+\Delta\rho/2}\dd{\rho} \mub'(\rho)\cos(\sqrt{\omega}(\rho-1/2)) = 0\,.\label{eq:lambda0_binodal_FH}
\end{equation}
This integral cannot be solved in closed form; however, one can show that it has a solution as long as $\Delta\rho < \pi/\sqrt{2\nuF^0/K}$. This results in a condition on $\nuF^0$; there are two cases to be distinguished: for $\chi<\chi_\ast$, one finds $\nuF^0/K > \pi^2/2$, so that the liquid binodal as one approaches the forbidden region lies at a finite distance from $\rho=1$; on the other hand, for $\chi\geq \chi_\ast$, one obtains $\nuF^0/K = \pi^2/2$: in this case, as $\nu\uparrow\nuF^0$, $\rhoL\to 1$ (see Fig.~\fref[a,b]{fig:flory}). The value of the Flory--Huggins parameter at which the behavior changes is found to be:
\begin{equation}
    \chi_\ast = \frac{\pi}{2}\int_0^\pi\frac{\sin x}{x}\dd{x} \approx 2.909\,.\label{eq:chiast}
\end{equation}
This is the value for which Eq.~\eqref{eq:lambda0_binodal_FH} for $\omega = \pi^2$ ceases to have a solution. 

For $\bar\lambda \neq 0$, the forbidden region cannot be found without explicitly computing the binodals, since $\bar\rho$ is unknown and cannot be eliminated from Eq.~\eqref{eq:binodals} as was done in App.~\ref{app:forbidden_region} to obtain Eq.~\eqref{eq:forbidden_omegapos}. However, we find numerically that $\chi_\ast$ defined in Eq.~\eqref{eq:chiast} governs the behavior for all $\bar\lambda$: For $\chi<\chi_\ast$, the binodals are finitely separated from $(0,1)$ as one approaches the forbidden region, while for $\chi>\chi_\ast$ one of them gets arbitrarily close to $(0,1)$. The boundary of the forbidden region is described by the same value of $\omega$ that was found for $\bar\lambda = 0$, \emph{i.e.}:
\begin{equation}
    \nuF(\bar\lambda,\chi) = \nuF^0(\chi) + \frac{\bar\lambda^2}{2K}\,.\label{eq:nuF_fh}
\end{equation}
While we have no theoretical explanation for this fact, but the numerically obtained binodals agree with Eq.~\eqref{eq:nuF_fh} accurately (see Fig.~\fref[c,d]{fig:flory}). Note that, crucially, the forbidden region does not intersect with the line $\omega = 0$, in contrast to the quartic $f(\rho)$ case (cfr. Fig.~\ref{fig:constantcoeff_forbiddenregion}).

\begin{figure}[tbp]
    \centering
    \includegraphics[width=\columnwidth]{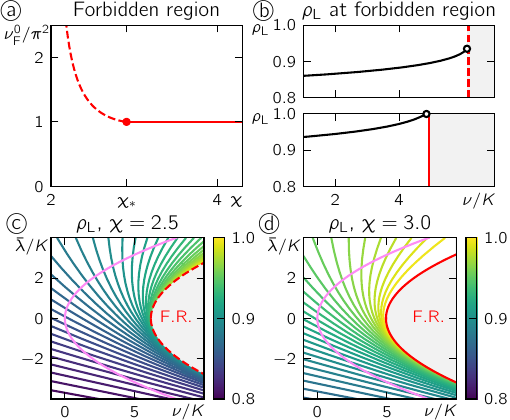}
    \caption{Binodals in Flory--Huggins $\ofour$ theory with constant coefficients. a) The value of $\chi$ determines where the forbidden region lies and the behavior of the binodals as its boundary is approached. For $\chi<\chi_\ast$ (dashed line), $\nuF^0>\pi^2$ and the binodals end at a finite distance from the values $(0,1)$. For $\chi>\chi_\ast$ (solid line), $\nuF^0=\pi^2$ and at least one binodal gets arbitrarily close to $(0,1)$ b) Top: Liquid binodal for $\bar\lambda=0,\chi=2.5<\chi_\ast$. Bottom: Liquid binodal for $\bar\lambda=0,\chi=3.0>\chi_\ast$. c,d) Liquid binodal $\rhoL$ for c) $\chi<\chi_\ast$ and d) $\chi>\chi_\ast$. The boundary of the forbidden region (dashed red line) is described by Eq.~\eqref{eq:nuF_fh}. Note that it never crosses the line $\omega = 0$ (solid pink line), in contrast to the quartic free energy case. The vapor binodal is given by $1-\rhoL$ upon taking $\bar\lambda \to -\bar\lambda$.\label{fig:flory}}
\end{figure}
\section{Numerical computation of binodals and interfacial tensions}
\label{app:bincalc}
We present here the numerical procedure to obtain the binodals using the framework presented in Sec.~\ref{sec:ofour_framework}. To obtain the binodals $\rhoVL$, the following steps are followed:
\begin{enumerate}
    \item Choose an interval $[\rho_\mathrm{min}, \rho_\mathrm{max}]$ within which the binodals lie.
    \item Obtain two linearly independent pseudodensities $\psi^{(1,2)}(\rho)$ as follows. To obtain $\psi^{(1)}(\rho)$, integrate Eq.~\eqref{eq:psi} with initial conditions $\psi^{(1)}(\rho_\mathrm{min})=0$ and $\psi^{(1)\prime}(\rho_\mathrm{min})=1$. Then, to obtain $\psi^{(2)}(\rho)$, integrate Eq.~\eqref{eq:psi} backwards with initial conditions $\psi^{(2)}(\rho_\mathrm{max})=0$ and $\psi^{(2)\prime}(\rho_\mathrm{max})=1$. The numerical integration is done by means of the solver \verb|solve_ivp()| from \verb|scipy| (or, if available, we use analytical solutions such as Eq.~\eqref{eq:psi_const}).
    \item Obtain the pseudopressures $\Pi^{(1,2)}(\rho)$ by numerically integrating Eq.~\eqref{eq:pseudopressure}.
    \item Define a nonnegative function $S(\rhoV,\rhoL) = \sum_{i=1,2}\left(\Delta\Pi^{(i)}(\rhoV,\rhoL)\right)^2$; the binodals $\rhoV$ and $\rhoL$ minimize this function by virtue of Eq.~\eqref{eq:binodals}. This minimization is done numerically with the \verb|least_squares()| routine from \verb|scipy|.
\end{enumerate}
Once the binodals are found, the interfacial tensions $\sigmaVL$ are calculated as follows:
\begin{enumerate}
    \item Obtain phase-anchored pseudodensities $\psiVL(\rho)$ by integrating Eq.~\eqref{eq:psi} with initial conditions $\psiVL(\rhoVL)=0$ and $\psiVL'(\rhoVL)=1$ (the latter being arbitrary), by means of \verb|solve_ivp()| (or use analytical solutions if available, \emph{e.g.}, Eq.~\eqref{eq:psi_const}).
    \item Obtain interfacial profile $w(\rho)$ by solving Eq.~\eqref{eq:w} using finite differences with boundary conditions $w(\rhoVL)=0$ (or use analytical solutions if available, \emph{e.g.}, Eq.~\eqref{eq:w_constcoeff}). Due to numerical error, $w(\rho)$ can be negative in a small interval close to the binodals that depends on the discretization; since the square root has to be taken, we set $w(\rho)=0$ wherever it is negative.
    \item Calculate the interfacial tensions $\sigmaVL$ by numerically integrating Eq.~\eqref{eq:sigma_rho}.
\end{enumerate}
\section{Multiple-scale expansions for AOUP in external potential}
\label{app:aoup}
In this Appendix, we demonstrate how the MS procedure can be employed for a single AOUP in an external potential. This amounts to deriving Eq.~\eqref{eq:result aoups external potential} in the limit of $\tau \downarrow 0$.
Starting from the Fokker--Planck equation~\eqref{eq:fpe-aoup} and projecting out the orientation, we write the time evolution of the single-particle orientational moments of the joint PDF~$\jointpdf$ that are defined as
\begin{subequations}\label{eq:moments-aoup}
\begin{align}
    \rho(\xv,t) &= \int  \dd{\uv} \jointpdf(\xv, \uv,t) 
    \,, 
    \\
     \pol (\xv,t) &= \int \dd{\vb{U}} \jointpdf(\xv, \uv,t)\,\uv
    \,,
    \\
    \Qv (\xv,t) &= \int \dd{\uv} \jointpdf(\xv, \uv,t)\,\uv \otimes\uv 
    \,,
    \\
    {\Tv}^{[n]}(\xv,t) &= \int \dd{\uv} \jointpdf(\xv, \uv,t) \bigotimes_{m=1}^n \uv 
    \,.
\end{align}
\end{subequations}
Note that we do not make the second moment and higher-order ones traceless. The time evolutions of the orientational moments then read  (all gradients are now in space): 
\begin{subequations}
\label{eq:dt_moments-aoup}
\begin{align}
    \partial_t \rho&= - v_0 \grad \vdot \pol + T \nabla^2 \rho + \aoupmob \grad \vdot  \left(\rho\grad  \externalPotential   \right) 
    \label{eq:dt_density-aoup}
    \,, \\ 
     \partial_t \pol&= -\frac{\pol}{\tau}
     - v_0 \grad \vdot \Qv + T \nabla^2 \pol 
     \nonumber\\
     &\phantom{=}+ \aoupmob \grad \vdot  \left(\pol  \otimes \grad  \externalPotential   \right) 
     \,, 
     \label{eq:dt_pol-aoup}\\ 
     \partial_t \Qv&= -\frac{2\Qv}{\tau}
     - v_0 \grad \vdot  {\Tv}^{[3]} + T \nabla^2 \Qv \nonumber\\
     &\phantom{=}+ \aoupmob \grad \vdot  \left(\Qv \otimes \grad  \externalPotential   \right) + \frac{2}{\tau} \rho \id
     \,, \label{eq:dt_Q-aoup}\\ 
     \partial_t {\Tv}^{[n]}&= -\frac{n{\Tv}^{[n]}}{\tau}
     - v_0 \grad \vdot  {\Tv}^{[n+1]} + T \nabla^2 {\Tv}^{[n]} \nonumber \\
     &\phantom{=}+ \aoupmob \grad \vdot  \left({\Tv}^{[n]} \otimes \grad  \externalPotential   \right)\nonumber \\
     &\phantom{=}+ \frac{n(n-1)}{\tau}\left(\id\otimes{\vb{T}}^{[n-2]}\right)^\symm
     \,,  
     \label{eq:dt_generalmoment-aoup}
\end{align}
\end{subequations}
where for this section,  $\symm$ indicates total symmetrization with respect to all spatial indices, and the divergence always acts on the last index: $(\grad \vdot \vb{A})_{i} = \nabla_j A_{ij}$, etc.).
From the structure of Eqs.~\eqref{eq:dt_moments-aoup}, we see that the polarization~$\pol$ contributes terms of at least order $\mathcal{O}(\tau)$ in the density equation, $\Qv$ terms of order $\mathcal{O}(1)$ and higher, and ${\Tv}^{[3]}$ terms of at least order $\mathcal{O}(\tau^3)$. To obtain all terms of quadratic order in $\tau$ in the density equation, it is therefore justified to set ${\Tv}^{[n]} = 0, n\ge 3$. 
\par 
We now proceed by making the following perturbative ansatz in the limit $\tau \downarrow 0$ in Eqs.~\eqref{eq:dt_moments-aoup}: 
\begin{subequations}
\begin{align}
    \rho &= \rho^{(0)} + \tau \rho^{(1)} + \tau^2 \rho^{(2)} + \mathcal{O}(\tau^3) \,, 
    \\
    \pol &= \pol^{(0)} + \tau \pol^{(1)} + \tau^2 \pol^{(2)} + \mathcal{O}(\tau^3)  \,,
     \\
    \Qv &= \Qv^{(0)} + \tau \Qv^{(1)}  + \mathcal{O}(\tau^2)  \,,
\end{align}
\end{subequations}
and by introducing the following times: 
\begin{align}
    t_0 = t\,, \quad t_1 = \tau t\,,\quad t_2 = \tau^2 t
    \,, 
\end{align}
which implies that the time derivatives in Eqs.~\eqref{eq:dt_moments-aoup} are replaced by
\begin{align}
    \frac{\partial}{\partial t}
    \to \frac{\partial}{\partial t_0}
    + \tau\frac{\partial}{\partial t_1}
    + \tau^2 \frac{\partial}{\partial t_2}
    + \cdots
    \, . 
\end{align}
We now solve the resulting equations order by order in the perturbation parameter.
At order $1/\tau$, we get from Eq.~\eqref{eq:dt_pol-aoup}: 
\begin{align}
    \pol^{(0)} = 0 \,, 
    \label{eq:pol-tau0-aoup}
\end{align}
and from Eq.~\eqref{eq:dt_Q-aoup}: 
\begin{align}
    \Qv^{(0)} =\rho^{(0)} \id \,, 
    \label{eq:Q-tau0-aoup}
\end{align}
The terms of order $\mathcal{O}(1)$ in Eq.~\eqref{eq:dt_density-aoup} give, using Eq.~\eqref{eq:pol-tau0-aoup}: 
\begin{align}
    \frac{\partial}{\partial t_0} \rho^{(0)} 
    &= T \nabla^2 \rho^{(0)} + \aoupmob \grad \vdot  \left(\rho^{(0)}\grad  \externalPotential  \right)\,,
    \label{eq:rho-tau0-aoup}
\end{align}
and Eq.~\eqref{eq:dt_pol-aoup} becomes, substituting Eqs.~\eqref{eq:pol-tau0-aoup} and \eqref{eq:Q-tau0-aoup}: 
\begin{align}
    \pol^{(1)} = - v_0 \grad \rho^{(0)} 
    \, .
    \label{eq:pol-tau1-aoup}
\end{align}
Likewise, Eq.~\eqref{eq:dt_Q-aoup} yields upon substitution of Eq.~\eqref{eq:rho-tau0-aoup}: 
\begin{align}
    \Qv^{(1)} =\rho^{(1)} \id \,. 
    \label{eq:Q-tau1-aoup}
\end{align}
At order $\mathcal{O}(\tau)$,  
the polarization becomes upon substitution of Eqs.~\eqref{eq:pol-tau0-aoup}, \eqref{eq:pol-tau1-aoup} and \eqref{eq:Q-tau1-aoup} into Eq.~\eqref{eq:dt_pol-aoup}: 
\begin{align}
    \pol^{(2)} &= v_0 \grad \left( \frac{\partial}{\partial t_0} \rho^{(0)} \right) - v_0 \grad \rho^{(1)} - T v_0 \nabla^2 \grad \rho^{(0)} 
    \nonumber\\
    &\phantom{=}- v_0 \aoupmob \grad \vdot  \left(\grad\rho^{(0)} \otimes  \grad  \externalPotential    \right)
    \, .
\end{align}
Using Eq.~\eqref{eq:rho-tau0-aoup}, this gives
\begin{equation}
\begin{aligned}
    \pol^{(2)} &= - v_0 \grad \rho^{(1)} 
    + v_0 \aoupmob  \grad \grad \vdot  \left(\rho^{(0)}  \grad  \externalPotential  \right)
    \\
    &\phantom{=}- v_0 \aoupmob \grad \vdot  \left(\grad\rho^{(0)} \otimes  \grad  \externalPotential  \right)
    \, .
\end{aligned}
\end{equation}
Substituting $\pol^{(0)}, \pol^{(1)}, \pol^{(2)}$ into our perturbative expansion of Eq.~\eqref{eq:dt_density-aoup}, the effective time evolution of the density reads:
\begin{equation}
\begin{aligned}
    \partial_t\marginaldistro
    &=  \left(v_0^2 \tau + T \right) \laplacian \marginaldistro
    + 
     \aoupmob \div \left[\left(\grad \externalPotential  \right) \marginaldistro \right]
     \\[0.4em]
     &\phantom{=}- v_0^2 \tau^2 \aoupmob \grad\grad\boldsymbol{:} \left[\left(\grad\grad \externalPotential  \right)     \marginaldistro 
     \right]
     \,, 
\end{aligned}
\end{equation}
which is Eq.~\eqref{eq:result aoups external potential} in the main text. Hence, at order $\tau^2$, we have an effective Fokker--Planck equation. 
Note that $\tau$ has to be small with respect to the inverse of the largest eigenvalue of the Hessian~$\grad\grad \externalPotential$ of the external potential~\cite{baek-2023} to have a positive definite diffusion coefficient, so that $\marginaldistro$ can be interpreted as a probability distribution. In principle, higher-order corrections in $\tau$ involve higher-order derivatives of~$\externalPotential$ with corresponding criteria~\cite{baek-2023}. 
\par 
On the other hand, the UCNA result in our notations reads (following the standard procedure explained in Ref.~\cite{jung-1987} and reviewed in Refs.~\cite{martin-2021,baek-2023})
\begin{equation}
\begin{aligned}
    \partial_t \marginaldistro
    &= - \sum_{\alpha,\beta}\nabla_\alpha \left[B_{\alpha\beta} \aoupmob \left(\nabla_\beta \externalPotential\right) \marginaldistro \right]
    \\
    &\phantom{={}}+\sum_{\alpha,\beta,\gamma} \left(v_0^2 \tau + T \right) \nabla_\alpha \left[B_{\alpha\beta} \nabla_\gamma \left(B_{\beta\gamma} \marginaldistro \right) \right]
    \,, 
    \label{eq:ucna_dtrho}
\end{aligned}
\end{equation}
where $\left(B\right)^{-1}_{\alpha\beta} = \delta_{\alpha\beta} + \tau \aoupmob \nabla_\alpha \nabla_\beta \externalPotential$. Comparing Eqs.~\eqref{eq:result aoups external potential} and \eqref{eq:ucna_dtrho} it is then clear that UCNA is, as expected, incorrect beyond order $\mathcal{O}(\tau^0)$ at dynamical level.
\section{Multiple-scale expansions for \MakeLowercase{t}QSAPs}
\label{app:coarsegraining_QS}
We now give details on the moment equations associated with QS active particles in App.~\ref{app:moment-eqs_qs-aoups}, 
deriving Eqs.~\eqref{eq:qs_many-result} and Eqs.~\eqref{eq:o4-coeffs_dimful} of the main text. Finally, in App.~\ref{app:abps} we derive via multiple scale analysis the evolution for $\hat\rho$ from ABPs (rather than for AOUPs as done in the rest of the paper), and provide the corresponding coefficients of the $\ofour$ theory.
\subsection{Derivation of moment equations}
\label{app:moment-eqs_qs-aoups}
We now provide details on the derivation of the moment equations in tQSAPs. 
We obtain the time evolution of the $n$-th orientational moment by multiplying the Fokker--Planck equation~\eqref{eq:Nbody_FPE_QS_dimless} by $\bigotimes_{m=1}^n \uv_{i_m}$, integrating over $\allus$, and assuming that~$\jointpdf$ and its derivatives vanish for $|\uv| \to \infty$. By that, we obtain the time evolutions given in Eqs.~\eqref{eq:dt_moments}. For convenience in the next section, we explicitly write out the time evolution of the moments $\hat\Tv^{[3]}, \hat\Tv^{[4]}$ (with $\hat\Tv^{[n]} =0, n \ge 5$, as discussed in Sec.~\ref{sec:fpe_qsap}): 
\begin{subequations}\label{eq:dt_moments-3-4}
\begin{align}
    \epsfac\partial_t \hat\Tv^{[3]}_{ijk} &= -\frac{3}{ \epslvap^2}\hat\Tv^{[3]}_{ijk} 
     -\frac{1}{\epslvap}\grad_l\vdot\left(v_l \hat\Tv^{[4]}_{ijkl}\right)
    \nonumber
    \\
    &\phantom{{}={}}+  \frac{6}{\epslvap^2}\left(\delta_{ij}\id\otimes\hat\pol_k\right)^\symm 
    + \frac{\etaTsq}{A^2}\laplacian_\allxs \Tv^{[3]}_{ijk}
    \label{eq:dt_T3-qs}
    \,,
    \\
    \epsfac\partial_t \hat\Tv^{[4]}_{ijkl} &= -\frac{4}{\epslvap^2}\hat\Tv^{[4]}_{ijkl} 
    + \frac{12}{\epslvap^2}\left(\delta_{ij}\id\otimes\hat\Qv_{kl}\right)^\symm
    \nonumber\\
     &\phantom{{}={}} + \frac{\etaTsq}{A^2}\laplacian_\allxs \Tv^{[4]}_{ijkl}
     \label{eq:dt_T4-qs}
    \,. 
\end{align}
\end{subequations}
Note that, in contrast to Refs.~\cite{cates-2013, cates-2015, solon-2015, solon-2018, solon-2018-njp, martin-2021,dinelli-2024}, these are $N$-particle and not single-particle moments.  
\subsection{Multiple-scale expansion}\label{app:QSAPs-eq80}
To derive a closed equation for $\hat\rho$ from the hierarchy composed of Eqs.~\eqref{eq:dt_density-qs}--\eqref{eq:dt_Q-qs} and  \eqref{eq:dt_moments-3-4}, we now formulate the ansatz for the MS expansion: 
\begin{subequations}\label{eq:multi-field-ansatz-qs}
\begin{align}
    \hat\rho &=\hat\rho^{(0)}+\epslvap \hat\rho^{(1)}+\epslvap^2 \hat\rho^{(2)} + \epslvap^3 \hat\rho^{(3)} +\mathcal{O}(\epslvap^4)\,, 
     \\ 
    \hat\pol_i &=\hat\pol^{(0)}_i + \epslvap\hat\pol_i^{(1)} + \epslvap^2 \hat\pol_i^{(2)} + \epslvap^3 \hat\pol_i^{(3)} +\mathcal{O}(\epslvap^4)\,,
    \\
    \hat\Qv_{ij} &=  \hat\Qv_{ij}^{(0)} + \epslvap \hat\Qv_{ij}^{(1)} +  \epslvap^2 \hat\Qv_{ij}^{(2)} +\mathcal{O}(\epslvap^3)\,,
    \\
    \hat\Tv^{[3]}_{ijk} &= \hat{\Tv}^{[3](0)}_{ijk} + \epslvap \hat{\Tv}^{[3](1)}_{ijk}  + \mathcal{O}(\epslvap^2)\,, 
    \\
    \hat\Tv^{[4]}_{ijkl} &=  \hat{\Tv}^{[4](0)}_{ijkl} + \mathcal{O}(\epslvap) 
    \,, 
\end{align}
\end{subequations}
introduce multiple time scales as in the toy model (cf.~Eq.~\eqref{eq:qs-multi_time-scales}) and replace the time derivatives as in Eq.~\eqref{eq:multiple scale time derivative ansatz minimal model}. We now solve for $\hat\pol^{(i)}, i\le 3$ in the density equation~\eqref{eq:dt_density-qs} by inserting the MS ansatz into Eqs.~\eqref{eq:dt_density-qs}--\eqref{eq:dt_Q-qs} and  \eqref{eq:dt_moments-3-4}.
\par 
At order $\mathcal{O}(1/\epslvap^2)$, we obtain
\begin{subequations}\label{eq:multi-qs_eps0}
\begin{align}
    \hat\pol^{(0)}_i &= 0\,, \\
    \hat\Qv_{ij}^{(0)} &= \delta_{ij}\id \hat\rho^{(0)}\,,\\
    \hat{\Tv}^{[3](0)}_{ijk} &= 2 \left(\delta_{ij}\id\otimes\hat\pol_k^{(0)}\right)^\symm = 0\,, \\
    \hat{\Tv}^{[4](0)}_{ijkl}
    &= 3\left(\delta_{ij}\id\otimes\delta_{kl} \id \hat\rho^{(0)}\right)^\symm 
    \, . 
\end{align}
\end{subequations}
The next order gives, by virtue of Eqs.~\eqref{eq:multi-qs_eps0}: 
\begin{subequations}\label{eq:multi-qs_eps1}
\begin{align}
    \hat\pol^{(1)}_i &= - \grad_j \vdot \left(v_j \delta_{ij} \id \hat \rho^{(0)} \right)\,, \\
    \hat\Qv_{ij}^{(1)} &= \delta_{ij}\id \hat\rho^{(1)}\,,\\
    \hat{\Tv}^{[3](1)}_{ijk} &= 2 \left(\delta_{ij}\id\otimes\hat\pol_k^{(1)}\right)^\symm \nonumber  \\
    &\phantom{=}- \grad_l\vdot\left[v_l \left(\delta_{ij}\id\otimes\delta_{kl} \id \hat\rho^{(0)}\right)^\symm\right]
    \, . 
\end{align}
\end{subequations}
Using both Eqs.~\eqref{eq:multi-qs_eps0} and \eqref{eq:multi-qs_eps1}, the order $\mathcal{O}(\epslvap^0)$ yields
\begin{subequations}\label{eq:multi-qs_eps2}
\begin{align}
    \epsfac \frac{\partial}{\partial t_0}
    \hat \rho^{(0)} &= \grad_i \vdot\left[v_i \grad_i\left(v_i \hat\rho^{(0)}\right)\right] 
    +\frac{\etaTsq}{A^2} \nabla_i^2 \hat\rho^{(0)}\,,
    \\
    \hat\pol^{(2)}_i &= - \grad_j \vdot \left(v_j \delta_{ij} \id \hat \rho^{(1)} \right)\,,
    \\
    \hat\Qv_{ij}^{(2)} &= \delta_{ij}\id \hat\rho^{(2)} + \delta_{ik}\delta_{jl}\left\{\grad_k\left[v_k \grad_l\left(v_l \hat\rho^{(0)}\right)\right]\right\}^\symm
    \, . 
\end{align}
\end{subequations}
Substituting Eqs.~\eqref{eq:multi-qs_eps0},  \eqref{eq:multi-qs_eps1} and \eqref{eq:multi-qs_eps2} into Eq.~\eqref{eq:dt_pol-qs} leads to
\begin{align}
    \hat\pol^{(3)}_i &= - \grad_j \vdot \left(v_j \hat \Qv_{ij}^{(2)} \right)\nonumber\\
    &\phantom{{}={}}+ \delta_{ik}\grad_k\left\{v_k \grad_j \vdot\left[v_j \grad_j\left(v_j  \hat\rho^{(0)}\right)\right]\right\}\nonumber
    \\
    &\phantom{{}={}}-  \frac{\etaTsq}{A^2}\delta_{ik} \grad_k\left[\laplacian_\allxs( v_k \hat\rho^{(0)}) -v_k \laplacian_\allxs \hat\rho^{(0)}\right] 
    \,.
    \label{eq:multi-qs_eps3}
\end{align}
Substituting Eqs.~\eqref{eq:multi-qs_eps0}, \eqref{eq:multi-qs_eps1}, \eqref{eq:multi-qs_eps2}, and \eqref{eq:multi-qs_eps3} into the density Eq.~\eqref{eq:dt_density-qs}, we obtain the effective time evolution of the $N$-particle density field in dimensionless variables:
\begin{equation}
\begin{aligned}
     \epsfac\partial_t \hat\rho &=  
    \grad_i \vdot\left[v_i \grad_i\left(v_i \hat\rho\right)\right] 
    +\frac{\etaTsq}{A^2} \nabla_i^2 \hat\rho
    \\
    &+\frac{\epslvap^2\etaTsq}{A^2} \grad_i \vdot\left\{v_i \grad_i\left[\laplacian_\allxs\left(v_i \hat\rho\right) -v_i \laplacian_\allxs \hat\rho\right]\right\} \\
    & -\epslvap^2\grad_i \vdot\left(v_i \grad_i\left\{v_i \grad_j \vdot\left[v_j \grad_j\left(v_j  \hat\rho\right)\right]\right\}\right)
    \\
    &+\epslvap^2 \grad_i \vdot\left(v_i \grad_j \vdot\left\{v_j\left[\grad_i(v_i \grad_j\left(v_j \hat\rho)\right)\right]^\symm\right\}\right)
    \,,
\end{aligned}
\end{equation}
which is Eq.~\eqref{eq:qs_many-result}
of the main text. 
\subsection{Derivation of \texorpdfstring{$\ofour$}{O(grad 4)} theory}
\label{app:ofour_qsap_derivation}
We now show how to derive the $\ofour$ theory in dimensionful variables for tQSAPs given in Eq.~\eqref{eq:o4-coeffs_dimful} in the main text. 
For clarity, we first report here both the mean-field results for DD and MS in the subsequent equation:
\begin{equation}
\begin{aligned}
    &\epsfac\partial_t \rho =  
    \div\left[\qsspeed \grad\left(\qsspeed \rho\right)\right] 
    +\frac{\etaTsq}{A^2} \laplacian \rho
    \\
    &+ \cms \, \frac{\epslvap^2\etaTsq}{A^2}  \div\left\{\qsspeed \grad \left[\laplacian( \qsspeed \rho) -\qsspeed \laplacian \rho\right]\right\} \\
    & - \cms \,\epslvap^2\div\left(\qsspeed \grad\left\{\qsspeed \div\left[\qsspeed \grad(\qsspeed  \rho)\right]\right\}\right)
    \\
    &+ \cms \, \epslvap^2 \div\left(\qsspeed \div\left\{\qsspeed[\grad (\qsspeed \grad(\qsspeed \rho))]^\symm\right\}\right)
    \,,
    \label{eq:qs_all-result_ms-dd_resc}
\end{aligned}
\end{equation}
where $\cms$ is a bookkeeping parameter that equals unity in our MS derivation and vanishes for the DD approximation. 
In Eq.~\eqref{eq:qs_all-result_ms-dd_resc}, we now undo the rescaling in time and space in terms of $\ellr$ and $\tir$, reinstating the original variables of the particle model of Eqs.~\eqref{eq:qs_langevin}. This leads to 
\begin{equation}
\begin{aligned}
    &\partial_t \rho =  
    \tau \div\left[\qsspeed \grad\left(\qsspeed \rho\right)\right] 
    +T \laplacian \rho
    \\
    &+ \cms T \tau^2 \div\left\{\qsspeed \grad \left[\laplacian( \qsspeed \rho) -\qsspeed \laplacian \rho\right]\right\} \\
    & - \cms \tau^3\div\left(\qsspeed \grad\left\{\qsspeed \div\left[\qsspeed \grad(\qsspeed  \rho)\right]\right\}\right)
    \\
    &+ \cms \tau^3 \div\left(\qsspeed \div\left\{\qsspeed[\grad (\qsspeed \grad(\qsspeed \rho))]^\symm\right\}\right)
    \,.
    \label{eq:qs_all-result_ms-dd}
\end{aligned}
\end{equation}
As explained in Sec.~\ref{sec:ofour_qsap}, we now perform a gradient expansion of the QS speed $\qsspeed$. Equation~\eqref{eq:vqs_grad-expansion_resc} in dimensionful variables reads
\begin{align}
    \qsspeed[\rho] = \vloc(\rho) + \gamma^2 \vloc'(\rho)\laplacian \rho
    \,,
    \label{eq:vqs_grad-expansion}
\end{align}
with $\gamma$ defined in Eq.~\eqref{eq:gamma}. 
We now substitute Eq.~\eqref{eq:vqs_grad-expansion} into Eq.~\eqref{eq:qs_all-result_ms-dd}, and only keep terms up to fourth order in gradients. 
The first two terms in Eq.~\eqref{eq:qs_all-result_ms-dd} read
\begin{align}
    I_1 \coloneqq {}& \tau \div\left[\qsspeed \grad\left(\qsspeed \rho\right)\right] 
    +T \laplacian \rho
    \nonumber\\
    ={}& \nabla^2  \left( T \rho +  \tau \int \mathrm{d} \rho \,  
    (\vloc \vloc' \rho + \vloc^2)  + 
     \gamma^2 \tau\vloc \vloc' \rho \nabla^2 \rho\right) 
    \nonumber\\
     &- \div[- \tau \gamma^2 \vloc' \left( \vloc' \rho + \vloc\right)  \laplacian\rho   \grad \rho ] 
    \nonumber\\[0.4em]
     & 
    -  \div[ \gamma^2 \tau \vloc'^2 \rho \laplacian\rho   \grad \rho] 
     + \osix \, . 
     \label{eq:ofour_qsap_tau01}
\end{align}
The $\tau^2$ terms in Eq.~\eqref{eq:qs_all-result_ms-dd} read
\begin{align}
    I_2 \coloneqq {}& \cms T \tau^2 \div\left\{\qsspeed \grad \left[\laplacian( \qsspeed \rho) -\qsspeed \laplacian \rho\right]\right\} 
    \nonumber\\
    ={}&  \cms  T \tau^2
    \nabla^2 \left[\vloc   
       \left(\vloc'' \rho + 2\vloc'\right)  \left(\grad\rho\right)^2
    + \vloc \vloc' \rho \nabla^2 \rho 
    \right]
    \nonumber\\ 
    &-  \cms  T \tau^2 \div \left[ \vloc' \left(\vloc'' \rho + 2\vloc'\right) \left(\grad\rho\right)^2 \grad \rho \right.
    \nonumber\\&\quad \left. 
    {}+{}\vloc'^2 \rho \laplacian\rho  \grad \rho \right]
    + \osix \,. 
    \label{eq:ofour_qsap_tau2}
\end{align}
The $\tau^3$ terms in Eq.~\eqref{eq:qs_all-result_ms-dd} read:
\begin{align}
      I_3 \coloneqq {}& - \cms \tau^3\div\left(\qsspeed \grad\left\{\qsspeed \div\left[\qsspeed \grad(\qsspeed  \rho)\right]\right\}\right)
    \nonumber\\
    & + \cms \tau^3 \div\left(\qsspeed \div\left\{\qsspeed[\grad (\qsspeed \grad(\qsspeed \rho))]^\symm\right\}\right)
    \nonumber\\
    ={}& - \cms \tau^3\div(\tilde g(\rho) \laplacian \rho \grad \rho)
   \nonumber\\
   &+ \cms \tau^3   \laplacian[\frac{1}{2}\tilde g(\rho)  \left(\grad \rho\right)^2 ] 
   \nonumber\\
   &-  \cms \tau^3   \div[\frac{1}{2}\tilde g'(\rho) \left(\grad \rho\right)^2 \grad \rho ]
   + \osix 
   \, ,
   \label{eq:ofour_qsap_tau3}  
\end{align}
where $\tilde g(\rho) =  \vloc (\vloc \vloc'^2 \rho + \vloc^2 \vloc')$ is an auxiliary function. 
From Eqs.~\eqref{eq:ofour_qsap_tau01}, \eqref{eq:ofour_qsap_tau2} and \eqref{eq:ofour_qsap_tau3}, using $\partial_t \rho = I_1 + I_2 + I_3$, Eqs.~\eqref{eq:continuity} and \eqref{eq:o4}, we find the $\ofour$ coefficients for tQSAPs given in Eq.~\eqref{eq:o4-coeffs_dimful} in the main text.
\subsection{Active Brownian particles}
\label{app:abps}
In this section, we show that the same MS approach can also be applied to tQSAPs whose orientation follows a Brownian dynamics on the unit sphere (the latter is also referred to as \enquote{active Brownian particles}). The coarse-graining procedure presented in the main text is therefore not restricted to Gaussian orientations. We follow the same steps laid out in Sec.~\ref{sec:coarsegraining_qsap}. For simplicity, we write the calculations for the two-dimensional case; the generalization to higher dimensions can be made by introducing spherical harmonics.  
\par 
We first start with the particle model: 
The dynamics of particle $i=1,\ldots,N$ with position $\xv_i$ and orientation vector $\uv_i$ follows: 
\begin{subequations}
\label{eq:qs_langevin-brownian}
\begin{align}
    \dot{\xv}_i  &=  \qsspeed[\rho_{\mathrm{d}}](\xv_i) \uv_i 
    + \sqrt{2T} \, \Lvt_i \,,
    \\[0.4em]
    \uv_i &= (\cos(\theta_i), \sin(\theta_i))^T\,,
    \\[0.4em]
    \dot{\theta}_i  &=   \sqrt{\frac{2}{\tau}} \, 
   \eta_i
   \,, 
\end{align}
\end{subequations}
where $\qsspeed$, $T, \tau$, and  $\Lvt$ are specified in Secs.~\ref{sec:particle-model_qsap} and \ref{sec:qsap_dimensionless} and $\eta_i$ is a Gaussian white noise with zero mean: 
\begin{equation}
    \langle \eta_i(t) \eta_j(t')\rangle 
    = \delta_{ij} \delta(t-t') 
    \, . 
\end{equation}
\subsubsection{Hierarchy for the orientational moments}
\label{sec:fpe_qsap-brownian}
The corresponding Fokker--Planck equation for the $N$-body PDF~$\jointpdf = \jointpdf(\allxs,\allthetas, t)$ with $\allthetas = (\theta_1, \hdots, \theta_N)$ reads
\begin{align}
    \partial_t \jointpdf  &= \tau^{-1} \partial_{\theta_i \theta_i}  \jointpdf
    + T \laplacian_\allxs \jointpdf - \grad_i \vdot(v_i(\allxs) \uv_i \jointpdf)
    \,,
    \label{eq:Nbody_FPE_QS-brownian}
\end{align}
where again $\grad_i = \grad_{\xv_i}$ and $\laplacian_\allxs \coloneqq \grad_i\vdot\grad_i$. 
In dimensionless variables, this becomes (in analogy to Eq.~\eqref{eq:Nbody_FPE_QS_dimless}): 
\begin{equation}
\begin{aligned}
    \epsfac\partial_t \jointpdf  &= \frac{1}{\epslvap^2} \partial_{\theta_i \theta_i}  \jointpdf
    + \frac{\etaTsq}{A^2} \laplacian_\allxs \jointpdf 
    \\&\phantom{=}
    - \frac{1}{\epslvap} \grad_i \vdot(v_i(\allxs) \uv_i \jointpdf)
    \,. 
    \label{eq:Nbody_FPE_QS_dimless-brownian}
\end{aligned}
\end{equation}
Because $\uv_i$ now lives on the unit sphere, the derivation of the QS active particles orientational moment equations is slightly more involved than for the Ornstein--Uhlenbeck case discussed in the main text. For this, we follow again the standard QS active particles references~\cite{cates-2013, cates-2015, solon-2015, solon-2018, solon-2018-njp,dinelli-2024}, and extend them to our $N$-body formulation. 
In accordance with the literature, we introduce the orientational moments as harmonic tensors, \emph{i.e.}, they are traceless and symmetric. The first three read
\begin{subequations}\label{eq:moments-brownian-0-2}
\begin{align}
     \hat\rho(\allxs,t) &= \int  \dd{\allthetas} \jointpdf(\allxs, \allthetas,t)
    \,, 
    \\
     \hat\pol_i(\allxs,t) &= \int \dd{\allthetas} \jointpdf(\allxs, \allthetas,t)\,\uv_i
    \,,
    \\
    \hat\Qv_{ij}(\allxs,t) &= \int \dd{\allthetas} \jointpdf(\allxs, \allthetas,t)\, 
    \times \nonumber \\
    &\phantom{=}\times \left(\uv_i\otimes\uv_j - \frac{\delta_{ij} \id}{2} \right)
    \,, 
\end{align}
\end{subequations}
and the next two are given by
\begin{subequations}\label{eq:moments-brownian-3-4}
\begin{align}
    &\hat\Tv^{[3]}_{ijk}(\allxs,t) = \int \dd{\allthetas} \jointpdf(\allxs, \allthetas,t)\, \times \nonumber \\
    &\phantom{=}\times \left[\uv_i\otimes\uv_j\otimes\uv_k - \frac{3}{4} \left(\delta_{ij} \id \otimes \uv_k\right)^\symm \right]
    \,,
    \\
    & \hat\Tv^{[4]}_{ijkm}(\allxs,t) = \int \dd{\allthetas} \jointpdf(\allxs, \allthetas,t)\, \times \nonumber \\
    &\phantom{=}\times \left[\uv_i\otimes\uv_j\otimes\uv_k \otimes \uv_m -\left(\delta_{ij} \id \otimes \uv_k \otimes \uv_m \right)^\symm \right. \nonumber
    \\
    &\phantom{=\times} \left.
    + \frac{1}{8} \left(\delta_{ij} \id \otimes \delta_{km} \id  \right)^\symm
    \right]
    \,. 
\end{align}
\end{subequations}
Here and in the following, all $\theta_i$ integrals are taken over the interval $[0, 2 \pi)$, and $\symm$ indicates again total symmetrization with respect to all spatial and particle indices. 
We proceed by projecting Eq.~\eqref{eq:Nbody_FPE_QS_dimless-brownian} on the moments of $\uv_i$, such that we obtain the time evolution of the moments (Eqs.~\eqref{eq:moments-brownian-0-2} and \eqref{eq:moments-brownian-3-4}).  
To this goal, we use the properties of Fourier (spherical harmonics in higher dimensions) functions. 
The evolution equations for the first two moments read
\begin{subequations}
\label{eq:dt_moments-brownian-0-1}
\begin{align}
    \epsfac\partial_t\hat\rho&= -\frac{1}{\epslvap}\grad_i\vdot\left(v_i \hat\pol_i \right) + \frac{\etaTsq}{A^2}\laplacian_\allxs \hat\rho\label{eq:dt_density-qs-brownian}
    \,,
    \\
    \epsfac\partial_t \hat\pol_i &= -\frac{1}{\epslvap^2}\hat\pol_i -\frac{1}{\epslvap}\grad_j\vdot\left(v_j \hat\Qv_{ij}\right)   
    \nonumber
    \\
    &\phantom{{}={}} -\frac{1}{2\epslvap}\grad_j\vdot\left(v_j \delta_{ij} \id \hat \rho \right) + \frac{\etaTsq}{A^2}\laplacian_\allxs \hat\pol_i\label{eq:dt_pol-qs-brownian}
    \,,
\end{align}
\end{subequations}
and for the third, we obtain
\begin{equation}
\begin{aligned}
    \epsfac\partial_t \hat\Qv_{ij} &= -\frac{4}{\epslvap^2}\hat\Qv_{ij} 
    -\frac{1}{\epslvap}\grad_k\vdot\left(v_k \hat\Tv^{[3]}_{ijk}\right)
    \\
    &\phantom{{}={}} -\frac{3}{4\epslvap}\grad_k\vdot\left[v_k \left( \hat\pol_k \otimes \delta_{ij} \id   \right)^\symm \right]
    \\
    &\phantom{{}={}} + \frac{1}{2 \epslvap} \grad_k\vdot\left[v_k \left(\delta_{ij} \id \otimes \hat\pol_k\right)  \right]
    \\
    &\phantom{{}={}}
     + \frac{\etaTsq}{A^2}\laplacian_\allxs \hat\Qv_{ij} 
    \,.
    \label{eq:dt_Q-qs-brownian}
\end{aligned}
\end{equation}
The time evolution of $\hat\Tv^{[3]}_{ijk}$ reads
\begin{equation}
\begin{aligned}
    \epsfac\partial_t \hat\Tv^{[3]}_{ijk} &= -\frac{9}{\epslvap^2}\hat\Tv^{[3]}_{ijk} 
    -\frac{1}{\epslvap}\grad_m\vdot\left(v_m \hat\Tv^{[4]}_{ijkm}\right)  
    \\
    &\phantom{{}={}} -\frac{1}{4\epslvap}
    \left[\delta_{ij} \id \otimes \grad_m \vdot \left(v_m \hat\Qv_{km} \right) \right]^\symm 
    \\
    &\phantom{{}={}} - \frac{1}{2 \epslvap} \left[\grad_m \vdot \left(v_m \hat\Qv_{ij} \otimes \delta_{km} \id  \right)^\symm  \right] 
    \\
    &\phantom{{}={}} 
     + \frac{\etaTsq}{A^2}\laplacian_\allxs \hat\Tv^{[3]}_{ijk} \label{eq:dt_T3-qs-brownian}
    \,.
\end{aligned}
\end{equation}
The time evolution of $\hat\Tv^{[4]}_{ijkm}$ follows
\begin{equation}
\begin{aligned}
    &\epsfac\partial_t \hat\Tv^{[4]}_{ijkm}  = -\frac{16}{\epslvap^2}\hat\Tv^{[4]}_{ijkm} 
    + \frac{\etaTsq}{A^2}\laplacian_\allxs \hat\Tv^{[4]}_{ijkm}
     \\
    & -\frac{1}{\epslvap}\grad_p\vdot \left[v_p \left\langle 
    \uv_p  \otimes \left(\uv_i \otimes \uv_j \otimes  \uv_k \otimes \uv_m \right)^\TS, P
    \right\rangle\right]
    \,,
    \label{eq:dt_T4-qs-brownian}
\end{aligned}
\end{equation}
where $\langle f ,g \rangle = \int \dd \allthetas f g$, and $\vb{A}^\TS$ denotes the symmetrized traceless version of tensor $\vb{A}$. As we argue in the next section, we do not need to simplify the third term on the RHS further.  
\subsubsection{Multiple-scale expansion}
By a similar power counting as in Sec.~\ref{sec:multiscales_qsap}, we conclude that we only need the first five moments of the orientational moment hierarchy to obtain all terms of fourth order in gradients in the density equation. In particular, we only need the $\mathcal{O}(\nabla^0)$ contribution of $\hat\Tv^{[4]}_{ijkm}$ to this order; this is however vanishing since, except for the decay term, all terms in its time evolution~\eqref{eq:dt_T4-qs-brownian} carry a gradient. 
Hence, we can set $\hat\Tv^{[4]}_{ijkm} = 0$ and perfom the multiple-scale analysis on Eqs.~\eqref{eq:dt_moments-brownian-0-1}, \eqref{eq:dt_Q-qs-brownian}, \eqref{eq:dt_T3-qs-brownian}, and \eqref{eq:dt_T4-qs-brownian}. 
\par 
We first make the ansatz given in Eqs.~\eqref{eq:multi-field-ansatz-qs} (with $\hat\Tv^{[4](0)}_{ijkm}   = 0$), and solve the system order by order. 
The terms of order $\mathcal{O}(1/\epslvap^2)$ give
\begin{subequations}\label{eq:multi-qs_abp-eps0}
\begin{align}    
    \hat \pol_i^{(0)} = 0 
    \label{eq:multi-pol_qs_abp-eps0}
    \,, \\
    \hat \Qv_{ij} ^{(0)} = 0
    \label{eq:multi-Q_qs_abp-eps0}
     \,, \\ 
    \hat\Tv^{[3](0)}_{ijk} = 0
    \, . 
\end{align}
\end{subequations}
At order $\mathcal{O}(1/\epslvap)$, we then obtain (using Eqs.~\eqref{eq:multi-qs_abp-eps0}): 
\begin{subequations}
\begin{align}
    \pol_i^{(1)} &= - \frac{1}{2}
    \grad_j \vdot \left(v_j \delta_{ij} \id \hat \rho^{(0)} \right)\,,
    \label{eq:multi-pol_qs_abp-eps1}
    \\
    \hat\Qv_{ij}^{(1)} &= 0\,,
    \label{eq:multi-Q_qs_abp-eps1}
    \\
    \hat\Tv^{[3](1)}_{ijk} &= 0 \,. 
    \label{eq:multi-T3_qs_abp-eps1}
\end{align}
\end{subequations}
At order $\mathcal{O}(\epslvap^0)$, the density equation~\eqref{eq:dt_density-qs-brownian} gives (substituting Eq.~\eqref{eq:multi-pol_qs_abp-eps1}): 
\begin{align}
    \epsfac \frac{\partial}{\partial t_0} \hat \rho^{(0)}
    = \frac{\etaTsq}{A^2} \nabla_{\allxs}^2 \hat \rho^{(0)} + \frac{1}{2}
    \grad_i \vdot \left[v_i \grad_i \left(v_i \hat \rho^{(0)}\right) \right]
    \, . 
    \label{eq:multi-rho_qs_abp-eps2}
\end{align}
From the polarization equation~\eqref{eq:dt_pol-qs-brownian}, we obtain, by virtue of Eqs.~\eqref{eq:multi-pol_qs_abp-eps0} and \eqref{eq:multi-Q_qs_abp-eps1}: 
\begin{align}
    \hat\pol_i^{(2)} &= - \frac{1}{2}
    \grad_j \vdot \left(v_j \delta_{ij} \id \hat \rho^{(1)} \right)
    \,,
    \label{eq:multi-pol_qs_abp-eps2}
\end{align}
and from Eqs.~\eqref{eq:dt_Q-qs-brownian}, \eqref{eq:multi-Q_qs_abp-eps0}, and \eqref{eq:multi-T3_qs_abp-eps1} the nematic tensor at this order is
\begin{align}
    \hat\Qv_{ij}^{(2)} 
    &=  - \frac{3}{16}\grad_k\vdot\left[v_k \left( \hat\pol_k^{(1)} \otimes \delta_{ij} \id   \right)^\symm \right]
    \nonumber
    \\
    &\phantom{{}={}} + \frac{1}{8}\delta_{ij} \id \grad_k\vdot \left(v_k  \hat\pol_k^{(1)}\right) 
    \, .
\end{align}
Substituting Eq.~\eqref{eq:multi-pol_qs_abp-eps1} and using that
\begin{equation}
\begin{aligned}
    &\grad_k\vdot\left[v_k \left( \hat\pol_k^{(1)} \otimes \delta_{ij} \id   \right)^\symm \right] \\
    &= \frac{1}{3} \delta_{ij} \id \grad_k \vdot \left(v_k \hat\pol_k^{(1)} \right)
    + \frac{2}{3}  \left[\grad_i\vdot\left(v_i  \hat\pol_j^{(1)} \right)\right] ^\symm
    \,,
\end{aligned}
\end{equation}
we find: 
\begin{equation}
\begin{aligned}
    \hat\Qv_{ij}^{(2)} 
    &= \frac{1}{16} \delta_{ik}\delta_{jl}\left\{\grad_k\left[v_k \grad_l\left(v_l \hat\rho^{(0)} \right)\right]\right\}^\symm
    \\
    &\phantom{=}- \frac{1}{32}\delta_{ij} \id \grad_k \vdot\left[v_k \grad_k\left(v_k  \hat\rho^{(0)}\right)\right]
    \, . 
    \label{eq:multi-Q_qs_abp-eps2}
\end{aligned}
\end{equation}
From the polarization equation~\eqref{eq:dt_pol-qs-brownian}, we obtain, by virtue of Eqs.~\eqref{eq:multi-rho_qs_abp-eps2}, \eqref{eq:multi-pol_qs_abp-eps0}, \eqref{eq:multi-pol_qs_abp-eps1}, and \eqref{eq:multi-Q_qs_abp-eps2}:
\begin{equation}
\begin{aligned}
    \hat\pol_i^{(3)} 
    &= - \delta_{ik}\frac{1}{2}
    \grad_k \vdot \left(v_k \hat \rho^{(2)} \right)
    \\
    &\phantom{=}-\frac{\etaTsq}{2A^2}\delta_{ik} \grad_k\left[\laplacian_\allxs\left(v_k \hat\rho^{(0)}\right) -v_k \laplacian_\allxs \hat\rho^{(0)}\right] 
    \\
    &\phantom{=}+\frac{1}{4}\delta_{ik}\grad_k\left\{v_k \grad_j \vdot\left[v_j \grad_j\left(v_j  \hat\rho^{(0)}\right)\right]\right\}
    \\
    &\phantom{=}-\frac{1}{16}\delta_{ik} \grad_j \left(v_j\left\{\grad_k\left[v_k \grad_j\left(v_j \hat\rho^{(0)} \right)\right]\right\}^\symm \right)
    \\
    &\phantom{=}+ \frac{1}{32}\delta_{ik}  \grad_k \left(v_k \grad_j \vdot\left[v_j \grad_j\left(v_j  \hat\rho^{(0)}\right)\right] \right)
    \, . 
    \label{eq:multi-pol_qs_abp-eps3}
\end{aligned}
\end{equation}
Substituting Eqs.~\eqref{eq:multi-pol_qs_abp-eps0}, \eqref{eq:multi-pol_qs_abp-eps1}, \eqref{eq:multi-pol_qs_abp-eps2}, and \eqref{eq:multi-pol_qs_abp-eps3} into the density Eq.~\eqref{eq:dt_density-qs-brownian}, its effective time evolution to order $\ofour$ is given by
\begin{align}
     \epsfac\partial_t &\hat\rho ={}  
    \frac{1}{2} \grad_i \vdot\left[v_i \grad_i\left(v_i \hat\rho\right)\right] 
    +\frac{\etaTsq}{A^2} \nabla_i^2 \hat\rho\nonumber
    \\
    &+\frac{\epslvap^2\etaTsq}{2A^2} \grad_i \vdot\left\{v_i \grad_i\left[\laplacian_\allxs\left(v_i \hat\rho\right) -v_i \laplacian_\allxs \hat\rho\right]\right\}\nonumber\\
    & -\epslvap^2 \left(\frac{1}{4} + \frac{1}{32}\right)\grad_i \vdot\left(v_i \grad_i\left\{v_i \grad_j \vdot\left[v_j \grad_j\left(v_j  \hat\rho\right)\right]\right\}\right)\nonumber
    \\
    &+\frac{\epslvap^2}{16} \grad_i \vdot\left(v_i \grad_j \vdot\left\{v_j\left[\grad_i(v_i \grad_j\left(v_j \hat\rho)\right)\right]^\symm\right\}\right)
    \,.\label{app:eq:ABP-O4}
\end{align}
Equation~\eqref{app:eq:ABP-O4} is the result obtained by MS analysis for ABPs that replaces Eq.~\eqref{eq:qs_all-result} for AOUPs. The $\ofour$ theory can be obtained from there by following the steps indicated in the main text; its coefficient functions, expressed in the original parameters of the model of Eq.~\eqref{eq:qs_langevin-brownian}, are given by
\begin{subequations}
\label{eq:abp-o4-coeffs_dimful}
    \begin{align}
        \mub(\rho) ={}& T \rho + \frac{\tau}{2}\int^\rho \dd{u}
        \left(\vloc \vloc' u + \vloc^2(u)\right)\,,
        \\
        K(\rho) ={}& - \frac{\gamma^2\tau}{2}\,\vloc \vloc' \rho
        - \cms \frac{T\tau^2}{2} \vloc \vloc' \rho\nonumber\\
        &{}+ \cms\frac{7}{32}\tau^3\vloc^2\tilde h(\rho)
        \,,\label{eq:abp_K}
        \\
        \lambda(\rho) ={}&  \cms \frac{T\tau^2}{2} \vloc   
        \left(\vloc'' \rho + 2\vloc'\right)\nonumber\\
        &{}+\cms \frac{\tau^3}{32} \left[\tilde g(\rho) - 7 \vloc^2\tilde h'(\rho)\right]\,,\\
        \zeta(\rho) ={}&  - \frac{\gamma^2\tau}{2}\, \vloc \vloc'  + \cms \frac{T\tau^2}{2}  \vloc'^2 \rho\nonumber\\
        &{}- \cms \frac{5}{32} \tau^3\tilde g(\rho)
        \,,
        \\
        \nu(\rho) ={}& \cms \frac{T\tau^2}{2} \vloc' \left(\vloc'' \rho + 2\vloc'\right)\nonumber\\
        &{}+ \cms \frac{\tau^3}{32} \left[\tilde g'(\rho) - 7 \vloc\vloc'\tilde h'(\rho)\right]
        \,,
    \end{align}
\end{subequations}
where $\tilde h(\rho) \coloneqq  \vloc \vloc' \rho + \vloc^2$, while $\tilde g(\rho)=\vloc\vloc'\tilde h(\rho)$ is as in the main text. The DD result corresponds to $\cms=0$ while the MS result is given by $\cms=1$. Interestingly for ABPs, and at variance with AOUPs, DD and MS theories differ in a quasi-1d state such as the one considered in the binodal calculation even at $T=0$.

We further note that the function $K(\rho)$ can change sign, due to the last term in Eq.~\eqref{eq:abp_K}, which is not always positive. This is in contrast with AOUPs (cf.~\eqref{eq:coeffs_dimful_mub}). Importantly, as long as $\vloc' < 0$ ($A>1$), $K(\rho)<0$ only between the spinodals. To see this, recall that outside of the spinodals $\mub'(\rho)=T+\frac{\tau}{2}\tilde h(\rho) > 0$. We distinguish two cases: For $\tilde h \geq 0$, it is clear that $K(\rho)>0$. The other case, $-2T/\tau< \tilde h < 0$, leads to
\begin{align}
    K(\rho) ={}&{} - \left[\frac{\gamma^2\tau}{2}+ \cms \frac{T\tau^2}{2}\right]\vloc \vloc'\rho+ \cms\frac{7}{32}\tau^3\vloc^2\tilde h(\rho)\nonumber\\
    >&{} - \left[\frac{\gamma^2\tau}{2}+ \cms \frac{T\tau^2}{2}\right]\vloc \vloc'\rho- \cms\frac{7}{16}T\tau^2\vloc^2\nonumber\\
    =&{} - \left[\frac{\gamma^2\tau}{2}+ \cms \frac{T\tau^2}{16}\right]\vloc \vloc'\rho- \cms\frac{7}{16}T\tau^2\tilde h(\rho)\nonumber\\
    >&{} - \left[\frac{\gamma^2\tau}{2}+ \cms \frac{T\tau^2}{16}\right]\vloc \vloc'\rho > 0
    \,.
\end{align}
Thus, $K(\rho)<0$ can only hold inside the spinodals, and numerical inspection reveals that the sign change occurs far from the critical point, where the MS expansion is not guaranteed to be valid. Because $K(\rho)<0$ only when $\mub(\rho)<0$, it does not incur any additional instability of a homogeneous state on top of the spinodal instability. Nevertheless, a full theory with $K(\rho)<0$ might require higher-order gradient terms, since for $K(\rho)<0$ the growth rate of the spinodal instability diverges at arbitrarily short wavelengths (cf.~Sec.~\ref{sec:linear-stability_qsap}).

Furthermore, because $K(\rho)$ changes sign, we cannot apply the numerical algorithm of App.~\ref{app:bincalc} to obtain the binodals, as it assumes that in the pseudodensity equation Eq.~\eqref{eq:psi} $K(\rho)\neq 0$ holds everywhere. We thus postpone the study of the binodals of ABPs and the physical significance of $K(\rho)$ changing sign in this model to future studies.
\section{Details on \texorpdfstring{$\ofour$}{O(grad 4)} theory of \MakeLowercase{t}QSAPs}\label{app:all_details}
\subsection{Mean-field critical point}
\label{app:ofour_qsap_crit}
\begin{figure}[tbp]
    \centering
    \includegraphics[width=\columnwidth]{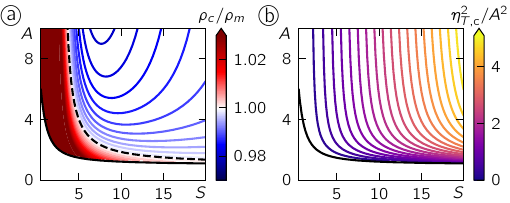}
    \caption{Critical point characterization. a) The mean-field critical density is given by $\rhoc\approx\rhom$ for large regions of parameter space (the dashed black line indicates $\rhoc=\rhom$). Under the solid black line, no phase separation is possible. b) For large $A$, the critical value of $\etaTsq$ is quadratic in $A$ while growing linearly with $S$. \label{fig:criticalpt}}
\end{figure}
In this section, we calculate the position of the mean-field critical point of tQSAPs with AOUP dynamics. To proceed, note the critical point is where the spinodal region, in which $\mub'(\rho)<0$ (leading to long-wavelength linear instability, cfr.~Eq.~\eqref{eq:mips_threshold}), disappears. The spinodal region is delimited by the spinodal lines where $\mub'(\rho)=0$; when they meet, the spinodal region disappears, so that $\mub'(\rho)\geq 0$ necessarily has a minimum, and $\mub''(\rho) = 0$. The critical point thus satisfies the following two conditions (in the dimensionless variables of Eqs.~\eqref{eq:coefficients}):
\begin{subequations}
\begin{align}
\mub'(\rhoc)&= \eta_{T,\mathrm{c}}^2 + \vloc(\vloc'\rho + \vloc)|_{\rho=\rhoc} = 0\,,\label{eq:etaTc}\\
\mub''(\rhoc) &= \vloc(3\vloc'+\vloc''\rho)+\vloc'^2\rho|_{\rho=\rhoc} = 0\label{eq:rhoc}\,.
\end{align}
\end{subequations}
The second condition gives the critical density $\rhoc$ for a given $\vloc(\rho)$ (\emph{i.e.}, for given $A$ and $S$). The first condition then gives the critical value $\eta^2_{T,\mathrm{c}}$. For a large range of $A$ and $S$, numerically solving Eq.~\eqref{eq:rhoc} shows that the critical density is $\rhoc \approx \rhom$ (see Fig.~\fref[a]{fig:criticalpt}). From Eq.~\eqref{eq:etaTc}, we then find that:
\begin{equation}
    \eta_{T,\mathrm{c}}^2 \approx A^2\left(1-\frac{A-1}{2A}\right)\left[(S+1)\frac{A-1}{2A}-1\right]\,.
\end{equation}
For large $A\gg 1$, the critical value of $\etaTsq$ grows quadratically in $A$ and linearly in $S$, $\eta_{T,\mathrm{c}}^2 \approx A^2(S-1)/4$ (see Fig.~\fref[b]{fig:criticalpt}).
\begin{figure}[tbp]
    \centering
    \includegraphics[width=\columnwidth]{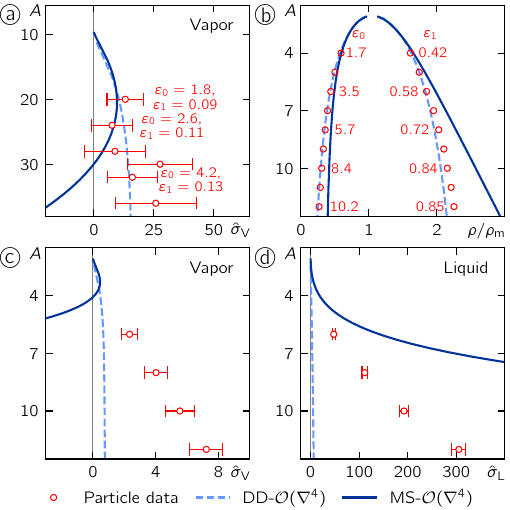}
    \caption{Failure of MS-$\ofour$ theory far from critical point. a) Vapor tension $\sigmaL$, rescaled according to Eq.~\eqref{eq:sigma_scaling_dimensions}, for $T = 1$, $v_1 = 1/8$, $\rhow = 50$, $\rhom = 200$, $\rcut = 1$, $\tau = 1$ (corresponding to varying $A$ at fixed $S = 4$, $\alphag=\tilde{\alpha}_\gamma/8$, $\etaT = 8$); see Fig.~\ref{fig:qsap_tensions_sim} for corresponding liquid tension. b) Binodal curves varying $v_0$ for $T = 1$, $v_1 = 1$, $\rhow = 5$, $\rhom = 20$, $\rcut = 1$, $\tau = 1$ (corresponding to varying $A$ at fixed $S = 4$, $\alphag=\alphagt$ and $\etaT = 1$). Estimates of $\varepsilon_{0,1}=v_{0,1}\tau/\ellr$ are shown using for $\ellr$ the interfacial width extracted from particle simulations. c,d) Vapor and liquid tensions, rescaled according to Eq.~\eqref{eq:sigma_scaling_dimensions}, for the same dimensionless parameters of panel b), with $\rhow = 50$, $\rhom = 200$.\label{fig:app_details}}
\end{figure}
\subsection{Failure of MS theory far from the critical point}
\label{app:ofour_qsap_failure}
\subsubsection{Negative vapor tension}
\label{app:ofour_qsap_tensions}
In Fig.~\fref[a]{fig:app_details}, we compare the vapor tension $\sigmaL$ obtained from particle simulations with the theoretical predictions of DD-$\ofour$ and MS-$\ofour$ theories, as a function of $A$. Unlike the liquid tension of Fig.~\ref{fig:qsap_tensions_sim}, which differs strongly between the two theories, the vapor tensions of DD-$\ofour$ and MS-$\ofour$ are very similar near the critical point. However, further away from the critical point, the two theories differ in the sign of the vapor tension, with MS-$\ofour$ notably predicting $\sigmaV<0$. While the average values for $\sigmaV$ obtained from direct measurement are positive, the associated errors suggest that longer or higher-density simulations would be needed before definitely ruling out negative values of $\sigmaV$. In any case, the region where $\sigmaV<0$ in MS-$\ofour$ theory corresponds $\epslvap \gtrsim 3$, far beyond its formal domain of validity ($\epslvap \ll 1$); the theory should not be relied upon in this regime.
\subsubsection{Binodals and tensions far from the critical point}
In Fig.~\fref[b]{fig:app_details}, we show the dependence of binodal densities on $A$ at fixed $\etaT$, $\alphag$, and $S$, choosing a larger value of $v_1 = 1$ compared to Fig.~\fref[b]{fig:qsap_binodals_quant}, where $v_1 = 1/8$; thus, $\etaT$ is smaller and $\alphag$ is larger. This results in smaller interfacial widths. Thus, the condition $\epslvap \ll 1$ under which our MS-$\ofour$ theory is valid is relegated to a small region close to the critical point. Moving away from this region, the theory fails at predicting the binodals from particle data. Interestingly, DD-$\ofour$ theory remains accurate in estimating the binodals even far from the critical point. We have no theoretical reason to explain this observation and we thus cannot rule out that this accuracy is purely accidental (holding solely for this choice of parameters).
\par
In Fig.~\fref[c,d]{fig:app_details}, we show the interfacial tensions extracted from particle simulations for the same dimensionless parameters. MS-$\ofour$ theory once more predicts a negative vapor tension $\sigmaV$, in contrast to a clearly positive $\sigmaV$. Notably, DD-$\ofour$ theory severely fails in predicting the tensions in this regime as well, significantly underestimating both tensions. This shows that the applicability of DD-$\ofour$ theory in this regime is limited to the binodals.
\section{Details on particle simulations}
\label{app:sims}
\begin{figure}[ht!]
    \centering
    \includegraphics[width=\linewidth]{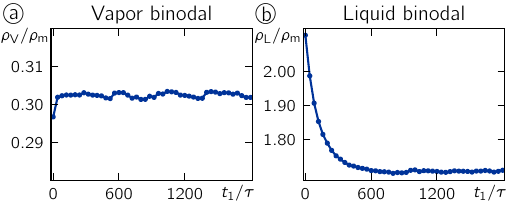}
    \caption{Example of a convergence check for the binodals extracted from particle simulations. We show the value of the a) vapor and b) liquid binodal, where each point has been computed from a time-averaged and projected interface $\expval{\rho}_{y,t}(x)$ obtained from $41$ system snapshots between $t_1$ and $t_2 = t_1 + 200$. The simulation parameters are those in Fig.~\fref[a]{fig:qsap_binodals_quant}, at $T = 0.5$ ($\etaT = \sqrt{32} \approx 5.66$).
    }
    \label{fig:time-convergence-check-qs-binodals}
\end{figure}
Below we give further details on the molecular dynamics simulations performed for tQSAPs with AOUP dynamics. We solved Eqs.~\eqref{eq:qs_langevin}, \eqref{eq:vqs}, \eqref{eq:qs_vloc}, and \eqref{eq:kernel} with a parallel code in 2d continuous space with periodic boundary conditions. 
For time-stepping, an Euler-Maruyama scheme was used with time step~$\Delta t = 0.005 \tau$ (for Fig.~\fref[a]{fig:qsap_binodals_quant}, $\Delta t = 0.001 \tau$). 
The box lengths are chosen to be $L_x =  60 \rcut, L_y = 30 \rcut$ for the binodals of the flat interface. 
To obtain the binodals for a curved interface (liquid droplet immersed in vapor), needed to estimate the Ostwald tensions in Figs.~\ref{fig:qsap_tensions_sim} and \fref[a,c,d]{fig:app_details}, we chose the box lengths to be $L_x =  L_y = 48\rcut$. 
\par 
To compute the binodal densities for a flat interface, we have initialized the system from a band with the interface extending parallel to the $y$ direction. We then carried out the following steps:
\begin{enumerate}
    \item At a fixed point $t$ in time, we compute the density of particles $\rho(x,y,t)$ on a grid of lattice spacing $\rcut$. 
    \item We then average $\rho(x,y,t)$ across the $y$ direction and time-average the interfacial profile over a time window~$[t_1, t_2]$  to obtain the one-dimensional projection $\expval{\rho}_{y,t}(x)$ of the density, the interfacial profile for the band configuration.
    The binodal densities $\rhoV$ and $\rhoL$ are obtained as the lower and upper values between which $\expval{\rho}_{y,t}(x)$ varies, respectively.
    Examples of time-averaged interfacial profiles are given in Fig.~\fref[a]{fig:qsap_RPS_sim}.
    \item We have ensured that the interfacial profile and hence the binodal estimates are stationary in time. To this end, we have varied $t_1$ and $t_2$ to verify that the choices of $t_1, t_2$ given above ensure that the binodal values have converged. In Fig.~\ref{fig:time-convergence-check-qs-binodals}, we display one example of these checks for one of the binodal pairs shown in Fig.~\fref[a]{fig:qsap_binodals_quant}. 
    \item To extract the interfacial width $\xi$ used to estimate $\varepsilon_{0,1}$, we fitted a $\tanh$ profile of the form $f(x) = \rhoV + (\rhoL - \rhoV)\left(1 + \tanh((x - x_c) / \xi)\right)/2$ to the time-averaged projected interface profiles~$\expval{\rho}_{y,t}(x)$.
    \item The same procedure has been followed to produce the averaged $x$ component of the polarization, $\expval{p_x}_{y,t}(x)$ in Fig.~\fref[a]{fig:qsap_RPS_sim}, for which we used the orientations $u_x$ of the particles.   
\end{enumerate}
For the binodals of a curved interface, the procedure is analogous except that the spatial average is now in the angle $\theta$. We also extract an estimate of the droplet radius $R$ from the interface profiles, defined as the radius at which $\expval{\rho}_{\theta,t}(R) = (\rhoL + \rhoV)/2$ holds. 
\end{document}